\documentclass[preprint,prd,tightenlines,nofootinbib,eqsecnum,superscriptaddress]{revtex4-1}

\usepackage{amsmath}
\usepackage{amssymb}
\usepackage{mathrsfs}
\usepackage{bm}
\usepackage{hyperref}
\usepackage{color}
\usepackage{mathtools}
\usepackage[normalem]{ulem}
\usepackage{soul}

\usepackage{booktabs, bigdelim, rotating}

\AtBeginDocument{
\heavyrulewidth=.08em
\lightrulewidth=.05em
\cmidrulewidth=.03em
\belowrulesep=.65ex
\belowbottomsep=0pt
\aboverulesep=.4ex
\abovetopsep=0pt
\cmidrulesep=\doublerulesep
\cmidrulekern=.5em
\defaultaddspace=.5em
}

\DeclareSymbolFontAlphabet{\mathrsfs}{rsfs}

\definecolor{CiteColor}{rgb}{0,0,0.35}
\definecolor{URLColor}{rgb}{0,0,0.35}
\hypersetup{colorlinks,citecolor=CiteColor,urlcolor=URLColor}

\allowdisplaybreaks

\newcommand{\beq}{\begin{equation}}
\newcommand{\eeq}{\end{equation}}
\newcommand{\ud}{\mathrm{d}}
\newcommand{\ui}{\mathrm{i}}
\newcommand{\ad}{\chi}
\newcommand{\pBL}{\varphi}
\newcommand{\pAdv}{\phi}
\DeclareMathOperator\const{const}

\begin{document}

\title{Tidal Love Numbers of Kerr Black Holes}

\author{Alexandre Le Tiec}
\affiliation{Laboratoire Univers et Th{\'e}ories, Observatoire de Paris, CNRS, Universit{\'e} PSL, Universit{\'e} de Paris, 92190 Meudon, France}
\affiliation{Centro Brasileiro de Pesquisas F{\'i}sicas (CBPF), Rio de Janeiro, CEP 22290-180, Brazil}

\author{Marc Casals}
\affiliation{Centro Brasileiro de Pesquisas F{\'i}sicas (CBPF), Rio de Janeiro, CEP 22290-180, Brazil}
\affiliation{School of Mathematics and Statistics, University College Dublin, Belfield, \\ Dublin 4, Ireland}

\author{Edgardo Franzin}
\affiliation{Centro Brasileiro de Pesquisas F{\'i}sicas (CBPF), Rio de Janeiro, CEP 22290-180, Brazil}
\affiliation{SISSA, International School for Advanced Studies, via Bonomea 265, 34136 Trieste, Italy}
\affiliation{INFN, Sezione di Trieste, via Valerio 2, 34127 Trieste, Italy}
\affiliation{IFPU, Institute for Fundamental Physics of the Universe, \\ via Beirut 2, 34014 Trieste, Italy}

\date{\today}

\begin{abstract}
The open question of whether a Kerr black hole can become tidally deformed or not has profound implications for fundamental physics and gravitational-wave astronomy. We consider a Kerr black hole embedded in a weak and slowly varying, but otherwise arbitrary, multipolar tidal environment. By solving the static Teukolsky equation for the gauge-invariant Weyl scalar $\psi_0$, and by reconstructing the corresponding  metric perturbation in an ingoing radiation gauge, for a general harmonic index $\ell$, we compute the linear response of a Kerr black hole to the tidal field. This linear response vanishes identically for a Schwarzschild black hole and for an axisymmetric perturbation of a spinning black hole. For a nonaxisymmetric perturbation of a spinning black hole, however, the linear response does not vanish, and it contributes to the Geroch-Hansen multipole moments of the perturbed Kerr geometry. As an application, we compute explicitly the rotational black hole tidal Love numbers that couple the induced quadrupole moments to the quadrupolar tidal fields, to linear order in the black hole spin, and we introduce the corresponding notion of tidal Love tensor. Finally, we show that those induced quadrupole moments are closely related to the well-known physical phenomenon of tidal torquing of a spinning body interacting with a tidal gravitational environment.
\end{abstract}


\maketitle

\section{Introduction}\label{sec:intro}

The existence of black holes is one of the key predictions of the general theory of relativity, and the Kerr family of solutions~\cite{Ke.63} to the Einstein field equation covers all possible stationary solutions in four-dimensional spacetime. For a Kerr black hole of mass $M$ and spin angular momentum $J$, the ``no hair'' theorem~\cite{Is.67,Ca.71,Ro.75} states that the mass-type multipole moments $M_\ell$ and the current-type multipole moments $S_\ell$  are uniquely determined as functions of $M$ and $J$ only, according to~\cite{Ge.70,Ha.74}
\beq\label{M_S}
	M_\ell + \ui S_\ell = M (\ui a)^\ell \, ,
\eeq
where $a \!\equiv\! J/M$ is the reduced angular momentum and $\ell$ a non-negative integer corresponding to the multipolar index. Realistically, however, astrophysical black holes are neither perfectly isolated nor strictly stationary. How, then, may the presence of an external tidal field  modify the remarkably simple structure \eqref{M_S} of the multipole moments of an isolated, spinning black hole in general relativity?

The tidal deformability of a self-gravitating body can be characterized by a set of so-called tidal Love numbers (TLNs)~\cite{Lov,PoWi}. Existing and future gravitational-wave measurements of TLNs in binary inspirals provide a novel way of testing compact objects (black holes and neutron stars) and general relativity in the strong-field regime~\cite{FlHi.08,Hi.08,Hi.al2.10,Ca.al.17}. The events GW170817 and GW190425 have already been used to set upper bounds on the tidal deformability of neutron stars, thereby constraining their radii and equation of state at nuclear densities~\cite{Ab.al3.17,Mo.al.18,De.al.18,Ab.al2.18,Ab.al.20}. Over the coming decades, the observation by the LISA mission~\cite{Am.al.17} of the gravitational-wave signals resulting from the inspiral of stellar mass compact objects into supermassive black holes might place constraints on the TLNs of the central object which are roughly eight orders of magnitude more stringent than current ones on neutron stars~\cite{PaMa.19}.

On the fundamental physics side, TLNs of black holes have triggered much recent work on alternatives to general relativity~\cite{Ca.al.18,CaWa.19} and scenarios with higher space dimensions~\cite{Ch.al.19,Ca.al.19,Ta.20}, quantum corrections on the horizon scale~\cite{Ad.al.19,Ma.al.19}, environmental effects due to matter~\cite{CaDu.20}, or ultracompact stars~\cite{Ch.al.20}. Black hole TLNs have been connected with the physics of Newtonian viscous fluids~\cite{Po.09} and to an emergent, near-horizon Carroll symmetry~\cite{Pe.18}.

Most importantly, several independent analyses have established that the static  TLNs of a \emph{non}-spinning black hole all vanish in four-dimensional general relativity~\cite{BiPo.09,DaNa2.09,KoSm.12,Ch.al2.13,Gu.15}. In particular, G{\"u}rlebeck~\cite{Gu.15} gave a nonperturbative proof that is valid for a static and axisymmetric tidal environment. More recently, Landry and Poisson~\cite{LaPo.15} established that the TLNs of a spinning black hole vanish as well, to linear order in spin, for a quadrupolar tidal field. Pani et al.~\cite{Pa.al.15} extended this conclusion to quadratic order in the black hole spin, for an  axisymmetric quadrupolar tidal field of electric type.

In contrast to the mentioned vanishing TLN results for black holes, in~\cite{LeCa.21}, two of the authors showed that the TLNs  do not vanish for the general case of an \emph{arbitrarily spinning} black hole immersed in a \emph{generic multipolar} static tidal field. This result appears to contradict that reached by Landry and Poisson~\cite{LaPo.15}, who found zero quadrupolar black hole TLNs to linear order in spin. Apart from the fact that Ref.~\cite{LaPo.15} uses a definition of TLNs different from ours, this disagreement is a consequence of the use of different tidal-response splits of the full physical solution, as explained in App.~\ref{app:splits}. In this paper we provide the full details of the calculation performed in Ref.~\cite{LeCa.21}.

Our analysis is performed in the context of linear black hole perturbation theory, in which a stationary and axisymmetric Kerr black hole is perturbed slightly by an applied tidal field. Moreover, following~\cite{BiPo.09} we shall restrict our analysis to parametric slowly-changing tidal fields; this means that while the tidal moments do depend on time, to reflect the changes in the external distribution of matter, the dependence is sufficiently slow that the response of the black hole presents only a \emph{parametric} dependence upon time. Conveniently, this allows us to ignore time-derivative terms in the field equations, because they are much smaller than the spatial-derivative terms, and to treat the time coordinate as an adiabatic parameter.

More precisely, we first solve the Teukolsky equation~\cite{Te.73, TePr.74} that governs linear perturbations of the modes of the curvature Weyl scalar $\psi_0$, corresponding to an external $\ell$-pole tidal perturbation of the background Kerr black hole. We solve the static (i.e., mode frequency $\omega=0$) equation for a generic $\ell\in\mathbb{R}$, and this allows us to unambiguously disentangle the contribution to $\psi_0$ that comes from the weak tidal perturbation from the contribution that corresponds to the linear response of the spinning black hole. We then apply the so-called Hertz potential formalism, originally developed for Kerr perturbations by Chrzanowksi, Cohen and Kegeles~\cite{CoKe.74,Ch.75,KeCo.79}, to construct a metric perturbation $h_{\alpha\beta}$ in an ingoing radiation gauge, through the intermediate use of a Hertz potential $\Psi$.

As a straightforward consequence of our general-$\ell$ analysis, the black hole linear response is found to vanish for a generic tidal perturbation of a Schwarzschild black hole, as well as for an axisymmetric tidal perturbation of a Kerr black hole (for an arbitrary spin), so that the corresponding contribution to the Geroch-Hansen multipole moments vanishes as well, yielding zero black hole TLNs in those two cases. This last result generalizes that previously obtained by Pani et al.~\cite{Pa.al.15}, which was restricted to a quadrupolar tidal perturbation of electric type, and to quadratic order in the black hole spin. For a \emph{non}-axisymmetric tidal perturbation of a \emph{spinning} black hole, however, the linear response is found \textit{not} to vanish, yielding in general nonzero black hole TLNs. We compute explicitly the quadrupolar TLNs of a Kerr black hole, to linear order in the spin.

Since~\cite{LeCa.21}, Ref.~\cite{Ch2.20} has  extended our result at the level of the Weyl scalar to the case of a tidal field with a ``small" nonzero $\omega$ by using the same generic-$\ell$ method that we use, thus vindicating it. (However, the author did not compute the multipole moments induced by the perturbing tidal field, so that he was not in a position to compute the general relativistic TLNs as defined by our  Eq.~\eqref{TLN}.) Reference~\cite{Ch2.20} refers to this effect as a {\it dissipative} effect (despite not actually proving that it is a purely dissipative effect) rather than calling it a TLN. Our viewpoint is that it is legitimate to refer to any  multipole moment of the body which is induced by a perturbing tidal field as a tidal gravitational deformation, i.e., as a TLN --- this is indeed the historical notion of tidal Love numbers~\cite{PoWi,FlHi.08,Hi.08,Pa.al.15}. This is the sense in which we shall use the term TLN in this paper. In fact, our definition of TLNs is precisely that introduced in Refs.~\cite{Pa.al.15,Pa.al2.15}. Regardless of the choice of terminology for what should be called a Love number, what ultimately matters is the type of physical effect that the result has on a compact binary system and on the gravitational waveform generated by its orbital motion, including determining whether it is a purely dissipative effect or not. We leave these important questions for future work.

The remainder of this paper is organized as follows. We begin by developing the Newtonian theory of TLNs in Sec.~\ref{sec:Newton}. In Sec.~\ref{sec:GR}, we give a geometrical definition of the TLNs of a generic compact object in general relativity. We then compute the curvature scalar of a Kerr black hole interacting with a weak and slowly varying tidal environment in Sec.~\ref{sec:Weyl}, and the corresponding Hertz potential in Sec.~\ref{sec:Hertz}. We reconstruct the Kerr metric perturbation from this Hertz potential in Sec.~\ref{sec:metric}, and then compute in Sec.~\ref{sec:multipoles} the Geroch-Hansen quadrupole moments associated with the black hole's response, and the corresponding nonzero TLNs, explicitly to first order in spin. Finally, in Sec.~\ref{sec:torquing} we relate those tidally-induced quadrupole moments to the tidal torquing of a spinning body interacting with a tidal environment.

Some technical aspects of this work are relegated to appendices. We recall the definition of the spin-weighted spherical harmonics in App.~\ref{app:SWSH}. In App.~\ref{app:NS_check} we consider a nonspinning black hole and relate our results, based on the Bardeen-Press equation~\cite{BaPr.73} (the Teukolsky equation for $a=0$), to those established in~\cite{DaNa2.09}, based on the Regge-Wheeler-Zerilli formalism. The delicate mathematical relation between the full physical solution and the contributions from the perturbing tidal field and the linear response of the black hole is clarified in App.~\ref{app:continuity}. The alternative tidal-response split of the full physical solution advocated in~\cite{LaPo.15} is explained in App.~\ref{app:splits}. Finally, the calculation of the perturbed scalar twist---a key ingredient entering the construction of the Geroch-Hansen multipole moments---is detailed in App.~\ref{app:twist}.

Our sign conventions are those of~\cite{Wal}. In particular, the metric signature is $(-,+,+,+)$ and we set $G = c = 1$. Greek indices $(\alpha,\beta,\gamma,\dots)$ run over all spacetime coordinates, while indices $(i,j,k,\dots)$ from the middle of the Latin alphabet run over spatial coordinates only. The letters $(a,b,c,\dots)$ denote abstract indices for tensors defined over a hypersurface $V$. In the calculations in Kerr spacetime, we use advanced Kerr coordinates $(v,r,\theta,\pAdv)$, which are regular throughout the future event horizon of the Kerr black hole. The Weyl scalar $\psi_0$ will be defined in terms of the Kinnersley tetrad~\cite{Ki.69}. We let $\mathbb{L} \equiv \mathbb{Z}^+ \setminus \{1\}$ denote the set of integers larger than one. Throughout the paper, an overbar denotes complex conjugation, we use the shorthand $\sum_{\ell m} \equiv \sum_{\ell=2}^\infty \sum_{m=-\ell}^\ell$, and $\mathbf{F}(a,b,c;z) \equiv F(a,b,c;z)/\Gamma(c)$ is the regularized hypergeometric function, $F(a,b,c;z)$ being the Gaussian hypergeometric function.

\section{Newtonian theory of tidal Love numbers}\label{sec:Newton}

In this section, we consider the problem of a Newtonian body perturbed by an external tidal field: we first review it at the level of the gravitational potential in Sec.~\ref{subsec:U_Newt}, and then extend it at the level of the curvature Weyl scalar in Sec.~\ref{subsec:psi0_Newt}, for a nonspinning, spherically symmetric body. We generalize those results to an axisymmetric, spinning body in Sec.~\ref{subsec:U-psi0_spin}.

\subsection{Gravitational potential}\label{subsec:U_Newt}

The Newtonian theory of tidally deformed bodies has been formulated by Love~\cite{Lov} and is presented in modern language in e.g.  Ref.~\cite{PoWi}. Consider an isolated and spherical Newtonian body of mass $M$ and equilibrium radius $R$, as well as a Cartesian-type coordinate system $(t,\mathbf{x})$, whose origin coincides with the center of mass. Now consider embedding this body in an external gravitational potential $U_\text{ext}(t,\mathbf{x})$. Performing a Taylor series expansion of this external potential about the origin $\bm{0}$, the tidal environment of the body can be characterized, for an arbitrary multipolar index $\ell \in \mathbb{L}$, by a set of time-dependent, symmetric and trace-free (STF) tidal moments
\beq\label{E_L}
	\mathcal{E}_L(t) \equiv - \frac{1}{(\ell-2)!} \; \partial_{\langle L \rangle} U_\text{ext} (t,\bm{0}) \, ,
\eeq
where $L \equiv i_1 \cdots i_\ell$ is a multi-index, while the brackets $\langle \cdot \rangle$ denote the STF part. Accordingly, the perturbed body develops a contribution $U_\text{body}(t,\mathbf{x})$ to the total gravitational potential, which can be characterized by a set of time-dependent, STF multipole moments,
\beq\label{eq:Newt multip}
	I_L(t) \equiv \int_{\mathbb{R}^3} \! \delta\rho(t,\mathbf{x}) \, x^{\langle L \rangle} \, \ud^3 x \, ,
\eeq
where $\delta\rho(t,\mathbf{x})$ is the mass density perturbation induced by the applied tidal field \eqref{E_L}, and we introduced the notation $x^L \!\equiv\! x^{i_1} \cdots  x^{i_\ell}$. In the center-of-mass frame, in which the dipole moment $I_i$ vanishes, the total gravitational potential $U \equiv U_\text{ext} + U_\text{body}$ thus reads~\cite{PoWi}
\beq\label{U_Newt}
	U(t,\mathbf{x}) = - \sum_{\ell = 2}^\infty \frac{(\ell-2)!}{\ell!} \, x^L \mathcal{E}_L(t) + \frac{M}{r} + \sum_{\ell = 2}^\infty \frac{(2\ell-1)!!}{\ell!} \, \frac{n^L I_L(t)}{r^{\ell+1}} \, ,
\eeq
where $r \equiv \Vert \mathbf{x} \Vert$ and $\mathbf{n} \equiv \mathbf{x}/r$, and we define $n^L \!\equiv\! n^{i_1} \cdots  n^{i_\ell}$. For a given value of the multipolar index $\ell$, the first term scales as $r^\ell$ and corresponds to the \emph{applied} $\ell$-pole tidal field $\mathcal{E}_L(t)$, while the third term scales as $r^{-(\ell+1)}$ and corresponds to the \emph{induced} $\ell$-pole moment $I_L(t)$ of the body. Let $\mathcal{R}$ denote the typical length scale of variation of the tidal environment. The tidal field \eqref{E_L} then typically scales as $U_\text{ext} / \mathcal{R}^\ell$. Consequently, the multipolar expansion of the external potential $U_\text{ext}$ [the first sum in \eqref{U_Newt}] is only valid in the domain $0 \leqslant r \ll \mathcal{R}$. On the other hand, the multipolar expansion of the body's response [the second sum in \eqref{U_Newt}] is valid for all $r \geqslant R$. Therefore, the form \eqref{U_Newt} of the total gravitational potential is valid in the region $R \leqslant r \ll \mathcal{R}$.

For a weak and slowly varying external tidal field (adiabatic approximation), the induced mass multipole moments are proportional to the applied tidal multipoles~\cite{PoWi},
\beq\label{lambda}
	I_L(t) = \lambda_\ell \, \mathcal{E}_L(t) \, ,
\eeq
where the tidal deformability parameter, or simply TLN, $\lambda_\ell$ is a real-valued constant that depends only on the internal structure of the body and that scales as $R^{2\ell+1}$. Introducing the dimensionless TLNs $k_\ell$, which are defined via the relation
\beq\label{really?}
    \lambda_\ell \equiv - \frac{2(\ell-2)!}{(2\ell-1)!!} \, k_\ell \, R^{2\ell+1} \, ,
\eeq
the Newtonian gravitational potential \eqref{U_Newt} then becomes
\beq\label{U_Newt_bis}
	U(t,\mathbf{x}) = \frac{M}{r} - \sum_{\ell = 2}^\infty \frac{(\ell-2)!}{\ell!} \, n^L \mathcal{E}_L(t) \, r^\ell \left[ 1 + 2 k_\ell \left(\frac{R}{r}\right)^{2\ell+1} \right] .
\eeq
A Newtonian body that would not develop any multipole moments \eqref{lambda} under the influence of a generic, external tidal field would thus have vanishing TLNs $k_\ell$, for all $\ell \in \mathbb{L}$.

Instead of working with STF tensors such as $\mathcal{E}_L$ and $I_L$, it may be convenient to introduce spherical polar coordinates $(\theta,\pBL)$ and to expand the scalar field $n^L \mathcal{E}_L$ over the orthonormal basis of spherical harmonics $Y_{\ell m}(\theta,\pBL)$ (see App.~\ref{app:SWSH}). Such expansion is given by
\beq\label{nLEL}
	n^L(\theta,\pBL) \mathcal{E}_L(t) = \sum_{m=-\ell}^\ell \mathcal{E}_{\ell m}(t) \, Y_{\ell m}(\theta,\pBL) \, , \quad \text{where} \quad \mathcal{E}_{\ell m}(t) = \mathcal{E}_L(t) \oint_{\mathbb{S}^2} n^L \bar{Y}_{\ell m} \, \ud\Omega \, .
\eeq
Here, we introduced the solid angle element $\ud\Omega \equiv \sin{\theta} \, \ud\theta\ud\pBL$ of the unit two-sphere $\mathbb{S}^2$. For a given $\ell \in \mathbb{L}$, the $2\ell+1$ degrees of freedom of the real-valued STF tensor $\mathcal{E}_L(t)$ are encoded in the $2\ell+1$ complex-valued multipoles $\mathcal{E}_{\ell m}(t)$, which satisfy $\mathcal{E}_{\ell \, -m} = (-)^m \bar{\mathcal{E}}_{\ell m}$. According to Eq.~\eqref{nLEL}, these multipoles are themselves linear combinations of Cartesian components of the STF tensor $\mathcal{E}_L(t)$. For instance, for $\ell = 2$ and $\ell = 3$ we have
\begin{subequations}\label{pilou}
	\begin{align}
		\mathcal{E}_{2,0}	  &= - 2\sqrt{\frac{\pi}{5}} \, (\mathcal{E}_{xx} + \mathcal{E}_{yy}) \, , \label{E_20} \\
		\mathcal{E}_{2,\pm 1} &= \mp 2 \sqrt{\frac{2\pi}{15}} \, (\mathcal{E}_{xz} \mp \ui \mathcal{E}_{yz}) \, , \\
		\mathcal{E}_{2,\pm 2} &= \sqrt{\frac{2\pi}{15}} \, (\mathcal{E}_{xx} - \mathcal{E}_{yy} \mp 2 \ui \mathcal{E}_{xy}) \, , \\
		\mathcal{E}_{3,0}	  &= - 2\sqrt{\frac{\pi}{7}} \, (\mathcal{E}_{xxz} + \mathcal{E}_{yyz}) \, , \label{E_30} \\
		\mathcal{E}_{3,\pm 1} &= \pm \sqrt{\frac{3\pi}{7}} \, (\mathcal{E}_{xxx} + \mathcal{E}_{xyy} \mp \ui (\mathcal{E}_{xxy} + \mathcal{E}_{yyy})) \, , \label{E_31} \\
		\mathcal{E}_{3,\pm 2} &= \sqrt{\frac{6\pi}{35}} \, (\mathcal{E}_{xxz} - \mathcal{E}_{yyz} \mp 2\ui \mathcal{E}_{xyz}) \, , \\
		\mathcal{E}_{3,\pm 3} &= \mp \sqrt{\frac{\pi}{35}} \, (\mathcal{E}_{xxx} - 3\mathcal{E}_{xyy} \mp \ui (3\mathcal{E}_{xxy} - \mathcal{E}_{yyy})) \, ,
	\end{align}
\end{subequations}
where we have used the STF nature of $\mathcal{E}_{ij}$ and $\mathcal{E}_{ijk}$ in order to simplify Eqs.~\eqref{E_20}, \eqref{E_30} and \eqref{E_31}. Similar formulas are easily deduced at higher multipolar orders. By substituting  Eq.~\eqref{nLEL} into \eqref{U_Newt_bis}, the Newtonian gravitational potential finally reads as
\beq\label{U_Newt_ter}
	U(t,r,\theta,\pBL) = \frac{M}{r} - \sum_{\ell m} \frac{(\ell-2)!}{\ell!} \, \mathcal{E}_{\ell m}(t) \, r^\ell \left[ 1 + 2 k_\ell \left(\frac{R}{r}\right)^{2\ell+1} \right] Y_{\ell m}(\theta,\pBL) \, .
\eeq

\subsection{Curvature scalar}\label{subsec:psi0_Newt}

Starting in Sec.~\ref{sec:Weyl}, we shall compute the TLNs of a Kerr black hole perturbed by a weak and slowly varying tidal field. This general relativistic calculation will rely on the curvature Weyl scalar $\psi_0 = C_{\alpha\beta\gamma\delta} \ell^\alpha m^\beta \ell^\gamma m^\delta$, where $C_{\alpha\beta\gamma\delta}$ is the Weyl tensor and $\ell^\alpha$ and $m^\alpha$ are two elements of a null tetrad (defined in \eqref{tetrad} below), as per the Newman-Penrose formalism~\cite{NePe.62}. In the Newtonian limit, the components of the metric with respect to the Cartesian-type coordinates $(ct,\mathbf{x})$ are given in terms of the gravitational potential $U$ according to\footnote{In this subsection we keep explicit all occurrences of the vacuum speed of light $c$ in order to clarify the calculation of the Newtonian limit of the curvature Weyl scalar $\psi_0$.}
\begin{subequations}\label{eq:g-U}
    \begin{align}
        g_{00} &= - 1 + \frac{2U}{c^2} + O(c^{-4}) \, , \\
        g_{0i} &= O(c^{-3}) \, , \\
        g_{ij} &= \delta_{ij} \left( 1 + \frac{2U}{c^2} \right) + O(c^{-4}) \, ,
    \end{align}
\end{subequations}
where $\delta_{ij}$ is the Kronecker symbol, so that $\delta_{ij} = 1$ if $i=j$ and zero otherwise. In vacuum, the components of the Weyl tensor $C_{\alpha\beta\gamma\delta}$ are then related in a simple manner to the Newtonian quadrupolar tidal field, as
\begin{subequations}\label{C_Newt}
    \begin{align}
        C_{0i0j} &= - \frac{1}{c^2} \, \partial_i \partial_j U + O(c^{-4}) \, , \\
        C_{0ijk} &= O(c^{-3}) \, , \\
        C_{ijkl} &= \frac{1}{c^2} \left( \delta_{il} \, \partial_j \partial_k U - \delta_{jl} \, \partial_i \partial_k U + \delta_{jk} \, \partial_i \partial_l U - \delta_{ik} \, \partial_j \partial_l U \right) + O(c^{-4}) \, .
    \end{align}
\end{subequations}

To compute the Weyl scalar $\psi_0 = C_{\alpha\beta\gamma\delta} \ell^\alpha m^\beta \ell^\gamma m^\delta$ in the Newtonian limit, we may simply consider the flat spacetime limit of the null vectors $\ell^\alpha$ and $m^\beta$. Using Cartesian coordinates, their components read $\ell^\alpha = (1, n^i)$ and $m^\alpha = (0,m^i)$, where we recall that $\mathbf{n} = \mathbf{x} / r$, while $\mathbf{m} = (\boldsymbol{\partial}_\theta + \ui \csc{\theta} \, \boldsymbol{\partial}_\pAdv)/(r\sqrt{2})$. Consequently, by using the Euclidean scalar products $\mathbf{n} \cdot \mathbf{n} = 1$ and $\mathbf{n} \cdot \mathbf{m} = 0 = \mathbf{m} \cdot \mathbf{m}$, the Newtonian limit of the Weyl scalar reduces to
\beq\label{psi0_Newt_def}
	\lim_{c\to\infty} c^2 \psi_0 = c^2 C_{0i0j} m^i m^j + c^2 C_{ijkl} \ell^i m^j \ell^k m^l = - 2 m^i m^j \partial_i \partial_j U = - 2 m^i m^j \nabla_i \nabla_j U \, ,
\eeq
where $\nabla_i$ is the covariant derivative compatible with the three-dimensional Euclidean metric. Since the formula \eqref{psi0_Newt_def} is covariant, we may evaluate it while using spherical coordinates $(r,\theta,\pBL)$ instead of Cartesian-type coordinates. Substituting for the expression \eqref{U_Newt_ter} of the Newtonian gravitational potential, we find
\beq\label{psi0_Newt}
	\lim_{c\to\infty} c^2 \psi_0 = - \frac{1}{r^2} \left(\partial^2_{\theta\theta} + 2\ui \csc{\theta} \, \partial^2_{\theta\pBL} - \csc^2{\theta} \, \partial^2_{\pBL\pBL} - \cot{\theta} \, \partial_\theta + 2\ui \csc'{\theta} \, \partial_\pBL\right) U 
	\equiv \lim_{c\to\infty} c^2\sum_{\ell m}\psi_0^{\ell m},
\eeq
with the modes given by
\begin{align}\label{psi0_Newt mode}
    \lim_{c\to\infty} c^2\psi_0^{\ell m}
    &= 
	\frac{(\ell-2)!}{\ell!} \, \mathcal{E}_{\ell m}(t) \, r^{\ell-2} \left[ 1 + 2 k_\ell \left(\frac{R}{r}\right)^{2\ell+1} \right] \eth_1 \eth_0  Y_{\ell m}(\theta,\pBL) \nonumber \\
	&= 
	\alpha_{\ell m}(t) \, r^{\ell-2} \left[ 1 + 2 k_\ell \left(\frac{R}{r}\right)^{2\ell+1} \right] {}_{2}Y_{\ell m}(\theta,\pBL) \, .
\end{align}
Throughout this paper we use the prime notation $f'(x) \equiv \ud f / \ud x$ to denote the derivative of any function $f(x)$ of a variable $x$. The spin-$s$ raising operator $\eth_s$ and the spin-weighted spherical harmonics ${}_{s}Y_{\ell m}$ are defined in App.~\ref{app:SWSH}.\footnote{Equation \eqref{psi0_Newt} can be rewritten in the compact and elegant form $\lim_{c\to\infty} c^2 \psi_0 = - r^{-2} \, \eth_1 \eth_0 U$.} In the last line of \eqref{psi0_Newt mode} we also introduced the time-dependent coefficients
\beq\label{alpha}
	\alpha_{\ell m}(t) \equiv \sqrt{\frac{(\ell+2)(\ell+1)}{\ell(\ell-1)}} \, \mathcal{E}_{\ell m}(t) \, .
\eeq
Despite its simplicity, and to the best of our knowledge, the connection established between the Weyl scalar and the Newtonian potential in the Newtonian limit is a new result.

Note that given the definition \eqref{nLEL} of the modes $\mathcal{E}_{\ell m}$ of the tidal tensors \eqref{E_L}, the coefficients \eqref{alpha} obey the relation $\alpha_{\ell \, -m} = (-)^m \bar{\alpha}_{\ell m}$, so that the Newtonian $\psi_0$ in Eq.~\eqref{psi0_Newt} is \emph{real}, despite the relationship \eqref{psi0_Newt_def} involving a complex $\mathbf{m}$. The first term in the square brackets in \eqref{psi0_Newt mode} corresponds to the applied $\ell$-pole tidal field, while the second term corresponds to the linear response of the central body. Far from it, in the intermediate region $R \ll r \ll \mathcal{R}$, the second term in the square brackets in Eq.~\eqref{psi0_Newt mode} can be neglected compared to the first term, so that the Newtonian Weyl scalar modes take on the asymptotic form
\beq\label{psi0_Newt_asym}
	\lim_{c\to\infty} c^2 \psi_0^{\ell m} \sim \alpha_{\ell m}(t) \, r^{\ell-2} \, {}_{2}Y_{\ell m}(\theta,\pBL) \, .
\eeq

\subsection{Extension to a spinning body}\label{subsec:U-psi0_spin}

The previous results were derived in the simplest case of a spherically symmetric, nonspinning Newtonian body. For a generic, axisymmetric, spinning body, however, the rotational flattening breaks the spherical symmetry of the unperturbed configuration. Consequently, we must generalize the proportionality relationship \eqref{lambda} between the perturbing tidal moments $\mathcal{E}_L(t)$ and the induced mass multipole moments $I_L(t)$ into a \textit{tensorial} relation of the general form
\beq\label{lambda_tensor}
    I_L(t) = \sum_{\ell' = 2}^\infty \lambda_{LL'} \, \mathcal{E}^{L'}\!(t) \, , 
\eeq
where the various constant tensors $\lambda_{LL'}$ will be referred to as the tidal deformability tensors of the spinning body, or more simply as the body's \textit{tidal Love tensors} (TLTs). The multi-indices $L = a_1 \cdots a_\ell$ and $L' = b_1 \cdots b_{\ell'}$ can possibly have different numbers $\ell$ and $\ell'$ of indices. Given that $I_L$ and $\mathcal{E}_{L'}$ are both STF tensors, we have  $\lambda_{LL'} = \lambda_{\langle L \rangle L'}$ and we only need to control $\lambda_{L \langle L' \rangle}$. Similarly to the nonspinning case, the formula \eqref{lambda_tensor} holds as long as the tidal field is weak and slowly varying. By computing the spherical-harmonic modes $I_{\ell m}$ of the induced multiple moments \eqref{lambda_tensor}, we readily obtain
\beq\label{plop}
    I_{\ell m}(t) = \sum_{\ell'm'} \lambda_{\ell m, \ell'm'} \, \mathcal{E}_{\ell'm'}(t) \, , \quad \text{where} \quad \lambda_{\ell m, \ell'm'} = \frac{4\pi\ell!}{(2\ell+1)!!} \, \lambda_{LL'} \mathscr{Y}^L_{\ell m} \bar{\mathscr{Y}}^{L'}_{\ell'm'} \, .
\eeq
As shown in App.~\ref{app:SWSH}, the constant STF tensors $\mathscr{Y}^L_{\ell m} \equiv (2\ell+1)!!/(4\pi\ell!) \oint_{\mathbb{S}^2} n^{\langle L \rangle} \bar{Y}_{\ell m} \, \ud \Omega$ can be used to expand STF tensors such as $\mathcal{E}^{L'}$ over their spherical-harmonic modes $\mathcal{E}_{\ell'm'}$, according to $\mathcal{E}^{L'} = \sum_{m'} \bar{\mathscr{Y}}^{L'}_{\ell'm'} \, \mathcal{E}_{\ell'm'}$. Importantly, contrary to the spherically symmetric case where the constants $\lambda_\ell$ are real-valued, the tidal deformability parameters, or simply TLNs, $\lambda_{\ell m, \ell'm'}$ are \textit{complex} numbers in general. Moreover, the property $\mathscr{Y}^L_{\ell,-m} = (-)^m \bar{\mathscr{Y}}^L_{\ell m}$ implies $\lambda_{\ell,-m,\ell',-m'} = (-)^{m+m'} \bar{\lambda}_{\ell m, \ell'm'}$. Given the variety of ``Love quantities" defined in this paper, we summarize them all in Tab.~\ref{tab:quantities} for the benefit of the reader.

\renewcommand{\arraystretch}{1.5}
\begin{table}[t]
 \begin{center}
 \begin{tabular}{rccccc} 
 \toprule
 \textbf{Context}  & \textbf{Name} && \textbf{Symbol} && \textbf{Equation}
 \\ [0.5ex] 
\midrule
\hspace{-1em}\ldelim\{{2}{*}[N, nonrot.]
&
  TLNs   && $\lambda_\ell \in\mathbb{R}$ && \eqref{lambda}
  \\
 &
 dimensionless TLNs && $k_\ell \in\mathbb{R}$ && \eqref{really?}
\\
\cline{2-6}
  \hspace{-1em}\ldelim\{{3}{*}[N, rot.]
  &
TLTs && $\lambda_{LL'} \in\mathbb{R}$ && \eqref{lambda_tensor}
 \\
 &
 TLNs  && $\lambda_{\ell \ell' m} \in\mathbb{C}$ && \eqref{plop}, \eqref{dis-donc}
 \\
 &
dimensionless TLNs && $k_{\ell \ell' m} \in\mathbb{C}$ && \eqref{zoom}, \eqref{dis-donc}
 \\
 \cline{2-6}
   \hspace{-1em}\ldelim\{{4}{*}[GR]
 &
TLNs && $\lambda_{\ell\ell'm}^{M/S\ \mathcal{E}/\mathcal{B}} \in\mathbb{C}$ && \eqref{TLN}
 \\
&
``Newtonian TLNs" && $k_{\ell m} \in\mathbb{C}$ && \eqref{k_phys} 
 \\
&
quadrupolar TLT && $\lambda_{ijkl} \in\mathbb{R}$ && \eqref{M=lambdaE}
 \\
 &
dim. relat. TLNs && $k_{\ell m}^{M/S\ \mathcal{E}/\mathcal{B}} \in\mathbb{C}$ && \eqref{pouet-pouet}
  \\
 \bottomrule
\end{tabular}
\end{center}
\caption{Table of the names, symbols (specifying whether it is a real or a complex quantity) and defining equations of the various ``Love quantities" used in this paper in the context of a Newtonian nonrotating body (N, nonrot.), a Newtonian rotating body (N, rot.)  and a compact body within General Relativity (GR). In  Newtonian gravity, the TLTs and the TLNs are related via \eqref{plop} and, in the case of no mode coupling its ``inverse" \eqref{TLT_exp}; these relationships also hold in GR under $\lambda_{\ell m, \ell' m'} \to \lambda_{\ell \ell' m}^{M \mathcal{E}}\delta_{\ell\ell'}\delta_{mm'}$ and $\lambda_{\ell \ell' m}^{S \mathcal{B}}\delta_{\ell\ell'}\delta_{mm'}$ and similarly for the TLTs (defined  as in \eqref{lambda_tensor} and its spin-magnetic equivalent), in the case of no mode coupling nor parity mixing, which is the case for a quadupolar ($\ell = 2$) tidal perturbation of a Kerr black hole, at least to linear order in spin. For a Kerr black hole, we conjectured that the TLNs $\lambda_{\ell\ell'm}^{M/S\ \mathcal{E}/\mathcal{B}}$ are related to the ``Newtonian TLNs" $k_{\ell m}$ via \eqref{legal!}. We proved this relation for $\ell=2$, at least to leading order in spin.}
\label{tab:quantities}
\end{table}

Substituting for Eq.~\eqref{plop} into \eqref{eq:Newt multip} yields the Newtonian gravitational potential of the tidally perturbed spinning body as
\beq\label{U_spinning}
    U = \frac{M}{r} - \sum_{\ell m} \frac{(\ell-2)!}{\ell!} \, r^\ell \left[ \mathcal{E}_{\ell m}(t) + 2 \left(\frac{R}{r}\right)^{2\ell+1} \sum_{\ell'm'} k_{\ell m, \ell'm'} \, \mathcal{E}_{\ell'm'}(t) \right] Y_{\ell m}(\theta,\pBL) \, ,
\eeq
where, following the nonspinning case, we introduced the dimensionless TLNs $k_{\ell m, \ell' m'}$, which are defined via the relationship
\beq\label{zoom}
    \lambda_{\ell m, \ell'm'} \equiv - \frac{2(\ell-2)!}{(2\ell-1)!!} \, k_{\ell m, \ell'm'} \,R^{2\ell+1} \, ,
\eeq
still involving the equilibrium radius $R$ of the body in the \textit{nonspinning} limit. Similarly, the Newtonian curvature scalar of a tidally-perturbed, spinning body reads
\beq\label{psi0_Newt_spinning}
	\lim_{c\to\infty} c^2 \psi_0 = \sum_{\ell m} r^{\ell-2} \left[ \alpha_{\ell m}(t) + 2 \left(\frac{R}{r}\right)^{2\ell+1} \sum_{\ell'm'} k_{\ell m, \ell'm'} \, \alpha_{\ell'm'}(t) \right] {}_{2}Y_{\ell m}(\theta,\pBL) \, .
\eeq

The sums over the multipolar and azimuthal numbers $\ell'$ and $m'$ in \eqref{U_spinning} and \eqref{psi0_Newt_spinning} show that, in general, the body's response involves $(\ell m)$--$(\ell'm')$ mode couplings. In fact, it can be argued on a fairly general basis that the linear response of an axisymmetric spinning body to a perturbing tidal field cannot display any $m$-mode coupling. This is immediately clear in the calculation up to the perturbed metric: only one $m$-sum of the metric perturbation can appear since we  calculate the {\it linearly} perturbed metric. While the calculation of the induced multipole moments from the (linearly) perturbed metric involves various nonlinearities which could introduce $m$-mode couplings, that calculation is only carried out to {\it linear} order in the metric perturbation and so no $m$-mode couplings may appear at any stage. Therefore, we necessarily have
\beq\label{dis-donc}
    \lambda_{\ell m, \ell' m'} = \lambda_{\ell \ell' m} \, \delta_{mm'} \quad \text{and} \quad k_{\ell m, \ell' m'} = k_{\ell \ell' m} \, \delta_{mm'} \, ,
\eeq
so that the sums over $-\ell' \leqslant m' \leqslant \ell'$ in Eqs.~\eqref{U_spinning} and \eqref{psi0_Newt_spinning} collapse.
Although in general the linear response of a spinning body will display $\ell$-mode coupling, even at linear order in the metric perturbation, we shall now focus on the special case in which it is not present, i.e., such that $k_{\ell m, \ell' m'} = k_{\ell m} \, \delta_{\ell\ell'} \, \delta_{mm'}$. This restriction will prove relevant while studying the TLNs of a Kerr black hole. Then the formulas \eqref{U_spinning} and \eqref{psi0_Newt_spinning} simply reduce to Eqs.~\eqref{U_Newt_ter} and \eqref{psi0_Newt}--\eqref{psi0_Newt mode} with $k_\ell \to k_{\ell m}$. In particular,
\beq\label{psi0_Newt_spin}
    \lim_{c\to\infty} c^2 \psi_0 = \sum_{\ell m} \alpha_{\ell m}(t)  \, r^{\ell-2} \left[ 1 + 2 k_{\ell m} \left(\frac{R}{r}\right)^{2\ell+1} \right] {}_{2}Y_{\ell m}(\theta,\pBL) \, ,
\eeq
where the complex-valued TLNs must satisfy the property $k_{\ell,-m} = \bar{k}_{\ell m}$, as a consequence of the general property $k_{\ell,-m,\ell',-m'} \!=\! (-)^{m+m'} \bar{k}_{\ell m, \ell'm'}$. Hence $k_{\ell m}$ must depend on the azimuthal number $m$ only through the imaginary multiples $\ui m$. If $\ad$ denotes a dimensionless measure of the body's rotation, with $\ad = 0$ in the nonspinning limit, it can easily be shown that the TLNs must admit a power series expansions of the type 
\beq\label{k_exp_spin}
    k_{\ell m} = k_\ell^{(0)} + \ad \sum_{n = 1}^\infty k_\ell^{(n)}(\chi) \, (\ui m)^n \, ,
\eeq
where $k_\ell^{(0)}$ are the real-valued TLNs introduced earlier in the nonspinning limit $\chi \to 0$. Here, the $m$-independent and dimensionless coefficients $k_\ell^{(n)}(\chi)$ are real-valued, smooth functions of the ``spin'' variable $\chi$. They also depend on the internal structure of the body, just like the constants $k_\ell^{(0)}$. Other definitions of the TLNs of a Newtonian, spinning body~\cite{Wa.al.17} share the feature of the degeneracy with respect to the azimuthal number $m$ being lifted as the spin increases.

Finally, let us derive a necessary condition on the TLT $\lambda_{LL'}$ for such a mode decoupling to occur. Equations \eqref{lambda_tensor} and \eqref{plop} with the condition $\lambda_{\ell m, \ell' m'} = \lambda_{\ell m} \, \delta_{\ell\ell'} \, \delta_{mm'}$ successively imply\footnote{The symbol $\mathscr{Y}_L^{\ell m}$
denotes exactly the same as $\mathscr{Y}^L_{\ell m}$: the tensorial indices $L$ are lowered with a Kronecker delta and then we raise the indices $\ell m$ just for clarity.}
\beq\label{eq:lambdaE}
     \lambda_{LL'} \, \mathcal{E}^{L'} = \sum_{m=-\ell}^\ell \bar{\mathscr{Y}}_L^{\ell m} I_{\ell m} = \sum_{m=-\ell}^\ell \lambda_{\ell m} \bar{\mathscr{Y}}_L^{\ell m} \mathcal{E}_{\ell m} = \frac{4\pi\ell!}{(2\ell+1)!!} \sum_{m=-\ell}^\ell \lambda_{\ell m} \bar{\mathscr{Y}}_L^{\ell m} \mathscr{Y}_{L'}^{\ell m} \mathcal{E}^{L'} \, .
\eeq
Since $\mathcal{E}^{L'}$ is a STF tensor, the tensor difference that multiplies it when bringing the left and right hand sides of \eqref{eq:lambdaE} together must be a linear combination of terms either antisymmetric in the indices of the multi-index $L'$ or proportional to traces with respect to those indices. By considering the STF part of that tensor difference, we thus obtain
\begin{align}\label{TLT_exp}
     \lambda_{L \langle L' \rangle} &= \frac{4\pi\ell!}{(2\ell+1)!!} \sum_{m=-\ell}^\ell \lambda_{\ell m} \bar{\mathscr{Y}}_L^{\ell m} \mathscr{Y}_{L'}^{\ell m} \nonumber \\ &= \lambda_\ell^{(0)} \delta_{LL'} + \ad \sum_{n=1}^\infty \, (-)^n \! \left[ \lambda_{\ell}^{(2n-1)}(\chi) \, I^{(2n-1)}_{LL'} + \lambda_{\ell}^{(2n)}(\chi) \, R^{(2n)}_{LL'} \right] ,
\end{align}
where we used an expansion of $\lambda_{\ell m}$ akin to Eq.~\eqref{k_exp_spin}, using some functions $\lambda_{\ell}^{(n)}(\chi)$ and constant $\lambda_{\ell}^{(0)}$, and we used $\sum_{m=-\ell}^\ell \bar{\mathscr{Y}}_L^{\ell m} \mathscr{Y}_{L'}^{\ell m} = \frac{(2\ell+1)!!}{4\pi\ell!} \delta_{LL'}$ in the second equality. We recover the proportionality relation \eqref{lambda} in the nonspinning limit. Equation \eqref{plop} gives the TLNs in terms of the TLTs; similarly, as long as there are no mode couplings, Eq.~\eqref{TLT_exp} gives the TLTs in terms of the TLNs. 

Remarkably, the tensor structure of the tidal deformability tensor \eqref{TLT_exp} is encoded in the two families of real-valued tensors
\begin{subequations}\label{STF_tensors}
    \begin{align}
        R^{(2n)}_{LL'} &\equiv \frac{8\pi\ell!}{(2\ell+1)!!} \sum_{m = 1}^\ell m^{2n} \, \Re{\left( \bar{\mathscr{Y}}_L^{\ell m} \mathscr{Y}_{L'}^{\ell m} \right)} , \label{R's} \\
        I^{(2n-1)}_{LL'} &\equiv \frac{8\pi\ell!}{(2\ell+1)!!} \sum_{m = 1}^\ell m^{2n-1} \, \Im{\left( \bar{\mathscr{Y}}_L^{\ell m} \mathscr{Y}_{L'}^{\ell m} \right)} . \label{I's}
    \end{align}
\end{subequations}
Notice that these tensors are symmetric and antisymmetric by exchange of the multi-indices $L$ and $L'$, respectively. These properties trigger the following observation on the TLTs \eqref{TLT_exp} of two special classes of tidally perturbed, spinning, Newtonian bodies:
\beq\label{eq:symm TLN}
    \begin{dcases}
       \lambda_{L' \langle L \rangle} = - \lambda_{L \langle L' \rangle} & \mbox{if } \lambda_\ell^{(2n)} = 0 \mbox{ for all } n \in \mathbb{N} \, , \\
        \lambda_{L' \langle L \rangle} = \lambda_{L \langle L' \rangle} & \mbox{if } \lambda_\ell^{(2n-1)} = 0  \mbox{ for all } n \in \mathbb{N}^* \, .
    \end{dcases}
\eeq

The algebraic reduction of the tensor product of the STF tensors $\bar{\mathscr{Y}}_L^{\ell m}$ and $\mathscr{Y}_{L'}^{\ell m}$ in \eqref{STF_tensors} shows that $R^{(2n)}_{LL'}$ and $I^{(2n-1)}_{LL'}$ can only depend on the Kronecker symbol $\delta_{ij}$, the totally anti-symmetric Levi-Civita symbol $\varepsilon_{ijk}$ (with $\varepsilon_{123} = +1$) and the unit spatial direction $s^i$ of the body's spin~\cite{BlDa.86}. For instance, for $\ell=2$ we have
\begin{subequations}\label{yesmyfriend}
    \begin{align}
        I^{(1)}_{ijkl} &= 2 \delta_{(i|\langle k} \, \varepsilon_{l \rangle|j)q} s^q \, , \label{I1} \\
        I^{(3)}_{ijkl} &= 8 \delta_{(i|\langle k} \, \varepsilon_{l \rangle|j)q} s^q - 6 s_{(i} \varepsilon_{j) q \langle k} s_{l \rangle} s^q \, , \label{I3} \\
        I^{(5)}_{ijkl} &= 32 \delta_{(i|\langle k} \, \varepsilon_{l \rangle|j)q} s^q - 30  s_{(i} \varepsilon_{j) q \langle k} s_{l \rangle} s^q \, . \label{I5}
    \end{align}
\end{subequations}

In Secs.~\ref{sec:multipoles} and \ref{sec:torquing} we will compute the quadrupolar tidal Love tensors of a Kerr black hole. It will then prove convenient to have in hand explicit expressions for the tensors \eqref{STF_tensors} in the quadrupolar case ($\ell=\ell'=2$), in matrix form. Introducing the four STF matrices
\begin{align}\label{matrices}
    \mathbf{M}_{11} \equiv
        \left( \begin{array}{ccc}
            1 & 0 & 0 \\
            0 & -1 & 0 \\
            0 & 0 & 0 \\
        \end{array} \right) , \;
    \mathbf{M}_{12} \equiv
        \left( \begin{array}{ccc}
            0 & 1 & 0 \\
            1 & 0 & 0 \\
            0 & 0 & 0 \\
        \end{array} \right) , \;
    \mathbf{M}_{13} \equiv 
        \left( \begin{array}{ccc}
            0 & 0 & 1 \\
            0 & 0 & 0 \\
            1 & 0 & 0 \\
        \end{array} \right) , \;
    \mathbf{M}_{23} \equiv 
        \left( \begin{array}{ccc}
            0 & 0 & 0 \\
            0 & 0 & 1 \\
            0 & 1 & 0 \\
        \end{array} \right) ,
\end{align}
we have $(\bar{\mathscr{Y}}_{ij}^{21}) \!=\! - \sqrt{15/(32\pi)} \, (\mathbf{M}_{13} + \ui \mathbf{M}_{23})$ and $(\bar{\mathscr{Y}}_{ij}^{22}) \!=\! \sqrt{15/(32\pi)} \, (\mathbf{M}_{11} + \ui \mathbf{M}_{12})$. For any integer $n \in \mathbb{N}$, the tensors \eqref{STF_tensors} with $L=ij$ and $L'=kl$ can then be expressed in terms of the STF matrices \eqref{matrices} according to
\begin{subequations}\label{TLT_matrices}
\begin{align}
\Big( R^{(2n)}_{ijkl} \Bigr) &= \frac{1}{2}
\left(
\begin{array}{ccc}
 2^{2n} \, \mathbf{M}_{11} & 2^{2n} \, \mathbf{M}_{12} & \mathbf{M}_{13} \\
 2^{2n} \, \mathbf{M}_{12} & - 2^{2n} \, \mathbf{M}_{11} & \mathbf{M}_{23} \\
 \mathbf{M}_{13} & \mathbf{M}_{23} & \mathbf{0} \\
\end{array}
\right) , \label{TLT_matrix_Re} \\
\Bigl( I^{(2n-1)}_{ijkl} \Bigr) &= 
\frac{1}{2}
\left(
\begin{array}{ccc}
 - 2^{2n-1} \, \mathbf{M}_{12} & 2^{2n-1} \, \mathbf{M}_{11} & - \mathbf{M}_{23} \\
 2^{2n-1} \, \mathbf{M}_{11} & 2^{2n-1} \, \mathbf{M}_{12} & \mathbf{M}_{13} \\
 - \mathbf{M}_{23} & \mathbf{M}_{13} & \mathbf{0} \\
\end{array}
\right) . \label{TLT_matrix_Im}
\end{align}
\end{subequations}
In order to avoid any misunderstandings, we note that the component $ijkl$ corresponds to the element in the $k$-th row and $l$-th column of the $3 \times 3$ matrix which is in the $i$-th row and $j$-th column of the larger matrix in Eqs.~\eqref{TLT_matrices}.

We observe \vspace{-0.13cm} based on \eqref{TLT_matrices} that only two tensors out of all the members $\{ R_{ijkl}^{(2n)}, n \in \mathbb{N}^* \}$ are linearly independent, and similarly for the tensors $\{ I_{ijkl}^{(2n-1)}, n \in \mathbb{N}^* \}$, as can be noticed in the particular case of the formulas \eqref{yesmyfriend}.

\section{General relativistic tidal Love numbers}\label{sec:GR}

We now move to the problem of a generic compact object perturbed by a weak and slowly varying tidal field, in the context of general relativity. We start by discussing in Sec.~\ref{subsec:split} how the physical contribution of the external perturbing tidal field can be disentangled from that of the linear response of the central object. We then recall the operational definition of the Geroch-Hansen multipole moments of a stationary spacetime in Sec.~\ref{subsec:multipoles}, and the description of the tidal environment of a compact object in Sec.~\ref{subsec:environment}, out of which we define the TLNs of the perturbed compact body in Sec.~\ref{subsec:defTLN}.

\subsection{Tidal field and linear response}\label{subsec:split}

In Newtonian gravity, the geometrical structure provided by the 3-dimensional Euclidean space, as well as the linearity of the Poisson equation, allows one to identify unambiguously the piece of the gravitational potential that corresponds to the applied tidal field from the piece (proportional to the TLNs $k_\ell$) that corresponds to the response of the central body. As can be read off from Eq.~\eqref{U_Newt_ter}, the former scales as $r^\ell$, while the latter scales as $r^{-(\ell+1)}$. In general relativity, however, the question of whether such an unambiguous split can be performed or not is much more subtle. In particular, Gralla~\cite{Gr.18} (see also Ref.~\cite{FaLo.05}) argued that the concept of TLNs of a compact object is intrinsically coordinate dependent. This can most easily be understood by focusing, for instance, on a quadrupolar ($\ell=2$) tidal field in Eq.~\eqref{U_Newt_ter}. A radial coordinate transformation of the form $r \to r \, [1 + C (M/r)^5]$, with $C$ a constant, would leave the leading behavior of the quadrupolar tidal field unchanged, but it would shift the value of the quadrupolar TLN according to
\beq
	k_2 \to k_2 - C \left(\frac{M}{R}\right)^5 \, .
\eeq

However, as argued by Binnington and Poisson~\cite{BiPo.09}, TLNs can be defined in an invariant manner for \emph{infinitesimal} coordinate (i.e.\ perturbative gauge) transformations. More importantly, as argued by the authors of Refs.~\cite{KoSm.12,Pa.al.15,Ca.al.19}, the issue of dependence of TLNs under a \emph{generic} (i.e.\ not necessarily infinitesimal) coordinate transformation can be circumvented by performing an analytic continuation in the number $d$ of spacetime dimensions. Let $h_{\alpha\beta}$ denote the linear perturbation to the background metric of the compact object, induced by an external, slowly varying, quadrupolar ($\ell = 2$) tidal field. By focusing on the asymptotic behavior of the relevant coordinate component of the metric perturbation, say $h_{00}$ as $r \to \infty$, we then have, schematically,
\beq\label{h00_dim}
	h_{00} \sim \alpha \, r^{2} \left( 1 + \cdots \right) + \beta \, r^{-d+1} \left( 1 + \cdots \right) ,
\eeq
where the dots represent series expansions in powers of $M/r$ (relativistic corrections), which may also include logarithms, while $\alpha$ and $\beta$ are $r$-independent quantities. Because a generic coordinate transformation cannot depend on the number $d$ of spacetime dimensions, the form \eqref{h00_dim} of the metric perturbation provides a unique, unambiguous identification  of the applied quadrupolar tidal field [first term in Eq.~\eqref{h00_dim}] and the linear response of the central body [second term in Eq.~\eqref{h00_dim}].

Alternatively, one might consider an \emph{arbitrary} multipolar index $\ell \in \mathbb{L}$ in $d = 4$ spacetime dimensions~\cite{BiPo.09,DaNa2.09}, and perform an analytic continuation\footnote{We note that taking $\ell\in\mathbb{R}$ is a  mathematical trick which is not uncommon  within black hole perturbation literature, such as in the Appendix of Ref.~\cite{Pa.76}.} to $\ell \in \mathbb{R}$, so that, schematically, each mode goes like
\beq\label{h00}
	h_{00} \sim \alpha \, r^{\ell} \left( 1 + \cdots \right) + \beta \, r^{-\ell-1} \left( 1 + \cdots \right) .
\eeq
Since a generic coordinate transformation cannot depend on the multipolar index $\ell$, the form \eqref{h00} of the metric perturbation provides a unique, unambiguous identification of the applied tidal field [first term in Eq.~\eqref{h00}] and the linear response of the central body [second term in Eq.~\eqref{h00}]. Once the contribution of the tidal field has been identified, it can be subtracted off the metric perturbation $h_{\alpha\beta}$. The remaining metric being a stationary, asympotically flat, source-free solution of the linearized Einstein equation, its Geroch-Hansen multipole moments can be extracted at spatial infinity.

\subsection{Multipole moments of a stationary spacetime}\label{subsec:multipoles}

Several equivalent definitions of the multipole moments of a stationary spacetime have been given, e.g. in~\cite{Ha.74,Th.80,SiBe.83}. Here we shall review the definition introduced by Hansen~\cite{Ha.74}, which generalizes that given earlier by Geroch~\cite{Ge.70} for a static spacetime. We thus consider a stationary spacetime $(\mathcal{M},g_{\alpha\beta})$ with a timelike Killing field $\xi^\alpha$, and let $\lambda \equiv - \xi^\alpha \xi_\alpha$ denote its norm.\footnote{The Killing field $\xi^\alpha$ is specified uniquely by the requirement that $\lambda \to 1$ at spatial infinity. It is timelike everywhere, except possibly close to the central body.} In vacuum general relativity, $R_{\alpha\beta} = 0$, the Bianchi identity and Killing's equation can be combined to establish that the twist $\omega^\alpha \equiv \varepsilon^{\alpha\beta\gamma\delta} \xi_\beta \nabla_\gamma \xi_\delta$ derives from a scalar potential, in the sense that there exists a scalar field $\omega$ such that $\omega_\alpha = \nabla_\alpha \omega$~\cite{Wal}. Here,  $\varepsilon_{\alpha\beta\gamma\delta}$ and $\nabla_\alpha$ are, respectively, the canonical volume form and covariant derivative associated with $g_{\alpha\beta}$.

On any spacelike hypersurface $V$ orthogonal to $\xi^\alpha$, the spacetime metric $g_{\alpha\beta}$ induces a positive-definite 3-metric
\beq\label{gamma}
    \gamma_{ab} \equiv \lambda g_{ab} + \xi_a \xi_b \, .
\eeq
To define the Geroch-Hansen multipole moments of $(\mathcal{M},g_{\alpha\beta})$, the hypersurface $V$ is required to be asymptotically flat, in the sense that there exists a point $\Lambda$, a manifold $\tilde{V}$, as well as a conformal factor $\Omega$, such that
\begin{itemize}
    \item[(i)] $\tilde{V} = V \cup \Lambda$,
    \item[(ii)] $\tilde{\gamma}_{ab} = \Omega^2 \gamma_{ab}$ is a smooth metric on $\tilde{V}$,
    \item[(iii)] $\Omega|_\Lambda = 0$, $\tilde{D}_a \Omega|_\Lambda = 0$ and $\tilde{D}_a \tilde{D}_b \Omega|_\Lambda = 2 \tilde{\gamma}_{ab}|_\Lambda$,
\end{itemize}
where $\tilde{D}_a$ is the covariant derivative in $\tilde{V}$ associated with the metric $\tilde{\gamma}_{ab}$. We then introduce a complex scalar potential $\Phi$, defined out of the norm $\lambda$ and twist $\omega$ of the timelike Killing field $\xi^\alpha$, according to
\beq\label{Phi}
    \Phi \equiv \Phi_M + \ui \Phi_S \, , \quad \text{where} \quad \Phi_M \equiv \frac{1 - \lambda^2 - \omega^2}{4\lambda} \quad \text{and} \quad \Phi_S \equiv \frac{\omega}{2\lambda} \, .
\eeq
The complex multipole moments ($P, P_a, P_{ab}, P_{abc}, \dots$) are defined in a recursive manner, out of the monopole $P \equiv \tilde{\Phi} \equiv \Phi / \sqrt{\Omega}$, through the formula
\beq\label{P's}
    P_{a_1 \cdots a_\ell} = \tilde{D}_{\langle a_1} P_{a_2 \cdots a_\ell \rangle} - \frac{1}{2} \, (\ell-1)(2\ell-3) \tilde{R}_{\langle a_1 a_2} \, P_{a_3 \cdots a_\ell \rangle} \, ,
\eeq
where $\tilde{R}_{ab}$ is the Ricci tensor on $\tilde{V}$ and we recall that the angular brackets $\langle \cdot \rangle$ denote the STF part of a multi-index $L = a_1 \cdots a_\ell$. The mass-type and current-type multipole moments $M_L$ and $S_L$ are then given by the real and imaginary parts of $P_L$ evaluated at $\Lambda$, respectively:\footnote{To ensure that the normalization of the general relativistic multipole moments $M_L$ in \eqref{ML_SL_PL} is consistent with that of the Newtonian multipole moments $I_L$ in \eqref{eq:Newt multip}, we introduced an extra factor of $1/(2\ell-1)!!$ with respect to the conventional definition given in Refs.~\cite{Ge.70,Ha.74}.}
\beq\label{ML_SL_PL}
    M_L + \ui S_L = \frac{P_L|_\Lambda}{(2\ell-1)!!} \, .
\eeq
The STF multipole moments $M_L$ and $S_L$ so-defined are manifestly coordinate-independent. It can be shown that the degree of freedom in the choice of the conformal factor $\Omega$ reflects the choice of an origin ``at infinity,'' with respect to which the multipole moments \eqref{ML_SL_PL} are taken~\cite{Ge.70,Ha.74}. Other choices of generating potential $\Phi$ that yield the same multipole moments $M_L$ and $S_L$ are also possible~\cite{SiBe.83}.

Following the Newtonian calculations carried out in Sec.~\ref{sec:Newton} above, we find it convenient to introduce the complex-valued, spherical-harmonic modes $M_{\ell m}$ and $S_{\ell m}$ of the mass-type and current-type multipole moments \eqref{ML_SL_PL}, defined according to
\beq\label{MlmSlm}
    M_{\ell m} \equiv M_L \oint_{\mathbb{S}^2} n^L \bar{Y}_{\ell m} \, \ud\Omega \quad \text{and} \quad S_{\ell m} \equiv S_L \oint_{\mathbb{S}^2} n^L \bar{Y}_{\ell m} \, \ud\Omega \, ,
\eeq
where $n^a$ in \eqref{MlmSlm} is understood to be evaluated at the point $\Lambda$, where it is the unit radial direction such that $\tilde{\gamma}_{ab} n^a n^b|_\Lambda = 1$.

\subsection{Tidal environment of a compact object}\label{subsec:environment}

Next, we provide a short overview of the description of the tidal environment of a generic compact object, as extensively developed by Poisson and his collaborators~\cite{Po2.04,PoVl.10,Ch.al.13,Po4.15} in the case of a black hole. When viewed on the large scale $\mathcal{R}$ of variation of the tidal environment, the compact object occupies a small region in the physical spacetime, which can be idealized as a timelike worldline $\gamma$. That is, as viewed from the perspective of the far region $r\gg M$, the compact object reduces to a worldline $\gamma$ in the external gravitational field characterized by tidal moments; in that region, the Weyl tensor $C_{\alpha\beta\gamma\delta}$ of the full physical spacetime is approximately equal to the Weyl tensor of the tidal field. This approximately ``external'' Weyl tensor is then matched in the region $M \ll r \ll \mathcal{R}$ to the Weyl tensor of the perturbed compact object. (This idea can be made precise and rigorous by using matched asymptotic expansions~\cite{Po.al.11}.) Let $u^\alpha$ denote the normalized tangent vector to $\gamma$. This vector is parallel-transported along $\gamma$ up to a good degree of approximation~\cite{Po.al.11}. Additionally, let $(e^\alpha_i)$ with $i \in \{1,2,3\}$ denote a set of three unit spacelike vectors defined along $\gamma$, orthogonal to $u^\alpha$ and to each other, and parallel-transported along $\gamma$. The orthonormal tetrad $(u^\alpha,e_i^\alpha)$ then defines a reference frame in a neighborhood of $\gamma$, known as the ``local asymptotic rest frame'' of the compact object.

In vacuum general relativity, the tidal environment of the compact body is fully characterized by the Weyl tensor $C_{\alpha\beta\gamma\delta}$. The local asymptotic rest frame components of the covariant derivatives of the Weyl tensor are simply given by projecting the covariant derivatives of the Weyl tensor, evaluated along $\gamma$, with respect to the tetrad $(u^\alpha,e_i^\alpha)$ . For instance, we write 
\begin{subequations}\label{eq:Weyl world}
	\begin{align}
		C_{0i0j}(v)  &\equiv u^\alpha e^\beta_i u^\gamma e^\delta_j\, C_{\alpha\beta\gamma\delta}(\gamma) \, , \\
		C_{0ijk;l}(v) &\equiv u^\alpha e^\beta_i e^\gamma_j e^\delta_k e^\lambda_l\, \nabla_\lambda C_{\alpha\beta\gamma\delta}(\gamma) \, ,
	\end{align}
\end{subequations}
and similarly for other frame components, where $v$ denotes the proper time elapsed along $\gamma$. In the black hole case, $v$ can be identified, on the event horizon of the black hole, with the advanced time coordinate in Kerr coordinates. The information on the tidal environment of the compact body can then be encoded in two sets of Cartesian-type STF tensors associated with the local asymptotic rest frame components of the (covariant derivatives of the) Weyl tensor, namely the electric-type and magnetic-type tidal tensors~\cite{Zh.86,BiPo.09,Pa.al.15}
\begin{subequations}\label{EL_BL_GR}
	\begin{align}
		\mathcal{E}_L(v) &\equiv {[(\ell-2)!]}^{-1} \, C_{0 \langle i_1 |0| i_2;i_3 \cdots i_\ell \rangle} (v) \, , \label{E_L_GR} \\
		\mathcal{B}_L(v) &\equiv \frac{3}{2} \, {[(\ell+1)(\ell-2)!]}^{-1} \, \varepsilon_{jk\langle i_1} C_{i_2|0jk|;i_3 \cdots i_\ell \rangle}(v) \, . \label{B_L}
	\end{align}
\end{subequations}
The electric-type tidal tensors \eqref{E_L_GR} are the general relativistic analogues of the Newtonian tidal tensors \eqref{E_L} introduced in Sec.~\ref{sec:Newton}. The magnetic-type tidal tensors \eqref{B_L} have no counterpart in Newtonian gravity, as they carry an additional factor of $1/c$ [recall \eqref{C_Newt}].

Following the Newtonian calculations performed in Sec.~\ref{sec:Newton}, we find it convenient to introduce the complex-valued, spherical-harmonic modes $\mathcal{E}_{\ell m}(v)$ and $\mathcal{B}_{\ell m}(v)$ of the tidal tensors \eqref{EL_BL_GR}, defined according to
\beq\label{ElmBlm}
    \mathcal{E}_{\ell m}(v) \equiv \mathcal{E}_L(v) \oint_{\mathbb{S}^2} n^L \bar{Y}_{\ell m} \, \ud\Omega \quad \text{and} \quad \mathcal{B}_{\ell m}(v) \equiv \mathcal{B}_L(v) \oint_{\mathbb{S}^2} n^L \bar{Y}_{\ell m} \, \ud\Omega \, ,
\eeq
where $\mathbb{S}^2$ now denotes the unit 2-sphere in the body's local asymptotic rest frame at any point along the worldline $\gamma$, covered by spherical polar coordinates $(\theta,\phi)$, with $\ud\Omega = \sin{\theta} \, \ud\theta\ud\phi$, and $n^i = (\sin{\theta} \cos{\phi}, \sin{\theta} \sin{\phi}, \cos{\theta})$ is the unit radial direction.

\subsection{Tidal Love numbers of a compact object}\label{subsec:defTLN}

The TLNs of a compact body perturbed by a weak tidal field can now be defined as follows. Let $\mathcal{E}_L$ and $\mathcal{B}_L$ denote the perturbing electric-type and magnetic-type tidal fields. Let also $M_L$ and $S_L$ denote the mass-type and current-type Geroch-Hansen multipole moments of the perturbed spacetime geometry, obtained after having subtracted the contribution of the applied tidal field from the metric perturbation $h_{\alpha\beta}$. Finally, let $M_{\ell m}$, $S_{\ell m}$, $\mathcal{E}_{\ell m}$ and $\mathcal{B}_{\ell m}$ denote the spherical-harmonic modes of the STF tensors $M_L$, $S_L$, $\mathcal{E}_L$ and $\mathcal{B}_L$, respectively, following a multipolar decomposition over spherical harmonics, as per Eqs.~\eqref{MlmSlm} and \eqref{ElmBlm}. Then, four discrete families of TLNs can be defined as the rates of change of $M_{\ell m}$ and $S_{\ell m}$ under small variations of $\mathcal{E}_{\ell'm}$ and $\mathcal{B}_{\ell'm}$, according to\footnote{These four families of TLNs coincide with the TLNs $\lambda^{(\ell\ell'm)}_{E,+}$, $\lambda^{(\ell\ell'm)}_{E,-}$, $\lambda^{(\ell\ell'm)}_{M,-}$ and $\lambda^{(\ell\ell'm)}_{M,+}$ in Eqs.~(2)--(3) of Ref.~\cite{Pa.al2.15}.} [recall e.g. Eqs.~\eqref{plop} and \eqref{dis-donc}]
\beq\label{TLN}
    \lambda_{\ell\ell'm}^{M\mathcal{E}} \equiv \frac{\partial M_{\ell m}}{\partial \mathcal{E}_{\ell'm}} \, , \quad \lambda_{\ell\ell'm}^{M\mathcal{B}} \equiv \frac{\partial M_{\ell m}}{\partial \mathcal{B}_{\ell'm}} \, , \quad \lambda_{\ell\ell'm}^{S\mathcal{E}} \equiv \frac{\partial S_{\ell m}}{\partial \mathcal{E}_{\ell'm}} \, , \quad \lambda_{\ell\ell'm}^{S\mathcal{B}} \equiv \frac{\partial S_{\ell m}}{\partial \mathcal{B}_{\ell'm}} \, .
\eeq
Indeed, as argued in~\cite{Pa.al2.15} and in Sec.~\ref{subsec:U-psi0_spin} above, given the axisymmetry of the unperturbed configuration, the linear response of the compact object shares the same azimuthal behavior as the perturbing tidal field, which implies no $m$-mode coupling. On the other hand, the selection rules for $\ell$-mode couplings are extensively discussed in Refs.~\cite{Pa.13,Pa.al.15,Pa.al2.15}.

The TLNs in Eq.~\eqref{TLN} are clearly defined  in a geometrical, coordinate-invariant manner. However, the tidal tensors introduced in Eq.~\eqref{EL_BL_GR} are frame-dependant, as they depend on the choice of a tetrad $(e_0^\alpha,e_i^\alpha)$ defined along the worldline $\gamma$ associated with the central compact object. The timelike leg $e_0^\alpha = u^\alpha$ is geometrically defined as the unit tangent to $\gamma$, but there remains the freedom to perform an internal rotation of the triad $(e_i^\alpha)$. If the unperturbed compact object is axisymmetric, however, the rotational Killing vector of the background spacetime can be used to define an axis of symmetry, with respect to which the spherical-harmonic modes \eqref{ElmBlm}  of the perturbing tidal fields \eqref{EL_BL_GR} can be taken unambiguously.

Importantly, the coordinate-invariant, geometrical definition \eqref{TLN} of the TLNs of a compact object relies on the unambiguous split \eqref{h00} between the perturbing external tidal field and the corresponding linear response of the central body. As we are working to linear order in the perturbing tidal fields $\mathcal{E}_L$ and $\mathcal{B}_L$, the coefficients \eqref{TLN} are constants. Thus, if $\mathring{M}_{\ell m}$ and $\mathring{S}_{\ell m}$ denote the mass-type and current-type multipole moments of the background geometry, respectively, then the Geroch-Hansen multipole moments of the perturbed spacetime must necessarily take the general form\footnote{For the axisymmetric Kerr black hole solution, a nonvanishing multipole moment necessarily has $m = 0$, with $\mathring{M}_{\ell 0}$ and $\mathring{S}_{\ell 0}$ proportional to the multipole moments appearing in the Hansen formula \eqref{M_S}.}
\begin{subequations}\label{MLSL}
    \begin{align}
        M_{\ell m} &= \mathring{M}_{\ell m} + \sum_{\ell'=2}^\infty \left( \lambda_{\ell\ell'm}^{M\mathcal{E}} \, \mathcal{E}_{\ell'm} + \lambda_{\ell\ell'm}^{M\mathcal{B}} \,\mathcal{B}_{\ell'm} \right) \equiv \mathring{M}_{\ell m} + \delta M_{\ell m} \, , \\
        S_{\ell m} &= \mathring{S}_{\ell m} + \sum_{\ell'=2}^\infty \left( \lambda_{\ell\ell'm}^{S\mathcal{E}} \, \mathcal{E}_{\ell'm} + \lambda_{\ell\ell'm}^{S\mathcal{B}} \,\mathcal{B}_{\ell'm} \right) \equiv \mathring{S}_{\ell m} + \delta S_{\ell m} \, .
    \end{align}
\end{subequations}
As a consequence of the definition \eqref{MlmSlm}, the modes $M_{\ell m}$ of the mass-type multipole moments $M_L$ obey the general property $M_{\ell\,-m} \!=\! (-)^m \bar{M}_{\ell m}$, and similarly for the modes $S_{\ell m}$, $\mathcal{E}_{\ell m}$, $\mathcal{B}_{\ell m}$. Consequently, the definitions \eqref{TLN} of the TLNs combined with the expressions \eqref{MLSL} imply the important property
\beq\label{lambda_sym}
    \lambda_{\ell\ell' \, -m}^{M\mathcal{E}} = \bar{\lambda}_{\ell\ell'm}^{M\mathcal{E}} \, ,
\eeq
and similarly for the $M\mathcal{B}$, $S\mathcal{E}$ and $S\mathcal{B}$ couplings.

The general strategy summarized above was suggested by Pani et al.~\cite{Pa.al.15,Pa.al2.15}, who have developed a formalism to compute the TLNs of slowly spinning compact objects (black holes and neutron stars), up to second order in the spin angular momentum. Unfortunately, the mode couplings that occur in the metric perturbation equations prevented the authors from computing the metric perturbation of a given slowly rotating compact object for an arbitrary multipolar index $\ell \in \mathbb{L}$; their quadratic-in-spin analysis was thus restricted to a quadrupolar ($\ell = 2$), axisymmetric, pure electric-type tidal field. Landry and Poisson~\cite{LaPo.15} also performed a complementary analysis that includes nonaxisymmetric and magnetic-type tidal fields as well, still for $\ell = 2$, but which is restricted to linear order in the spin and relies on a different (gauge-invariant) definition of the TLNs, tied to their particular use of the light-cone gauge.

Remarkably, in the case of a Kerr black hole, the static Teukolsky equation that is obeyed by the Weyl scalar $\psi_0$ can be solved in closed form for a generic $\ell \in \mathbb{R}$, as shown explicitly in the next section.\footnote{This has been known since the seminal work of Teukolsky himself; see for instance Eqs.~(5.7) and (5.8) in Chap.~6 of Ref.~\cite{Te.74}.} This simple yet crucial observation is at the core of our computation of the linear response of a spinning black hole to a generic tidal perturbation, without any restriction on the black hole spin nor multipolar index $\ell$. Indeed, thanks to the analytic continuation to $\ell \in \mathbb{R}$, the asymptotic form of the mode $\psi_0^{\ell m}$ of $\psi_0$ as $r \to \infty$ will be shown to take the form
\beq\label{psi0_schema}
    \psi_0^{\ell m} \sim r^{\ell-2} \left( 1 + \cdots \right) + \kappa \, r^{-\ell-3} \left( 1 + \cdots \right) ,
\eeq
where the dots denote relativistic corrections to the Newtonian result \eqref{psi0_Newt_spin}, while $\kappa$ is an $r$- independent quantity. Unambiguously, the first term on the right-hand side corresponds to the tidal field, while the second term corresponds to the linear response of the Kerr black hole. As we shall prove in the next Sec.~\ref{sec:Weyl}, the linear response vanishes identically (i.e., $\kappa \equiv 0$) for a generic  tidal perturbation of a Schwarzschild black hole and for an axisymmetric tidal perturbation of a Kerr black hole, so that the corresponding contribution to the Geroch-Hansen multipole moments vanishes as well, yielding zero black hole (static) TLNs in those two particular cases. For a nonaxisymmetric perturbation of a spinning black hole, however, the linear response does not vanish ($\kappa \neq 0)$, yielding nonzero black hole (static) TLNs which we shall compute in Sec.~\ref{sec:multipoles}, at quadrupolar order, to linear order in the Kerr black hole spin. We will find, in particular, that the nonvanishing quadrupolar TLNs do not display any $\ell$-mode coupling, so that, for a general $(\ell,m)$ mode, we define
\beq\label{TLN_bis}
    \lambda_{\ell m}^{M\mathcal{E}} \equiv \frac{\partial M_{\ell m}}{\partial \mathcal{E}_{\ell m}} \, , \quad \lambda_{\ell m}^{M\mathcal{B}} \equiv \frac{\partial M_{\ell m}}{\partial \mathcal{B}_{\ell m}} \, , \quad \lambda_{\ell m}^{S\mathcal{E}} \equiv \frac{\partial S_{\ell m}}{\partial \mathcal{E}_{\ell m}} \, , \quad \lambda_{\ell m}^{S\mathcal{B}} \equiv \frac{\partial S_{\ell m}}{\partial \mathcal{B}_{\ell m}} \, .
\eeq

\section{Curvature Weyl scalar}\label{sec:Weyl}

The Newman-Penrose formalism~\cite{NePe.62} relies on the introduction of a null tetrad $(\ell^\alpha,n^\alpha,m^\alpha,\bar{m}^\alpha$), where $\ell^\alpha$ and $n^\alpha$ are real vectors whereas $m^\alpha$ and $\bar{m}^\alpha$ are complex vectors. By projecting the Weyl tensor $C_{\alpha\beta\gamma\delta}$ onto this tetrad, one can define two Weyl scalars of particular interest:
\beq\label{psi0_def}
    \psi_0 \equiv C_{\alpha\beta\gamma\delta} \ell^\alpha m^\beta \ell^\gamma m^\delta
    \quad \text{and} \quad
    \psi_4 \equiv C_{\alpha\beta\gamma\delta} n^\alpha \bar{m}^\beta n^\gamma \bar{m}^\delta \, .
\eeq
In principle, the curvature scalar $\psi_0$ contains the full information on the gravitational metric perturbation in vacuum, up to a few non-radiative degrees of freedom, namely infinitesimal changes in the black hole mass and spin~\cite{Wa.73,Me.al.16,vdM2.17}. Asymptotically, $\psi_0$ encodes all the relevant information regarding the tidal environment in which the spinning black hole is immersed. To compute the TLNs of a Kerr black hole, the main idea is to use this boundary condition far from the black hole and to solve for the perturbed  Weyl scalar $\psi_0$ in the region $r_+ \leqslant r \ll \mathcal{R}$, where $r_+$ is the coordinate location of the Kerr black hole horizon and $\mathcal{R}$ is the typical length scale of variation of the tidal environment. The linear response of the Kerr black hole to the applied tidal field can then be determined following a procedure of metric reconstruction, which we shall carry out in Secs.~\ref{sec:Hertz} and \ref{sec:metric} below.

\subsection{Radial Teukolsky equation}\label{subsec:radial}

Henceforth, we shall consider a Kerr black hole of mass $M$ and spin angular momentum $J$, perturbed by an external tidal field characterized by the tidal tensors \eqref{EL_BL_GR}. In advanced Kerr coordinates $(v,r,\theta,\pAdv)$, the Kerr metric $\mathring{g}_{\alpha\beta}$ reads~\cite{Cha}
\begin{align}\label{eq:metric}
    \mathring{g}_{\alpha\beta} \, \ud x^\alpha \ud x^\beta = &- \left( 1 - \frac{2Mr}{\Sigma} \right) \ud v^2 + 2 \ud v \ud r - \frac{4 M r}{\Sigma} \, a \sin^2{\theta} \, \ud v \ud \pAdv - 2a \sin^2{\theta} \, \ud r \ud \pAdv \nonumber \\ &+ \Sigma \, \ud \theta^2 + \left( r^2 + a^2 + \frac{2Mr}{\Sigma} \, a^2 \sin^2{\theta} \right) \sin^2{\theta} \, \ud \pAdv^2 \, ,
\end{align}
where $a \equiv J/M$ is the Kerr parameter and $\Sigma \equiv r^2 + a^2 \cos^2{\theta}$. It is regular on the \mbox{future event} horizon. Having characterized the tidal environment of a generic compact object in Sec.~\ref{subsec:environment} above, our objective is to compute the perturbed Weyl scalar $\psi_0$ in the Kinnersley tetrad~\cite{Ki.69}, which is given in advanced Kerr coordinates in Eq.~\eqref{tetrad} below, and such that $\mathring{\psi}_0 = 0$ in the background. The tidal environment of the black hole will be assumed to be \emph{slowly varying} and \emph{weak}, so that $\psi_0$ is a slowly-varying solution of the Teukolsky equation with spin weight $s=+2$~\cite{Te.73, TePr.74}.

For a generic curvature perturbation the angular part of a given $(\ell,m)$ mode with time factor $e^{-\ui \omega v}$, where $\omega$ is the frequency, is the spin-weighted spheroidal harmonic ${}_2S_{\ell m}^{a\omega}(\theta,\pAdv)$~\cite{Be.al2.06}. As we restrict our analysis to a slowly varying tidal field, we shall only solve for the \emph{static} modes (i.e.\ $\omega = 0$) of the Teukolsky equation, for which the spin-weighted spheroidal harmonics reduce to the spin-weighted \emph{spherical} harmonics ${}_2Y_{\ell m}(\theta,\pAdv)$. Following Refs.~\cite{Po2.04,YuGo.06,Ch.al.13}, we thus look for a perturbed Weyl scalar of the form
\beq\label{psi0}
	\psi_0 = \sum_{\ell m} z_{\ell m}(v) \, R_{\ell m}(r) \, {}_{2}Y_{\ell m}(\theta,\pAdv) \equiv \sum_{\ell m} \psi_0^{\ell m} \, ,
\eeq
where $z_{\ell m}(v)$ is a slowly varying (complex) function of advanced time $v$, which will be shown in the next subsection to be related in a simple manner to the tidal tensors \eqref{EL_BL_GR}. The radial function $R_{\ell m}(r)$ is itself a solution of the static and homogeneous (vacuum) radial Teukolsky equation~\cite{TePr.74,Ch.al.13},
\beq\label{Teuk}
	x(x+1) R''_{\ell m}(x) + (6x + 3 + 2 \ui m \gamma) R'_{\ell m}(x) + \biggl[ 4\ui m \gamma \, \frac{2x+1}{x(x+1)} - (\ell+3)(\ell-2) \biggr] R_{\ell m}(x) = 0 \, ,
\eeq
where we introduced the dimensionless radial variable $x \equiv (r - r_+) / (r_+ - r_-)$, together with the shorthand $\gamma \equiv a / (r_+ - r_-)$, where $r_\pm = M \pm \sqrt{M^2-a^2}$ are the coordinate radii of the outer/event ($+$) and inner/Cauchy ($-$)  horizons of the background Kerr black hole. Notice that if $R_{\ell m}(x)$ is a solution of Eq.~\eqref{Teuk} then $\bar{R}_{\ell \, -m}(x)$ is also a solution. Furthermore, if the boundary conditions that  $R_{\ell m}(x)$ obeys also satisfy that symmetry, then it follows that $R_{\ell m}(x)=\bar{R}_{\ell \, -m}(x)$.

Thereafter, we shall consider the cases $m\gamma = 0$ and $m\gamma \neq 0$ separately, as they will give rise to qualitatively different solutions to \eqref{Teuk}. Before doing so, however, we need to specify the boundary conditions obeyed by the perturbed curvature Weyl scalar $\psi_0$.

\subsection{Boundary condition at infinity}\label{subsec:infinity}

As explained in Sec.~\ref{sec:GR}, the asymptotic behavior as $r \to \infty$ of the solutions of the radial Teukolsky equation \eqref{Teuk} plays a central role in the calculation of the TLNs of a Kerr black hole. Far from the hole, i.e.\ when $r \gg r_+$, we have $x \gg 1$, and the radial Teukolsky equation \eqref{Teuk} reduces to
\beq \label{eq:ODE large-r}
	x^2 R''_{\ell m}(x) + 6x R'_{\ell m}(x) - (\ell+3)(\ell-2) R_{\ell m}(x) = 0 \, .
\eeq
Performing a Frobenius-type analysis of this second order, linear, homogeneous and ordinary differential equation, it follows that, for any $\ell \neq -1/2$, the two solutions are $x^{\ell-2}$ and $x^{-(\ell+3)}$. Since $x \propto r$ when $r \gg r_+$, the asymptotic behavior as $r \to \infty$ of the ($\ell$, $m$) mode $\psi_0^{\ell m}$ of the curvature scalar \eqref{psi0} is given by a linear combination of $r^{\ell-2}$ and $r^{-(\ell+3)}$. In the Newtonian limit, these correspond to an external $\ell$-polar tidal field and to the linear response of the central object, respectively, as was shown in Eq.~\eqref{psi0_Newt mode} or \eqref{psi0_Newt_spin}.

Hence, in the intermediate region $r_+ \ll r \ll \mathcal{R}$, and for any $\ell \in \mathbb{L}$, the Weyl scalar modes in Eq.~\eqref{psi0} asymptote as
\beq\label{psi0_asym}
	\psi_0^{\ell m} \sim z_{\ell m}(v) \, r^{\ell-2} \; {}_{2}Y_{\ell m}(\theta,\pAdv) \, ,
\eeq
with the arbitrary normalization choice of $R_{\ell m}\sim r^{\ell-2}$, as any overall normalization for $R_{\ell m}$ can be thought of as being absorbed by the coefficients $z_{\ell m}$. The $z_{\ell m}$ themselves, however, are determined by matching \eqref{psi0_asym} to an expansion for $r_+\ll r\ll \mathcal{R}$ of modes (considering different $\ell$ modes separately due to their different angular behavior) of the curvature scalar when written in terms of the Weyl tensor as in $\psi_0 = C_{\alpha\beta\gamma\delta} \ell^\alpha m^\beta \ell^\gamma m^\delta$, and so, equivalently, in terms of the $\mathcal{E}_L$ and $\mathcal{B}_L$ in Eq.~\eqref{EL_BL_GR}. As a result, the slowly varying external tidal field is encoded in the complex functions $z_{\ell m}(v)$ of advanced time $v$. More precisely, the authors of~\cite{Po2.04,Ch.al.13} have shown that, up to the next-to-leading ($\ell =3$) octupolar order, these functions read
\beq\label{z}
    z_{\ell m} = \alpha_{\ell m} + \ui \beta_{\ell m} \ ,
\eeq
where $\alpha_{\ell m}$ (resp. $\beta_{\ell m}$) are linear combinations of Cartesian components of the electric-type (resp. magnetic-type) tidal tensor $\mathcal{E}_L$ (resp. $\mathcal{B}_L$) defined in Eqs.~\eqref{EL_BL_GR}. For instance, at the leading ($\ell=2$) quadrupolar order and next-to-leading ($\ell=3$) octupolar order, they find\footnote{The normalization of the spin-weighted spherical harmonics ${}_{2}Y_{\ell m}(\theta,\pAdv)$ used in Ref.~\cite{Ch.al.13} is different from ours; see in particular their Eqs.~(30) and (31). We converted their expressions using our normalization, which is specified in App.~\ref{app:SWSH}.}
\begin{subequations}\label{alphas}
	\begin{align}
		\alpha_{2,0}     &= - 2\sqrt{\frac{6\pi}{5}} \, (\mathcal{E}_{xx} + \mathcal{E}_{yy}) \, , \\
		\alpha_{2,\pm 1} &= \mp 2 \sqrt{\frac{4\pi}{5}} \, (\mathcal{E}_{xz} \mp \ui \mathcal{E}_{yz}) \, , \\
		\alpha_{2,\pm 2} &= \sqrt{\frac{4\pi}{5}} \, (\mathcal{E}_{xx} - \mathcal{E}_{yy} \mp 2 \ui \mathcal{E}_{xy}) \, , \\
		\alpha_{3,0}     &= - 2\sqrt{\frac{10\pi}{21}} \, (\mathcal{E}_{xxz} + \mathcal{E}_{yyz}) \, , \\
		\alpha_{3,\pm 1} &= \pm \sqrt{\frac{10\pi}{7}} \, (\mathcal{E}_{xxx} + \mathcal{E}_{xyy} \mp \ui (\mathcal{E}_{xxy} + \mathcal{E}_{yyy})) \, , \\
		\alpha_{3,\pm 2} &= \sqrt{\frac{4\pi}{7}} \, (\mathcal{E}_{xxz} - \mathcal{E}_{yyz} \mp 2\ui \mathcal{E}_{xyz}) \, , \\
		\alpha_{3,\pm 3} &= \mp \sqrt{\frac{2\pi}{21}} \, \left(\mathcal{E}_{xxx} - 3\mathcal{E}_{xyy} \mp \ui (3\mathcal{E}_{xxy} - \mathcal{E}_{yyy})\right) \, ,
	\end{align}
\end{subequations}
together with
\begin{subequations}\label{betas}
	\begin{align}
		\beta_{2,0}     &= - 2\sqrt{\frac{6\pi}{5}} \, (\mathcal{B}_{xx} + \mathcal{B}_{yy}) \, , \\
		\beta_{2,\pm 1} &= \mp 2 \sqrt{\frac{4\pi}{5}} \, (\mathcal{B}_{xz} \mp \ui \mathcal{B}_{yz}) \, , \\
		\beta_{2,\pm 2} &= \sqrt{\frac{4\pi}{5}} \, (\mathcal{B}_{xx} - \mathcal{B}_{yy} \mp 2 \ui \mathcal{B}_{xy}) \, , \\
		\beta_{3,0} &= - \frac{8}{3} \sqrt{\frac{10\pi}{21}} \, (\mathcal{B}_{xxz} + \mathcal{B}_{yyz}) \, ,\label{beta30} \\
		\beta_{3,\pm 1} &= \pm \frac{4}{3} \sqrt{\frac{10\pi}{7}} \, (\mathcal{B}_{xxx} + \mathcal{B}_{xyy} \mp \ui (\mathcal{B}_{xxy} + \mathcal{B}_{yyy})) \, , \\
		\beta_{3,\pm 2} &= \frac{4}{3} \sqrt{\frac{4\pi}{7}} \, (\mathcal{B}_{xxz} - \mathcal{B}_{yyz} \mp 2\ui \mathcal{B}_{xyz}) \, , \\
		\beta_{3,\pm 3} &= \mp \frac{4}{3} \sqrt{\frac{2\pi}{21}} \, (\mathcal{B}_{xxx} - 3\mathcal{B}_{xyy} \mp \ui (3\mathcal{B}_{xxy} - \mathcal{B}_{yyy})) \, .
	\end{align}
\end{subequations}
Because the magnetic-type tidal tensors \eqref{B_L} carry an extra factor of $1/c$ compared to the electric-type tidal tensors \eqref{E_L_GR}, as can be seen from the Newtonian formula \eqref{C_Newt}, we have $z_{\ell m} \to \alpha_{\ell m}$ in the formal limit $c^{-1} \to 0$. Therefore the formula \eqref{psi0_asym} is consistent with the asymptotic form \eqref{psi0_Newt_asym} of the Newtonian limit of the curvature scalar $\psi_0$. More precisely, it can be checked that, up to $\ell = 3$, the asymptotic formula \eqref{psi0_asym} with the coefficients \eqref{alphas} is in perfect agreement with the Newtonian limit \eqref{psi0_Newt_asym} with the coefficients \eqref{alpha} and \eqref{pilou}.

The asymptotic form \eqref{psi0_asym} of the Weyl scalar $\psi_0$ far from the black hole is valid for all $(\ell,m)$ modes.  However, the precise relations between the coefficients $z_{\ell m}(v)$ and the tidal tensors \eqref{EL_BL_GR} have only been derived up to $\ell = 3$~\cite{Ch.al.13}. Given the Newtonian form \eqref{psi0_Newt_asym} of the asymptotic $\psi_0$ with the relation \eqref{alpha}, valid for all ($\ell,m$) modes, and given the definition \eqref{psi0_def} of $\psi_0$ in terms of the Weyl tensor, such relations must exist for all ($\ell,m$) modes, and not merely up to $\ell = 3$. With the help of Eqs.~\eqref{alpha}, \eqref{EL_BL_GR}, \eqref{pilou} and \eqref{alphas}--\eqref{betas} we are naturally led to \emph{conjecture} that Eqs.~\eqref{alphas}--\eqref{betas} generalize into
\beq\label{alphas-betas}
    \alpha_{\ell m} = \sqrt{\frac{(\ell+2)(\ell+1)}{\ell(\ell-1)}} \, \mathcal{E}_{\ell m} \quad  \text{and} \quad \beta_{\ell m} = \frac{\ell+1}{3} \sqrt{\frac{(\ell+2)(\ell+1)}{\ell(\ell-1)}} \, \mathcal{B}_{\ell m} \, ,
\eeq
where we recall that the modes $\mathcal{E}_{\ell m}(v)$ and $\mathcal{B}_{\ell m}(v)$ of the tidal moments $\mathcal{E}_L(v)$ and $\mathcal{B}_L(v)$ are defined in Eq.~\eqref{ElmBlm}.\footnote{In particular, it follows from Eq.~\eqref{ElmBlm} that the  relationship between $\mathcal{B}_{\ell m}$ and $\mathcal{B}_L$ for $\ell=2,3$ is the same as that between  $\mathcal{E}_{\ell m}$ and $\mathcal{E}_L$ given explicitly in Eqs.~\eqref{pilou}.} In the case of a \textit{nonspinning} black hole one may use the tidally perturbed Schwarzschild metric derived in~\cite{BiPo.09} to compute the asymptotic form of the perturbed curvature scalar $\psi_0$, by projecting the associated Weyl tensor onto the Kinnersley null tetrad, yielding the expression~\cite{Ch2.20}
\beq
    \psi_0^{\ell m} \sim 
	\sqrt{\frac{(\ell+2)(\ell+1)}{\ell(\ell-1)}} \, \Bigl[ \mathcal{E}_{\ell m}(v) + \ui \, \frac{\ell+1}{3} \, \mathcal{B}_{\ell m}(v) \Bigr] \, r^{\ell-2} \; {}_{2}Y_{\ell m}(\theta,\pAdv) \, .
\eeq
This expression coincides with the formula \eqref{psi0_asym}, where the slowly-varying coefficients \eqref{z} are precisely given by the formulas \eqref{alphas-betas}. We emphasize, however, that in the case of a Kerr background, the relationships \eqref{alphas-betas} have only been established up to octupolar ($\ell = 3$) order~\cite{Ch.al.13}. We leave a derivation valid for any $\ell \in \mathbb{L}$ to future work. The derivation of the quadrupolar multipole moments performed in Sec.~\ref{sec:multipoles} below will rely on the relations \eqref{alphas-betas} \textit{only} for the proven $\ell = 2$ (and $\ell=3$) modes.

\subsection{Boundary condition on the horizon}\label{subsec:boundary}

Additionally, the Weyl scalar $\psi_0$ must satisfy a smoothness condition on the future horizon of the Kerr background. Since  the Hartle-Hawking (HH) null tetrad~\cite{HaHa.72} is smooth on the future horizon ($r=r_+$), so should be the curvature scalar $\psi_0^\text{HH}$ defined from the HH tetrad. The two null vectors $\ell^\alpha$ and $n^\alpha$ in the Kinnersley tetrad are simply related to those $\ell^\alpha_\text{HH}$ and $n^\alpha_\text{HH}$ in the HH tetrad via $\ell^\alpha_\text{HH} = \Delta /(2(r^2+a^2)) \, \ell^\alpha$ and $n^\alpha_\text{HH} = \left(2(r^2+a^2)/\Delta\right) n^\alpha$~\cite{TePr.74}, where $\Delta\equiv (r-r_+)(r-r_-)$, while the other null vectors are the same in the two tetrads. It then follows that the corresponding Weyl scalars \eqref{psi0_def} are related as~\cite{Cha} 
\beq\label{reg}
	\psi_0^\text{HH} = \frac{\Delta^2}{4(r^2+a^2)^2} \, \psi_0
	\quad \text{and} \quad
	\psi_4^\text{HH} = \frac{4(r^2+a^2)^2}{\Delta^2} \, \psi_4 \, .
\eeq
Therefore, the product $\Delta^2 \psi_0$ must be smooth on the future event horizon in Kerr spacetime. This smoothness condition will play a central role in our calculation of the TLNs of a spinning black hole.

Interestingly, the smoothness condition following from \eqref{reg} can also be seen as the zero-frequency limit of a no-outgoing-wave boundary condition at the future horizon. Indeed, for a generic (nonstatic) perturbation of frequency $\omega$, the most general solution of the radial Teukolsky equation is a linear combination of two solutions, say $R_{\ell m}^\text{in}(r)$ and $R_{\ell m}^\text{out}(r)$, which have ingoing-wave and outgoing-wave behaviors at the horizon, respectively. The ingoing-wave solution is such that $\Delta^2 R_{\ell m}^\text{in}(r)$ is smooth at $r=r_+$, while the outgoing wave solution behaves as~\cite{TePr.74}
\beq\label{Rin}
	\Delta^2 R_{\ell m}^\text{out}(r) \sim (r-r_+)^2 e^{2\ui k r_*} \, ,
	\quad \text{as}\quad r \to r_+ \, ,
\eeq
where $k \equiv \omega - m \omega_\text{H}$, with $\omega_\text{H}$ the angular velocity of the event horizon of the black hole, and $r_*$ is the radial tortoise coordinate, which is such that $r_* \to -\infty$ as $r \to r_+$. From Eq.~\eqref{Rin} it is clear that $\Delta^2 R_{\ell m}^\text{out}(r)$ is not smooth at $r = r_+$, even in the static limit $\omega \to 0$.

\subsection{Nonspinning black hole and/or axisymmetric perturbation}\label{sec:Schw}

If the black hole is nonspinning $(a = 0)$ or if the perturbation is axisymmetric $(m = 0)$, then the product $m\gamma$ vanishes in the radial Teukolsky equation \eqref{Teuk}. In that case, the most general solution does not depend on $m$, and, for generic $\ell\in\mathbb{R}$, it takes the general form 
\beq\label{sol_NS}
	R_{\ell m}(x) = a_\ell \, \frac{P_\ell^2(1+2x)}{x(1+x)} + b_\ell \, \frac{Q_\ell^2(1+2x)}{x(1+x)} \, , 
\eeq
where $a_\ell$ and $b_\ell$ are constants, while $P_\ell^m(z)$ and $Q_\ell^m(z)$ are associated Legendre functions of, respectively, the first and second kind, of degree $\ell$ and order $m$. We note that, as $x>0$,  the argument of these functions is larger than $1$. In this regime, we specifically choose $P_\ell^m(z)$ and $Q_\ell^m(z)$ to refer to the functions defined in Eqs.~(14.3.9) and (14.3.7) in Ref.~\cite{NIST:DLMF}, respectively. The solution \eqref{sol_NS} is consistent with Eqs.~(58) and (59) of~\cite{Ch.al.13} for the $\ell \!=\! 2$ and $\ell \!=\! 3$ cases.

As we argued in Sec.~\ref{subsec:boundary}, the product $\Delta^2 \psi_0$ must be smooth on the future event horizon of the Kerr spacetime, which implies that $x^2 R_{\ell m}(x)$ must be smooth at $x = 0$. On the one hand, Eqs.~(14.3.6) and (15.8.12) in Ref.~\cite{NIST:DLMF} show that for all $\ell \in \mathbb{R}$ and for all $x \geqslant 0$,
\beq\label{eq:P x->0}
    \frac{x}{1+x} \, P_\ell^2(1+2x) = \mathbf{F}(\ell+1,-\ell,-1;-x) = (1+x)^{-2} \, \mathbf{F}(\ell-1,-\ell-2,-1;-x) \, .
\eeq
Since the regularized hypergeometric function $\mathbf{F}(a,b,c;z)$ is smooth at $z = 0$, the function $x P_\ell^2(1+2x)/(1+x)$ (corresponding to the first solution in \eqref{sol_NS}) is smooth at $x = 0$. On the other hand, an analysis as $x \to 0^+$, using e.g.\ Eqs.~(14.3.7), (15.8.12) and (15.8.11) in Ref.~\cite{NIST:DLMF}, shows that for all $\ell \in \mathbb{R}$,
\begin{align}\label{eq:Q x->0}
    \frac{x}{1+x} \, Q_\ell^2(1+2x) &= \frac{1}{2} - \frac{1}{2} \, \ell(\ell+1) \, x + \frac{1}{8} \, (\ell^4 + 10 \ell^3 + 11 \ell^2- 6\ell - 4) \, \times \nonumber \\ &\qquad\qquad \times \bigl[ 2\psi(\ell+3) + 2\gamma_\text{E} + \ln{x} \bigr] \, x^2 + o(x^2) \, ,
\end{align}
where $\psi(z) \equiv \Gamma'(z) / \Gamma(z)$ is the digamma function while $\gamma_\text{E}$ is the Euler-Mascheroni constant. Hence the function $x Q_\ell^2(1+2x)/(1+x)$ (corresponding to the second solution in \eqref{sol_NS}) is merely $C^1$ on the future event horizon, as its second derivative with respect to $x$ blows up logarithmically as $x \to 0$. Consequently, the physically relevant solution \eqref{sol_NS} to the radial Teukolsky equation \eqref{Teuk} with $m\gamma = 0$ must satisfy
\beq\label{b_l}
	b_\ell = 0 \, .
\eeq

Moreover, as pointed out in Sec.~\ref{subsec:infinity}, an additional condition is imposed by the asymptotic form \eqref{psi0_asym} of the curvature scalar $\psi_0$ for large $r$. So far, our calculation of the solution \eqref{sol_NS} with the condition \eqref{b_l} is valid for a generic $\ell\in \mathbb{R}$. This solution has two distinct asymptotic behaviors, one which goes like $r^{\ell-2}$ and the other one like $r^{-(\ell+3)}$. While for now we still keep $\ell\in \mathbb{R}$, we here require that the coefficient of the asymptotic behavior $r^{\ell-2}$ of our solution agrees with that in Eq.~\eqref{psi0_asym}. This condition determines the value of the remaining integration constant as (using e.g., Eqs.~(14.3.8), (15.8.2) and (15.2.1) of Ref.~\cite{NIST:DLMF})
\beq\label{a_l}
    a_\ell = (r_+-r_-)^{\ell-2} \, \frac{\Gamma(\ell-1)\Gamma(\ell+1)}{\Gamma(2\ell+1)} \, .
\eeq
Substituting for Eqs.~\eqref{eq:P x->0}, \eqref{b_l} and \eqref{a_l} into Eq.~\eqref{sol_NS}, and making use of the identity $\mathbf{F}(a,b,c;z) = \mathbf{F}(b,a,c;z)$, the physical solution to the radial Teukolsky equation \eqref{Teuk} with $m\gamma = 0$ is thus given by
\beq\label{sol_NS_bis}
	R_{\ell m}(x) = (r_+-r_-)^{\ell-2} \, \frac{\Gamma(\ell-1)\Gamma(\ell+1)}{\Gamma(2\ell+1)} \, \frac{\mathbf{F}(-\ell-2,\ell-1,-1;-x)}{x^2(1+x)^2} \, .
\eeq

The general solution \eqref{sol_NS} of the Bardeen-Press equation~\cite{BaPr.73} (the Teukolsky equation with $a=0$) is mapped, via the Chandrasekhar transformation~\cite{Ch2.75}, onto the general solution of the Cunningham-Price-Moncrief and Zerilli equations~\cite{ReWh.57,Ze3.70,Cu.al.78}.\footnote{The Regge-Wheeler and the Cunningham-Price-Moncrief equations are the same, but the Regge-Wheeler function is ill-defined in the static limit that we are considering. See also the related footnote \ref{footnote:RWCPM}.} We further checked that the condition \eqref{b_l}, i.e., $b_\ell=0$, is equivalent to the conditions $a_Q=0$ and $b_Q = 0$ derived in Ref.~\cite{DaNa2.09}, for a nonspinning black hole, based on the Regge-Wheeler-Zerilli formalism. See  App.~\ref{app:NS_check} for further details.

\subsection{Spinning black hole and nonaxisymmetric perturbation}\label{sec:spin nonaxi}

Next, we consider the case of a nonaxisymmetric ($m \!\neq\! 0$), linear perturbation of a spinning black hole ($a \neq 0$), so that $m\gamma \neq 0$. Two linearly independent solutions of the homogeneous, second-order, ordinary differential equation \eqref{Teuk} are
\begin{subequations}\label{sols}
	\begin{align}
		f_{\ell m}(x) &\equiv x^{-2}(1+x)^{-2} \, \mathbf{F}(-\ell-2,\ell-1,-1+2\ui m \gamma;-x) \, , \\
		g_{\ell m}(x) &\equiv (1+1/x)^{2\ui m \gamma} \, \mathbf{F}(-\ell+2,\ell+3,3-2\ui m \gamma;-x) \, . \label{g}
	\end{align}
\end{subequations}
The most general solution to the radial Teukolsky equation \eqref{Teuk} must be given by a linear combination of the two independent solutions \eqref{sols}, say
\beq\label{R}
	R_{\ell m}(x) = a_{\ell m} f_{\ell m}(x) + b_{\ell m} g_{\ell m}(x) \, ,
\eeq
where $a_{\ell m}$ and $b_{\ell m}$ are constants.

This radial function \eqref{R} must satisfy a smoothness condition at the Kerr event horizon. Indeed, as argued in Sec.~\ref{subsec:boundary} above, $\Delta^2 \psi_0$ must be smooth at $r = r_+$, which implies that $x^2 R_{\ell m}(x)$ must be smooth at $x = 0$. Using for instance Eq.~(15.2.1) in Ref.~\cite{NIST:DLMF}, one can see that $x^2 f_{\ell m}(x)$ is smooth as $x \to 0$, while $x^2 g_{\ell m}(x)$ is not smooth in that limit. Hence we conclude that the physically relevant solution to the radial Teukolsky equation with $m\gamma \neq 0$ must satisfy
\beq\label{eq:b=0 Kerr}
	b_{\ell m} = 0 \, .
\eeq
Consequently, like in the case of a nonspinning black hole, the smoothness condition on the future horizon---following from \eqref{reg}---is responsible for killing off one of the two homogeneous solutions to the radial Teukolsky equation. As discussed in Sec.~\ref{subsec:boundary}, this smoothness condition is equivalent to requiring no outgoing waves at the horizon.

Moreover, as pointed out in Sec.~\ref{subsec:infinity}, an additional condition is imposed by the asymptotic form \eqref{psi0_asym} of the Weyl scalar $\psi_0$ at large radii. Similarly to the Schwarzschild case, our calculation of the solution \eqref{R} with \eqref{sols} and condition \eqref{eq:b=0 Kerr} is valid for generic $\ell\in \mathbb{R}$. This solution $R_{\ell m} = a_{\ell m} f_{\ell m}$ has two distinct  asymptotic behaviors, one which goes like $r^{\ell-2}$ and the other one like $r^{-(\ell+3)}$. While for now we still keep $\ell\in \mathbb{R}$, we here require that the coefficient of the asymptotic behavior $r^{\ell-2}$ of our solution agrees with that in Eq.~\eqref{psi0_asym}. We then readily deduce that the integration constant $a_{\ell m}$ is given by (using e.g., Eqs.~(15.8.2) and (15.2.2) in Ref.~\cite{NIST:DLMF})
\beq\label{a}
	a_{\ell m} = (r_+-r_-)^{\ell-2} \, \frac{\Gamma(\ell-1)\Gamma(\ell+1+2\ui m \gamma)}{\Gamma(2\ell+1)} \, .
\eeq
This is consistent with Eqs.~(75) and (76) of Ref.~\cite{Ch.al.13} for the $\ell = 2$ and $\ell = 3$ cases. Finally, the physically relevant solution to the radial Teukolsky equation \eqref{Teuk} with $m\gamma \neq 0$ is given by Eqs.~\eqref{sols}--\eqref{a}, namely
\beq\label{sol_Kerr}
	R_{\ell m}(x) = (r_+-r_-)^{\ell-2} \, \frac{\Gamma(\ell-1)\Gamma(\ell+1+2\ui m \gamma)}{\Gamma(2\ell+1)} \, \frac{\mathbf{F}(-\ell-2,\ell-1,-1+2\ui m \gamma;-x)}{x^2(1+x)^2} \, .
\eeq
We notice that $R_{\ell \, -m} = \bar{R}_{\ell m}$, in agreement with the comment below Eq.~\eqref{Teuk}.

Having expressed the Schwarzschild solution \eqref{sol_NS_bis} and the Kerr solution \eqref{sol_Kerr} in terms of the regularized hypergeometric function $\mathbf{F}(a,b,c;z)$, the former is easily seen to be obtained as the limit when $\gamma \to 0$ of the latter.\footnote{The solution \eqref{g} was discarded because it is not smooth on the event horizon when $\gamma \neq 0$. In the limit where $\gamma \to 0$, however, this solution reduces to the same Eq.~\eqref{sol_NS_bis}. Consequently, for $\gamma=0$, Eq.~\eqref{R} does not represent two linearly independent solutions of the radial Teukolsky equation \eqref{Teuk}.} From now on, we shall thus consider the generic case of a spinning black hole, i.e., $\gamma \neq 0$. Moreover, the solution \eqref{sol_NS_bis} also describes an axisymmetric ($m=0$) tidal perturbation of a Kerr black hole. Since the formula \eqref{sol_Kerr} reduces to \eqref{sol_NS_bis} when $m=0$ as well, the radial solution \eqref{sol_Kerr} is also valid for $m=0$. In summary, we have established that the perturbed curvature scalar can be written as a sum $\psi_0 = \sum_{\ell m} \psi_0^{\ell m}$ [recall Eq.~\eqref{psi0}] of modes which, for generic $\ell\in\mathbb{R}$, are given by
\beq\label{psi0_lm}
	\psi^{\ell m}_0(v,r,\theta,\pAdv) = a_{\ell m} \, z_{\ell m}(v) \; \frac{\mathbf{F}(-\ell-2,\ell-1,-1+2\ui m \gamma;-x)}{x^2(1+x)^2} \; {}_{2}Y_{\ell m}(\theta,\pAdv) \, .
\eeq

\subsection{Linear response of a tidally perturbed Kerr black hole}\label{subsec:response}

The general relativistic formula \eqref{psi0_lm} should be compared to its Newtonian counterpart \eqref{psi0_Newt_spin}. In particular, the ($\ell$, $m$) mode of the Newtonian curvature Weyl scalar $\psi_0$ is the sum of two terms that scale as $r^{\ell -2}$ and $r^{-(\ell+3)}$, which correspond to the applied $\ell$-pole tidal field and to the body's linear response, respectively. Following the general strategy put forward in Sec.~\ref{sec:GR}, we shall now isolate from the result \eqref{psi0_lm} the part of $\psi^{\ell m}_0$ that corresponds to the tidal field from the part that corresponds to the linear response of the Kerr black hole. This relies crucially on the fact that our results so far are valid for generic $\ell \in \mathbb{R}$.

Keeping a generic $\ell\in\mathbb{R}$, our first task is to identify the two linearly independent solutions of the radial Teukolsky equation \eqref{Teuk} that carry \emph{separately} the asymptotic behaviors $x^{\ell-2}$ and $x^{-(\ell+3)}$ that were found from its asymptotic form \eqref{eq:ODE large-r} as $x \to \infty$. They are given by, respectively,\footnote{Notice that the solutions $R_{\ell m}^\text{tidal}(x)$ and $R_{\ell m}^\text{resp}(x)$ are exchanged under the transformation $\ell \to -\ell- 1$, which leaves the product $(\ell+3)(\ell-2)=\ell(\ell+1)-6$ invariant in the radial Teukolsky equation \eqref{Teuk}.}
\begin{subequations}\label{tidal-resp}
    \begin{align}
        R_{\ell m}^\text{tidal}(x) &\equiv (r_+-r_-)^{\ell-2} \, \frac{x^\ell}{(1+x)^2} \, F(-\ell-2,-\ell-2\ui m\gamma,-2\ell;-1/x) \, , \label{R_tidal} \\
        R_{\ell m}^\text{resp}(x) &\equiv (r_+-r_-)^{\ell-2} \, \frac{x^{-(\ell+1)}}{(1+x)^2} \, F(\ell-1,\ell+1-2\ui m\gamma,2\ell+2;-1/x) \, , \label{R_resp}
    \end{align}
\end{subequations}
where we used the common, dimensionful normalization factor $(r_+-r_-)^{\ell-2}$ for both solutions. Indeed, since both hypergeometric functions in \eqref{tidal-resp} are equal to $1$ in the limit as $x \to \infty$, the solutions \eqref{tidal-resp} have the asymptotic behaviors
\begin{subequations}\label{eq:tidal,resp}
    \begin{align}
        R_{\ell m}^\text{tidal} &\sim (r_+-r_-)^{\ell-2} \, x^{\ell-2} \sim r^{\ell-2} _, , \\
        R_{\ell m}^\text{resp} &\sim (r_+-r_-)^{\ell-2} \, x^{-(\ell+3)} \sim (r_+-r_-)^{2\ell+1} \, r^{-(\ell+3)} \, ,
    \end{align}
\end{subequations}
which correspond to the two distinct series found in the Newtonian result \eqref{psi0_Newt_spin}. Matching the relativistic solutions \eqref{tidal-resp} onto the Newtonian series \eqref{psi0_Newt_spin} for large radii ensures that, when $\ell \in \mathbb{L}$, the growing/dominant solution \eqref{R_tidal} has no ``hidden'' contribution from the decaying/subdominant solution \eqref{R_resp}. According to the discussion given in Sec.~\ref{sec:GR} above, the solutions \eqref{R_tidal} and \eqref{R_resp} thus correspond to a perturbing external tidal field and to the corresponding linear response of the Kerr black hole, respectively.

A few of words of caution about the solutions \eqref{tidal-resp} are in order, one on a purely mathematical aspect and the other on a more physical aspect. Regarding the mathematical word of caution, as explained in detail in App.~\ref{app:continuity}, the limit of $\ell$ to an element of $\mathbb{L}$ must be taken {\it simultaneously} in the first and third arguments of the hypergeometric function in  \eqref{R_tidal} (this amounts to using Eq.~(15.2.6), rather than Eq.~(15.2.5), in~\cite{NIST:DLMF}). We now turn to the physical word of caution. In Newtonian gravity, the growing series is indeed associated with a \textit{purely external} tidal field that derives from an external gravitational potential, irrespective of the Newtonian body that is being tidally perturbed. In general relativity, however, the tidal solution \eqref{R_tidal} can only be given this interpretation in the asymptotic limit as $r \to \infty$, where it reduces to a linear combination of the modes $\mathcal{E}_{\ell m}$ and $\mathcal{B}_{\ell m}$ of the tidal moments of the external tidal environment. Close to the black hole, however, the tidal solution \eqref{R_tidal} also depends on the properties of the compact body that is being tidally perturbed; here its mass $M$ and spin $a$. We also notice that the tidal and response solutions \eqref{tidal-resp} are both \text{regular} on the Kerr horizon  when expressed in terms of the HH tetrad; however neither of them is smooth on the horizon when $m\gamma \neq 0$. These behaviors on the horizon are a consequence of the following. As $x\to 0$, the hypergeometric function in \eqref{R_tidal} contains the two linearly independent asymptotic behaviors  $x^{-\ell-2}$ and $x^{-\ell-2im\gamma}$ while the hypergeometric function in \eqref{R_resp} contains $x^{\ell-1}$ and $x^{\ell+1-2im\gamma}$ (this may be seen from, e.g., Eqs.~(15.8.2) and (15.2.1) in~\cite{NIST:DLMF}). So, from Eqs.~\eqref{tidal-resp}, both $R^{\text{tidal}}_{\ell m}$ and $R^{\text{resp}}_{\ell m}$ contain the two asymptotic behaviors $x^{-2}$ and  $x^{-2im\gamma}$ as $x\to 0$. Now, as per Eq.~\eqref{reg}, as $x\to 0$, $\psi_0$ in the HH tetrad (which is regular on the horizon) has the same asymptotic behavior as $\psi_0$ in the Kinnersley tetrad times $x^2$. Thus, the radial asymptotics at the horizon of the tidal/response contributions to $\psi_0$ in the HH tetrad are $x^2 R^{\text{tidal/resp}}_{\ell m}$ and so they both contain the behaviors $1$ and $x^{2-2im\gamma}$. Hence, both contributions are regular but neither is smooth on the Kerr horizon.

Since the physical solution \eqref{sol_Kerr} is valid for $\ell\in\mathbb{R}$, it can be written as a linear combination of the two linearly independent, homogeneous solutions \eqref{tidal-resp} of the radial Teukolsky equation \eqref{Teuk}:
\beq\label{eq:tidal,resp comb}
    R_{\ell m}(x) = R_{\ell m}^\text{tidal}(x) + 2 \kappa_{\ell m} R_{\ell m}^\text{resp}(x) \, ,
\eeq
where $\kappa_{\ell m}$ is a dimensionless constant that will turn out to be closely related to the TLNs in \eqref{TLN} of a spinning black hole. Given the formulas \eqref{sol_Kerr} and \eqref{tidal-resp}, the identity (15.8.2) in Ref.~\cite{NIST:DLMF} shows that Eq.~\eqref{eq:tidal,resp comb} is satisfied if, and only if (see App.~\ref{app:continuity}),
\begin{align}\label{kappa}
    \kappa_{\ell m} &= \frac{\Gamma(\ell-1)\Gamma(-2\ell-1)\Gamma(\ell+1+2\ui m \gamma)}{2\Gamma(2\ell+1)\Gamma(-\ell-2)\Gamma(-\ell+2\ui m \gamma)} \nonumber \\
               &= \frac{|\Gamma(\ell+1+2\ui m \gamma)|^2 \, \Gamma(\ell-1)}{2\Gamma(2\ell+1)\Gamma(2\ell+2)} \, \frac{\sin{\left(\pi(\ell-2\ui m \gamma)\right)}}{\sin{\left(\pi(2\ell+1)\right)} \, \Gamma(-\ell-2)} \, .
\end{align}
In the second equality we used the identity $\Gamma(z)\Gamma(1-z) = \pi / \sin{(\pi z)}$, valid for all $z \in \mathbb{C} \setminus \mathbb{Z}$, as well as $\Gamma(\bar{z}) = \bar{\Gamma}(z)$. Our way of finding the unique and unambiguous tidal/response split \eqref{eq:tidal,resp comb}--\eqref{kappa} relied crucially on the analytic continuation to $\ell \in \mathbb{R}$. In App.~\ref{app:splits} we detail how the prescription adopted in Ref.~\cite{LaPo.15} (which does not identify the two distinct large-radius asymptotic behaviors that are exhibited by the Newtonian solution \eqref{psi0_Newt_spin}) would instead lead to the conclusion that the black hole response contribution to the physical solution \eqref{sol_Kerr} vanishes.

Having unambiguously identified the two different applied tidal field and black hole linear response behaviors, we can now restrict ourselves to the physical parameter space, i.e.\ $\ell \in \mathbb{L}$. Note that the $\sin{}$ function and the last gamma function in the denominator in the last line of \eqref{kappa} vanish and have a simple pole, respectively, whenever $\ell \in \mathbb{L}$, so that \eqref{kappa} is regular in that limit. For any $\ell \in \mathbb{L}$ and any $|m| \leqslant \ell$, we find
\begin{align}\label{kappa_phys}
    \kappa_{\ell m} &= - \frac{\ui}{4\pi} \, \sinh{(2 \pi m \gamma)} \, |\Gamma(\ell+1+2\ui m \gamma)|^2 \, \frac{(\ell-2)!(\ell+2)!}{(2\ell)!(2\ell+1)!} \nonumber \\ &= - \ui m \gamma \, \frac{(\ell-2)!(\ell+2)!}{2(2\ell)!(2\ell+1)!} \prod_{n=1}^\ell (n^2 + 4 m^2 \gamma^2) \, .
\end{align}
This is one of the most important results of this paper, and several comments are in order. First, the coefficients \eqref{kappa_phys} are purely imaginary and depend on the azimuthal number $m$. They couple the Kerr black hole spin $a$ to the $z$-component of angular momentum of the tidal field (proportional to $m$) through the product $m\gamma = ma / (r_+ - r_-) = m\ad/(2\sqrt{1-\ad^2})$, with $\ad \equiv a/M$ the dimensionless Kerr parameter. The purely imaginary nature of $k_{\ell m}$ and the specific $m$-dependence in Eq.~\eqref{kappa_phys} imply that
\beq\label{eq:symm k}
    \bar{\kappa}_{\ell m} = - \kappa_{\ell m} = \kappa_{\ell \, -m} \, .
\eeq
Second, these coefficients do fit the pattern \eqref{k_exp_spin} identified for the TLNs of an axisymmetric, spinning, Newtonian body. Moreover, the numbers \eqref{kappa_phys} are \textit{polynomials} of degree $2\ell+1$ in the product $m\gamma$, and only involve \textit{odd} powers of $m\gamma$. Third, in the nonspinning limit where $\gamma \to 0$, the coefficients $\kappa_{\ell m}$ all vanish, so that the linear response of a Schwarzschild black hole to a generic multipolar tidal field vanishes identically in the curvature Weyl scalar $\psi_0$.\footnote{We remind the reader that in this paper we are dealing with static (i.e., $\omega=0$) perturbing tidal fields; the Schwarzschild black hole response under non-static ($\omega\neq 0$) tidal fields is generally nonzero~\cite{Ch.al2.13,Ch2.20}.} Fourth, for an axisymmetric ($m=0$) tidal perturbation, the coefficients $\kappa_{\ell 0}$ vanish for all $\ell \in \mathbb{L}$, so that the linear response of a spinning black hole to a generic, axisymmetric tidal field vanishes as well in $\psi_0$. In summary, we established the key results
\beq\label{kappa=0}
    \lim_{a \to 0} \kappa_{\ell m} = 0 = \kappa_{\ell 0} \, . 
\eeq
These results will carry over to the contribution to the Geroch-Hansen multipole moments from the black hole response in the metric perturbation, so that the (static) TLNs of a Schwarzschild black hole all vanish identically, in agreement with Refs.~\cite{BiPo.09,DaNa2.09,KoSm.12,Gu.15}, as well as those of a Kerr black hole perturbed by a generic, axisymmetric tidal field. This generalizes to any spin $0 \leqslant a < M$ and to any multipolar order $\ell \in \mathbb{L}$ the result of Ref.~\cite{Pa.al.15}, which was restricted to quadratic order in the black hole spin, for a quadrupolar ($\ell = 2$) tidal perturbation. Fifth, the leading behavior of the coefficients \eqref{kappa_phys} is given, for small spins $0 < \ad \ll 1$, by
\beq\label{kappa_small_a}
    \kappa_{\ell m} = - \ui m \ad \; \frac{(\ell-2)! (\ell !)^2 (\ell+2)!}{4(2\ell)!(2\ell+1)!} + O(\ad^3) \, .
\eeq
Sixth, for a near-extremal black hole, the deviation from extremality ($\ad = 1$) is parameterized by the small number $\epsilon \equiv \sqrt{1-\ad^2}$, so that $\gamma = 1/(2\epsilon) + O(\epsilon)$ and the formula \eqref{kappa_phys} becomes (using Eq.~(5.11.7) in Ref.~\cite{NIST:DLMF}) 
\beq\label{kappa_extremal}
    \kappa_{\ell m} \sim - \ui \left( \frac{m}{\epsilon} \right)^{2\ell+1} \frac{(\ell-2)!(\ell+2)!}{4(2\ell)!(2\ell+1)!} \quad \text{as} \quad \epsilon \to 0^+ \, ,
\eeq
which blows up in the extremal limit. However the black hole response $\kappa_{\ell m} R_{\ell m}^\text{resp}$ itself remains finite in that limit. Indeed, $r_+ - r_- = 2M \epsilon$ and $x = (r - r_+)/(2M \epsilon)$ imply that $R_{\ell m}^\text{resp} \sim \epsilon^{2\ell+1}$ as $\epsilon \to 0$, which exactly compensates the singular behavior in \eqref{kappa_extremal}. Seventh, by using the Stirling-type formula $\Gamma(az+b) \sim \sqrt{2\pi} \, e^{-az} (az)^{az+b-1/2}$ as $|z| \to \infty$, valid for any fixed $a > 0$ and $b \in \mathbb{C}$, the large-$\ell$ asymptotics of Eq.~\eqref{kappa_phys} simply reads, for a generic spin,
\beq\label{kappa_large-l}
    \kappa_{\ell m} \sim - \frac{\ui}{2^{4\ell+3}} \, \sinh{(2\pi m\gamma)} \quad \text{as} \quad \ell \to \infty \, ,
\eeq
which shows that the black hole linear response vanishes quickly for large multipolar orders.

Most importantly, for a generic \emph{non}axisymmetric tidal perturbation of a Kerr black hole, the coefficients \eqref{kappa_phys} are nonzero, so that the linear response of a Kerr black hole to the applied tidal field does not vanish. This conclusion was established here at the level of the scalar $\psi_0$. By analyzing the operations that take the Weyl scalar to the metric perturbation, the result of nonvanishing black hole response will continue to hold at the level of the metric perturbation. More precisely, as will be shown in Secs.~\ref{sec:Hertz} and \ref{sec:metric} below, to linear order in the perturbing tidal field, the metric can be written as $g_{\alpha\beta} = \mathring{g}_{\alpha\beta} + h_{\alpha\beta}$, where the metric perturbation of the background Kerr metric $\mathring{g}_{\alpha\beta}$ is given by
\beq
    h_{\alpha\beta} = h^\text{tidal}_{\alpha\beta} + h^\text{resp}_{\alpha\beta} = \sum_{\ell m} \left( h^\text{tidal}_{\alpha\beta,\ell m} + 2\kappa_{\ell m} h^\text{resp}_{\alpha\beta,\ell m} \right) ,
\eeq
where both $h^\text{tidal}_{\alpha\beta}$ and $h^\text{resp}_{\alpha\beta}$ are source-free solutions of the linearized Einstein equation, which are unambiguously associated with the perturbing tidal field and the corresponding black hole response, respectively.

We also note that the tidal/response split in Eq.~\eqref{Teuk} which we carried out in the static (i.e., $\omega=0$) case can be trivially generalized to ``small" but nonzero $\omega$ by using the matched asymptotic result in the Appendix of~\cite{Pa.76}, as done in~\cite{Ch2.20}.

We end this subsection by discussing the form of the full physical solution \eqref{sol_Kerr}, the tidal field contribution \eqref{R_tidal}, and the black hole linear response contribution \eqref{R_resp}, whenever the multipolar index $\ell$ is restricted to be an integer larger than one. In particular, for any $\ell \in \mathbb{L}$, the hypergeometric function in Eq.~\eqref{sol_Kerr} reduces to a \emph{polynomial} of degree $\ell+2$ in the variable $x$, so that $(1+x)^2 R_{\ell m}(x)$ reduces to a polynomial of degree $\ell$ in $x$ in that limit. Moreover, it can be shown that part of the tidal field contribution $(1+x)^2 R_{\ell m}^\text{tidal}(x)$ will reduce to that same polynomial when $\ell \in \mathbb{L}$, while the remaining part cancels out precisely the response $2 (1+x)^2 \kappa_{\ell m} R_{\ell m}^\text{resp}(x)$ of the Kerr black hole, thus ensuring that the relationship \eqref{eq:tidal,resp comb} is indeed satisfied. See App.~\ref{app:continuity} for more details.

\subsection{Love numbers of a tidally perturbed Kerr black hole}\label{subsec:TLN}

Having determined the linear response of the black hole at the level of the Weyl scalar, we may relate this general relativistic result to the Newtonian formula \eqref{psi0_Newt_spin} derived in Sec.~\ref{sec:Newton}. Substituting for Eqs.~\eqref{tidal-resp}--\eqref{eq:tidal,resp comb} into \eqref{psi0}, we readily obtain
\beq\label{psi0_GR}
    \psi_0 = \sum_{\ell m} z_{\ell m}(v) \left[ r^{\ell-2} \left( 1 + \cdots \right) + 2\kappa_{\ell m} (r_+-r_-)^{2\ell+1} r^{-(\ell+3)} \left( 1 + \cdots \right) \right] {}_{2}Y_{\ell m}(\theta,\pAdv) \, ,
\eeq
where the dots refer to the relativistic corrections to the asymptotic behaviors \eqref{eq:tidal,resp}. Equation \eqref{psi0_GR} is a more precise version of the schematic equation \eqref{psi0_schema} written down earlier in Sec.~\ref{sec:GR}. The asymptotic form of the curvature scalar modes thus read, considering again $\ell\in\mathbb{R}$ so that it makes sense to keep both the tidal field and response contributions,
\beq\label{psi0_GR_asym}
    \psi^{\ell m} _0 \sim z_{\ell m}(v) \, r^{\ell-2} \left[ 1 + 2\kappa_{\ell m} \left(\frac{r_+ - r_-}{r}\right)^{2\ell+1} \right] {}_{2}Y_{\ell m}(\theta,\pAdv) \, .
\eeq
This expression is strikingly similar to the Newtonian formula \eqref{psi0_Newt_spin}. Recall however that the dimensionless TLNs of a Newtonian, spinning body were defined in terms of its equilibrium radius $R$ in the \textit{nonspinning} limit. Hence, motivated by the Newtonian limit we write the asymptotic formula \eqref{psi0_GR_asym} in the equivalent form
\beq\label{psi0_GR_asym_bis}
    \psi^{\ell m} _0 \sim z_{\ell m}(v) \, r^{\ell-2} \left[ 1 + 2k_{\ell m} \left(\frac{2M}{r}\right)^{2\ell+1} \right] {}_{2}Y_{\ell m}(\theta,\pAdv) \, ,
\eeq
where the areal radius $2M$ of a nonspinning (Schwarzschild) black hole plays the role of the equilibrium radius $R$ in Eq.~\eqref{psi0_Newt_spin}, and we introduced the numbers\footnote{In Ref.~\cite{LeCa.21}, we used the notation $k_{\ell m}$ for the coefficients $\kappa_{\ell m}$ in \eqref{psi0_GR_asym}, while in this paper the notation $k_{\ell m}$ is instead used for the coefficients in \eqref{psi0_GR_asym_bis}. As the formula \eqref{k_phys} shows, these two sets of numbers coincide up to quadratic order in spin.}
\beq\label{k_phys}
    k_{\ell m} \equiv (1-\ad^2)^{\ell+1/2} \, \kappa_{\ell m} = - \ui m \ad \; \frac{(\ell+2)! \, (\ell-2)!}{4(2\ell+1)!(2\ell)!} \prod_{n=1}^\ell \left[ n^2 (1-\ad^2) + m^2 \ad^2 \right] .
\eeq
For a small black hole spin $0 < \ad \ll 1$, we clearly have $k_{\ell m} = \kappa_{\ell m} + O(\chi^3)$.
In the extremal limit where $\ad \to 1$, however, the singular behavior of $\kappa_{\ell m}$ in \eqref{kappa_extremal} is exactly compensated by the factor of $(1-\ad^2)^{\ell+1/2}$ in Eq.~\eqref{k_phys}, in such a way that the coefficients $k_{\ell m}$ are regular functions of the black hole spin $\chi$. More precisely, they are mere \textit{polynomials} of degree $2\ell+1$ in the dimensionless Kerr parameter $\chi$, and only involve \textit{odd} powers of the spin.

Recall that the slowly varying coefficients in \eqref{psi0_GR_asym_bis} are given by $z_{\ell m} = \alpha_{\ell m} + \ui \beta_{\ell m}$, where $\alpha_{\ell m}$ and $\beta_{\ell m}$ are proportional to the modes $\mathcal{E}_{\ell m}$ and $\mathcal{B}_{\ell m}$ of the electric- and magnetic-type tidal moments $\mathcal{E}_L$ and $\mathcal{B}_L$ defined in \eqref{EL_BL_GR}, respectively. Since $z_{\ell m} \to \alpha_{\ell m}$ in the Newtonian limit, a comparison of Eqs.~\eqref{psi0_Newt_spin} and \eqref{psi0_GR_asym_bis} clearly reveals that the coefficients \eqref{k_phys} may be interpreted as the ``Newtonian TLNs" of a Kerr black hole. Consequently, we anticipate that the TLNs in \eqref{TLN} of a spinning black hole may only involve odd powers of the spin $\ad$.

Moreover, the relativistic formulas \eqref{psi0_GR_asym_bis}--\eqref{k_phys} suggest that the black hole spin $\chi$ might couple to the tidal multipoles $\mathcal{E}_{\ell m}$ and $\mathcal{B}_{\ell m}$ to correct the multipole moments \eqref{M_S} of the background Kerr black hole geometry. Such couplings can be characterized by means of relativistic TLNs, which were defined in Sec.~\ref{sec:GR} out of the contributions $\delta M_{\ell m}$ and $\delta S_{\ell m}$ of the black hole response to the Geroch-Hansen multipole moments of the perturbed spacetime.

Having defined the multipole moments \eqref{MLSL} associated with the Kerr black hole linear response to the perturbing tidal field, as well as the corresponding TLNs \eqref{TLN}, we shall now formulate a conjecture for their numerical values, based on an analogy with the Newtonian limit of the general relativistic problem. We first recall that for an axisymmetric, spinning, Newtonian body perturbed by an external tidal field, the Newtonian limit of the Weyl scalar modes read
\beq\label{psi0_Newt_IL}
    \lim_{c\to\infty}\psi_0^{\ell m} = \sqrt{\frac{(\ell+2)!}{(\ell-2)!}} \left( \frac{(\ell-2)!}{\ell!} \, \mathcal{E}_{\ell m} \, r^{\ell-2} - \frac{(2\ell-1)!!}{\ell!} \, \frac{I_{\ell m}}{r^{\ell+3}} \right) {}_{2}Y_{\ell m}(\theta,\pAdv) \, ,
\eeq
where $I_{\ell m}$ are the spherical-harmonic modes of the induced mass multipoles $I_L$. This formula simply follows from Eq.~\eqref{psi0_Newt_spinning} with \eqref{alpha} and \eqref{plop}.

Now, by analogy with the expression in \eqref{psi0_Newt_IL} for the Newtonian limit of the Weyl scalar modes, it is natural to conjecture that the black hole response contributions $\delta M_{\ell m}$ and $\delta S_{\ell m}$ to the Geroch-Hansen multipole moments [recall Eq.~\eqref{MLSL}] might appear in the asymptotic expansion of the general relativistic $\psi_0^{\ell m}$ according to
\beq\label{psi0_conjecture}
    \psi_0^{\ell m}\sim 
    \sqrt{\frac{(\ell+2)!}{(\ell-2)!}} \left( \frac{(\ell-2)!}{\ell!} \, \mathcal{T}_{\ell m} \, r^{\ell-2} - \frac{(2\ell-1)!!}{\ell!} \, \frac{Q_{\ell m}}{r^{\ell+3}} \right) {}_{2}Y_{\ell m}(\theta,\pAdv) \, ,
\eeq
where we defined the linear combinations $\mathcal{T}_{\ell m} \equiv \mathcal{E}_{\ell m} + \ui \, \frac{\ell+1}{3} \mathcal{B}_{\ell m}$ and $Q_{\ell m} \equiv \delta M_{\ell m} + \ui \, \delta S_{\ell m}$, the coefficient of $\mathcal{B}_{\ell m}$ arising from the ratio of coefficients in \eqref{alphas-betas}. Indeed, the tidal moments $\mathcal{E}_{\ell m}$ that appear in the tidal field contribution to the Newtonian result \eqref{psi0_Newt_IL} become promoted, in general relativity, to the complex combination $\mathcal{T}_{\ell m}$ of $\mathcal{E}_{\ell m}$ and $\mathcal{B}_{\ell m}$ in \eqref{psi0_conjecture}. By analogy, we may expect that the induced mass multipole moments \eqref{eq:Newt multip} in the Newtonian body's response in \eqref{psi0_Newt_IL} become promoted in \eqref{psi0_conjecture} to the complex combination $Q_{\ell m} = \delta M_{\ell m} + \ui \, \delta S_{\ell m}$ of the relativistic multipole moments associated with the Kerr black hole linear response, which is analogous to that appearing in the Hansen formula  \eqref{M_S}.

Finally, the independent asymptotic forms \eqref{psi0_GR_asym_bis} and \eqref{psi0_conjecture} of $\psi_0$ will both be satisfied if, and only if,
\beq\label{legal?}
    \delta M_{\ell m} + \ui \, \delta S_{\ell m} = - \frac{2(\ell-2)!}{(2\ell-1)!!} \, (2M)^{2\ell+1} \, k_{\ell m} \left( \mathcal{E}_{\ell m} + \ui \, \frac{\ell+1}{3} \mathcal{B}_{\ell m} \right) .
\eeq
Comparing with the formulas \eqref{MLSL}, this expression leads us to the following observation: \emph{if} the Geroch-Hansen multipole moments $\delta M_L$ and $\delta S_L$ associated with the black hole response $h_{\alpha\beta}^\text{resp}$ appear in the asymptotic form of the curvature Weyl scalar $\psi_0$ as in Eq.~\eqref{psi0_conjecture}, \emph{then} the TLNs \eqref{TLN_bis} of a Kerr black hole must obey the constraints
\beq\label{legal!}
    \lambda_{\ell m}^{M\mathcal{E}} + \ui \lambda_{\ell m}^{S\mathcal{E}} = \frac{3}{\ell+1} \left( \lambda_{\ell m}^{S\mathcal{B}} - \ui \lambda_{\ell m}^{M\mathcal{B}} \right) = - \frac{2(\ell-2)!}{(2\ell-1)!!} \, (2M)^{2\ell+1} \, k_{\ell m} \, ,
\eeq
where we recall that the constant $k_{\ell m}$ is given by \eqref{k_phys}. These two formulas can be combined with the general properties \eqref{lambda_sym} obeyed by the four families of TLNs, yielding individual expressions for $\lambda_{\ell m}^{M\mathcal{E}}$, $\lambda_{\ell m}^{S\mathcal{E}}$, $\lambda_{\ell m}^{S\mathcal{B}}$ and $\lambda_{\ell m}^{M\mathcal{B}}$. We readily obtain the following conjectured formulas, for any $\ell \in \mathbb{L}$ and $|m| \leqslant \ell$:
\begin{subequations}\label{TLN_conjecture}
    \begin{align}
        \lambda_{\ell m}^{M\mathcal{E}} &= \frac{3}{\ell+1} \, \lambda_{\ell m}^{S\mathcal{B}} = \ui m \ad \; \frac{(\ell+2)! \, [(\ell-2)!]^2 \, (2M)^{2\ell+1}}{2(2\ell+1)!(2\ell)!(2\ell-1)!!} \prod_{n=1}^\ell \left[ n^2 (1-\ad^2) + m^2 \ad^2 \right] , \label{zoubi} \\
        \lambda_{\ell m}^{S\mathcal{E}} &= \lambda_{\ell m}^{M\mathcal{B}} = 0 \, .
    \end{align}    
\end{subequations}
Remarkably, the two families of nonvanishing TLNs would take the form of \textit{polynomials} of degree $2\ell+1$ in the dimensionless Kerr parameter $\ad$, with only \textit{odd} powers of the spin and only \textit{odd} powers of $m$, and would display no $\ell$-mode coupling nor parity mixing. The absence of $\ell$-mode coupling is special as, generically, the linear response of a spinning body allows for such coupling; see Eq.~\eqref{dis-donc}. According to the analysis in Sec.~\ref{subsec:U-psi0_spin} (see Eqs.~\eqref{TLT_exp}--\eqref{eq:symm TLN} there), this would imply that the TLTs $\lambda_{LL'}^{M\mathcal{E}}$ and $\lambda_{LL'}^{S\mathcal{B}}$ are \textit{anti-symmetric} by exchange of the multi-indices $L$ and $L'$. Importantly, \eqref{TLN_conjecture} suggest that the mass-type (resp. current-type) multipole moments are sourced uniquely by the electric-type (resp. magnetic-type) tidal fields. This appealing conjecture will be tested in the remainder of this paper, where we shall compute the quadrupolar ($\ell=2$) TLNs of a slowly spinning Kerr black hole from the contribution of its linear response (proportional to $k_{\ell m}$) to the geometrically-defined Geroch- Hansen multipole moments.

For a small black hole spin $0 < \ad \ll 1$, Eq.~\eqref{zoubi} implies that the nonvanishing TLNs would then verify
\beq
    \lambda_{\ell m}^{M\mathcal{E}} = \frac{3}{\ell+1} \, \lambda_{\ell m}^{S\mathcal{B}} = \frac{\ui m \ad \, (\ell+2)! \, [\ell !(\ell-2)!]^2}{2(2\ell+1)!(2\ell)!(2\ell-1)!!} \, (2M)^{2\ell+1} + O(\ad^3) \, .
\eeq
In particular, at quadrupolar order this would imply $\lambda_{2m}^{M\mathcal{E}} = \lambda_{2m}^{S\mathcal{B}} = \ui m \ad \, (2M)^5 / 180 + O(\ad^3)$. For an extremal black hole ($\ad = 1$), the formula \eqref{zoubi} reduces to
\beq
    \lambda_{\ell m}^{M\mathcal{E}} = \frac{3}{\ell+1} \, \lambda_{\ell m}^{S\mathcal{B}} = \ui m^{2\ell+1} \, \frac{2^{2\ell} (\ell+2)! \, [(\ell-2)!]^2}{(2\ell+1)!(2\ell)!(2\ell-1)!!} \, M^{2\ell+1} \, .
\eeq

\section{Hertz potential}\label{sec:Hertz}

Using the Hertz potential formalism developed for gravity by Cohen and Kegeles~\cite{CoKe.74,KeCo.79}, Chrzanowksi~\cite{Ch.75}, Stewart~\cite{St.79} and Wald~\cite{Wa.78}, it is possible to construct a metric perturbation from source-free solutions to the Teukolsky equation. (See for instance Refs.~\cite{Ke.al2.10,vdMSh.15} for more recent discussions in the context of the gravitational self-force framework.) More precisely, the metric perturbation can be reconstructed in an ingoing radiation gauge (IRG), from the knowledge of a Hertz potential $\Psi$ which is a solution of the homogeneous Teukolsky equation with spin $s=-2$, and which obeys the fourth-order, linear differential equation\footnote{As pointed out by Keidl et al.~\cite{Ke.al2.10}, the papers~\cite{LoWh.02} and~\cite{Or.03} share an incorrect factor of two in each of their equations for $\Psi$, that is inherited from an error in~\cite{KeCo.79}. This mistake propagated in~\cite{YuGo.06} as well. \label{footnote}}~\cite{Wa.78,LoWh.02,Or.03,Ke.al2.10,vdMSh.15}
\beq\label{psi0_Psi}
	\bm{D} \circ \bm{D} \circ \bm{D} \circ \bm{D} (\bar{\Psi}) = 2\psi_0 \, , 
\eeq
where $\bm{D} \equiv \ell^\alpha \partial_\alpha = 2\bigl[(r^2+a^2)\partial_v + (\Delta/2)\partial_r + a\partial_\pAdv\big]/\Delta$ (see Eq.~\eqref{tetrad} below) is the directional derivative operator along the principal null outgoing direction $\ell^\alpha$ of the Kerr background. Since $\Psi$ has spin weight $s = -2$, the identity ${}_{s}\bar{Y}_{\ell m} = (-)^{m+s} {}_{-s}Y_{\ell \, -m}$ implies that $\bar{\Psi}$ can be decomposed over modes proportional to ${}_{2}Y_{\ell m}$. We thus look for a complex conjugated Hertz potential in the form
\beq\label{Psibar}
	\bar{\Psi} = \sum_{\ell m} z_{\ell m}(v) \, G_{\ell m}(r) \, {}_{2}Y_{\ell m}(\theta,\pAdv) \equiv \sum_{\ell m} \bar{\Psi}_{\ell m} \, ,
\eeq
where $G_{\ell m}(r)$ is a radial function to be determined, which must satisfy the radial Teukolsky equation for spin $s=-2$. Ori~\cite{Or.03} showed how the differential equation \eqref{psi0_Psi} can be solved in the frequency domain, with the use of the Teukolsky-Starobinsky identity (see e.g.,~\cite{Cha}), to express the modes of $\bar{\Psi}$ in terms of those of $\psi_0$, according to (correcting for the factor of two in Ref.~\cite{Or.03} mentioned in footnote \ref{footnote})
\beq\label{Psibar_lm}
	\bar{\Psi}_{\ell m} = \frac{2}{p_\ell} \, \Delta^2 \, \bm{D}^\dagger \circ \bm{D}^\dagger \circ \bm{D}^\dagger \circ \bm{D}^\dagger ( \Delta^2 \psi^{\ell m}_0) \, ,
\eeq
where $\bm{D}^\dagger$ is a differential operator given below and $p_\ell$ is a constant which, in the case $|s| = 2$ of interest, is given by $p_\ell = [(\ell+2)(\ell+1)\ell(\ell-1)]^2$. Using advanced Kerr coordinates and neglecting again partial time derivatives, the differential operator simply reduces to~\cite{Or.03,YuGo.06}
\beq\label{D}
	\bm{D}^\dagger = \partial_r = (r_+ - r_-)^{-1} \, \partial_x \, .
\eeq

Most remarkably, the factor of $\Delta^2 = (r_+-r_-)^4 x^2(1+x)^2$ that multiplies $\psi^{\ell m}_0$ in Eq.~\eqref{Psibar_lm} cancels out precisely the factor of $x^{-2}(1+x)^{-2}$ that appears in \eqref{psi0_lm}, in such a way that the derivative operator \eqref{D} acts solely on the (regularized) hypergeometric function. More precisely, substituting for \eqref{psi0_lm} and \eqref{D} into the formula \eqref{Psibar_lm}, and making use of elementary properties of the hypergeometric functions, we find\footnote{For the particular case $\ell = 2$, this general formula agrees with Eq.~(34) of Ref.~\cite{YuGo.06}, up to a factor of two, as explained in the footnote \ref{footnote} (and up to a sign, seemingly due to different sign conventions), namely $G_{2m}(r) = (r-r_+)^2(r-r_-)^2/12$, which does not depend on $m$.}
\begin{align}\label{G}
	G_{\ell m}(x) &= \frac{2}{p_\ell} \, \Delta^2 \, \frac{\ud^4}{\ud x^4} \bigl[x^2(1+x)^2 R_{\ell m}(x) \bigr] \nonumber \\
	              &= \frac{2a_{\ell m}}{p_\ell} \, \Delta^2 \, \frac{\ud^4}{\ud x^4}\mathbf{F}(-\ell-2,\ell-1,-1+2\ui m \gamma;-x) \nonumber \\
				  &= 2 (r_+-r_-)^{\ell+2} \, \frac{\Gamma(\ell+1+2\ui m \gamma) \Gamma^2(\ell+3) \, }{\Gamma(2\ell+1)\Gamma(\ell-1)} \, \frac{x^2 (1+x)^2 \, \mathbf{F}(-\ell+2,\ell+3,3+2\ui m \gamma;-x)}{[(\ell+2)(\ell+1)\ell(\ell-1)]^2} \, .
\end{align}
For a general ($\ell,m$) mode, we notice that $G_{\ell \, -m} = \bar{G}_{\ell m}$. We checked that the Hertz potential $\Psi$ that is constructed from the radial function \eqref{G} is a solution of the Teukolsky equation with spin $s=-2$. We further checked that the complex conjugated Hertz potential \eqref{Psibar} with \eqref{G} obeys the differential equation \eqref{psi0_Psi}.

Let us here look at the behavior of the physical Hertz potential on the horizon. Clearly from Eqs.~\eqref{Psibar}, \eqref{G} and Eq.~(15.2.1)~\cite{NIST:DLMF}, $\Psi_{\ell m}/x^2\propto \bar G_{\ell m}/x^2$ is a smooth function as $x\to 0$ when using the Kinnersley tetrad. As mentioned, $\Psi$ satisfies the Teukolsky equation with spin $s=-2$, which is also satisfied by ($\rho^{-4}$ times) the Weyl scalar $\psi_4$~\cite{Te.73}. It then  follows from \eqref{reg} that $\Psi_{\ell m}$ in the HH tetrad is a smooth function as $x\to 0$.

Next, we shall identify the contributions of the applied tidal field and of the response of the Kerr black hole to the Hertz potential \eqref{Psibar}, similarly to what has been done for the Weyl scalar $\psi_0$ in Sec.~\ref{subsec:response}. Substituting for Eqs.~\eqref{sol_Kerr}--\eqref{psi0_lm} and \eqref{eq:tidal,resp comb} into \eqref{Psibar}--\eqref{Psibar_lm}, we find that the radial part \eqref{G} of the (complex conjugated) physical Hertz potential can be decomposed into a linear combination of  tidal field and black hole response contributions, according to
\beq\label{G=Gtidal+Gresp}
    G_{\ell m}(x) = G_{\ell m}^\text{tidal}(x) + 2\kappa_{\ell m} G_{\ell m}^\text{resp}(x) \, ,
\eeq
where the coefficients $\kappa_{\ell m}$ are given in \eqref{kappa}--\eqref{kappa_phys}, while $G_{\ell m}^\text{tidal}(x)$ and $G_{\ell m}^\text{resp}(x)$ are obtained by applying the formula in the first line of \eqref{G} to, respectively, $R_{\ell m}^\text{tidal}(x)$ and $R_{\ell m}^\text{resp}(x)$ instead of $R_{\ell m}(x)$. Using the explicit expressions \eqref{tidal-resp}, we readily find
\begin{subequations}\label{Gtidal-Gresp}
    \begin{align}
        G_{\ell m}^\text{tidal}(x) &= 2 (r_+-r_-)^{\ell+2} \, \frac{\Gamma(\ell+3)}{\Gamma(\ell-1)} \, \frac{x^\ell (1+x)^2 \, F(-\ell+2,-\ell-2\ui m \gamma,-2\ell;-1/x)}{[(\ell+2)(\ell+1)\ell(\ell-1)]^2} \, , \label{G_tidal} \\
        G_{\ell m}^\text{resp}(x) &= 2 (r_+-r_-)^{\ell+2} \, \frac{\Gamma(\ell+3)}{\Gamma(\ell-1)} \, \frac{(1+x)^2 \, F(\ell+3,\ell+1-2\ui m \gamma,2\ell+2;-1/x)}{x^{\ell+1} \,  [(\ell+2)(\ell+1)\ell(\ell-1)]^2} \, . \label{G_resp}
    \end{align}
\end{subequations}
The asymptotic behaviors of those fields as $x \to \infty$ are $G_{\ell m}^\text{tidal}(x) \!\propto\! x^{\ell+2}$ and $G_{\ell m}^\text{resp}(x) \!\propto\! x^{-\ell+1}$. Alternatively, we could have directly identified the external tidal field and black hole response contributions in \eqref{G=Gtidal+Gresp} by applying the identity (15.8.2) in Ref.~\cite{NIST:DLMF} to the hypergeometric function that appears in Eq.~\eqref{G}.

Let us now assess the analytical behaviors of the physical Hertz potential \eqref{G}, the tidal field contribution \eqref{G_tidal} and the black hole response contribution \eqref{G_resp}, when the multipolar index is restricted to the physical space, i.e., when $\ell \in \mathbb{L}$. In that limit, the hypergeometric function in \eqref{G} reduces to a \emph{polynomial} of degree $\ell-2$ in the variable $x = (r-r_+) / (r_+-r_-)$. The contribution of multipolar order $(\ell,m)$ to the Hertz potential \eqref{Psibar} can thus be written in the form
\beq\label{Psi}
	\Psi_{\ell m} \propto x^2 (1+x)^2 \sum_{k=0}^{\ell-2} c_k^{\ell m} x^k \propto (r-r_+)^2 (r-r_-)^2 \sum_{k=0}^{\ell-2} d_k^{\ell m} r^k \propto \sum_{k=0}^{\ell+2} e_k^{\ell m} r^k \, ,
\eeq
where $c_k^{\ell m}$, $d_k^{\ell m}$, $e_k^{\ell m}$ are constant coefficients. In particular, the coefficients $c_k^{\ell m}$ are dimensionless functions of the product $\ui m\gamma$, and have a smooth limit as $\gamma \to 0$ (nonspinning black hole). Despite the fact that the Hertz potential in Eq.~\eqref{Psi} is proportional to a polynomial of degree $\ell +2$ in the radial coordinate $r$, it does \emph{not} physically correspond to an external $\ell$-pole tidal field perturbation that includes relativistic corrections, with a corresponding vanishing linear response of the spinning black hole. Such an argument of series truncation has been appealed to in previous work~\cite{BiPo.09,Pa.al.15} in order to identify the tidal field and linear response contributions to the metric perturbation. As our analysis reveals, most dramatically at the level of the Hertz potential $\Psi$, despite being intuitive and appealing, this prescription of series truncation is erroneous. Instead, as our work illustrates, an analytic continuation of the multipole index $\ell \in \mathbb{R}$ provides an unambiguous identification of the two physical contributions to the full solution. By contrast, for any $\ell \in \mathbb{L}$, neither the tidal field contribution \eqref{G_tidal}, nor the black hole response \eqref{G_resp} reduce to a polynomial in $x$. Rather, just like for the Weyl scalar $\psi_0$ (detailed in App.~\ref{app:continuity}), part of the tidal contribution $G_{\ell m}^\text{tidal}(x)$ cancels out identically the black hole response contribution $G_{\ell m}^\text{resp}(x)$, leaving a polynomial of degree $\ell+2$ in $x$ that coincides with the full Hertz potential $G_{\ell m}(x)$.

Equations \eqref{G} and \eqref{Gtidal-Gresp} have a smooth limit as $m \gamma \to 0$, which correspond either to a nonspinning black hole ($\gamma \!=\! 0$) or to an axisymmetric ($m \!=\! 0$) perturbation of a spinning black hole. In both cases, the constant $\kappa_{\ell m}$ in \eqref{G=Gtidal+Gresp} vanishes, so that $G_{\ell m} = G_{\ell m}^\text{tidal}$. Interestingly, when $m\gamma = 0$ the hypergeometric functions in Eqs.~\eqref{G} and 
\eqref{Gtidal-Gresp} can be expressed in terms of the ordinary Legendre functions of degree $\ell$ of the first and second kind, $P_\ell(z)$ and $Q_\ell(z)$, using Eqs.~(14.3.8) and (14.7.11) in Ref.~\cite{NIST:DLMF} for $G_{\ell m}(x)$, as well as Eqs.~(15.8.13), (14.3.7) and (14.7.12) for $G_{\ell m}^\text{resp}(x)$. Indeed, for a \emph{non}spinning black hole and for any $\ell \in \mathbb{L}$, we find from, respectively, Eqs.~\eqref{G} and \eqref{G_resp},
\begin{subequations}\label{G_tidal_resp_Schw}
    \begin{align}
	    G_{\ell m}(r) &= G_{\ell m}^\text{tidal}(r) = (2M)^{\ell+2} \, \frac{\ell! \, [(\ell-2)!]^2}{2(2\ell)!(\ell+2)!} \; (r/M)^2 (r/M-2)^2 P''_\ell(r/M-1) \, , \label{G_Schw} \\
	    G_{\ell m}^\text{resp}(r) &= (2M)^{\ell+2} \, \frac{(2\ell+1)!(\ell-2)!}{\ell!\, [(\ell+2)!]^2} \, (r/M)^2 (r/M-2)^2 Q''_\ell(r/M-1) \, , \label{G_resp_Schw}
    \end{align}
\end{subequations}
where we recall that $P'_\ell(z) \equiv \ud P_\ell / \ud z$. While the function $P_\ell(z)$ is a polynomial of degree $\ell$ in $z$, the function $Q_\ell(z)$ scales as $z^{-\ell-1}$ for large $z$, in agreement with the known asymptotic behaviors of $G_{\ell m}^\text{tidal}(r)$ and $G_{\ell m}^\text{resp}(r)$. The expressions \eqref{G_tidal_resp_Schw} will prove useful in the next Sec.~\ref{sec:metric}, as we shall reconstruct the metric perturbation out of the Hertz potential.

\section{Metric perturbation}\label{sec:metric}

Given a Newman-Penrose null and complex tetrad $(\ell^\alpha,n^\alpha,m^\alpha,\bar{m}^\alpha)$, normalized as $\ell^\alpha n_\alpha = -1$ and $m^\alpha \bar{m}_\alpha = 1$, all other inner products vanishing, a generic metric perturbation to the background Kerr metric $\mathring{g}_{\alpha\beta}$ can be written as
\begin{align}\label{h}
	h_{\alpha\beta} &= \ell_\alpha \ell_\beta h_{nn} + 2\ell_{(\alpha} n_{\beta)} h_{n\ell} - 2 \ell_{(\alpha} m_{\beta)} h_{n\bar{m}} - 2 \ell_{(\alpha} \bar{m}_{\beta)} h_{nm} + n_\alpha n_\beta h_{\ell\ell} \nonumber \\
					&- 2 n_{(\alpha} m_{\beta)} h_{\ell \bar{m}} - 2 n_{(\alpha} \bar{m}_{\beta)} h_{\ell m} + m_\alpha m_\beta \, h_{\bar{m}\bar{m}} + 2 m_{(\alpha} \bar{m}_{\beta)} h_{m\bar{m}} + \bar{m}_\alpha \bar{m}_\beta h_{mm} \, ,
\end{align}
where $h_{nn} \equiv n^\alpha n^\beta h_{\alpha\beta}$ and similarly for all the other tetrad components of the metric perturbation. One can consistently impose in the vacuum domain $r_+ \leqslant r \ll \mathcal{R}$ the \emph{five} transverse and traceless IRG conditions~\cite{Pr.al.07}
\beq\label{IRG}
	\ell^\alpha h_{\alpha\beta} = 0 \quad \text{and} \quad \mathring{g}^{\alpha\beta} h_{\alpha\beta} = 0 \, ,
\eeq
where $\mathring{g}^{\alpha\beta}$ is the inverse metric of the background Kerr spacetime. Substituting for Eq.~\eqref{h} into \eqref{IRG} yields the five constraints $h_{\ell\ell} = h_{\ell n} = h_{\ell m} = h_{\ell\bar{m}} = 0$ and $h_{m\bar{m}} = h_{\ell n} = 0$. In the IRG, the general formula \eqref{h} thus reduces to
\beq\label{h_IRG}
	h_{\alpha\beta} = \ell_\alpha \ell_\beta h_{nn} - 4 \Re{[\ell_{(\alpha} m_{\beta)} h_{n\bar{m}}]} + 2 \Re{[m_\alpha m_\beta h_{\bar{m}\bar{m}}]} \, ,
\eeq
so that the five remaining degrees of freedom in the perturbation $h_{\alpha\beta}$ are encoded in the real tetrad component $h_{nn}$ and the complex tetrad components $h_{n\bar{m}}$ and $h_{\bar{m}\bar{m}}$. These nonvanishing tetrad components are themselves obtained, up to parts for which $\psi_0$ vanishes (see the comment at the end of this section), by applying certain second-order differential operators on the Hertz potential $\Psi$, namely~\cite{LoWh.02,YuGo.06}
\begin{subequations}\label{metric}
	\begin{align}
		h_{nn} &= - (\bm{\delta} + 2\beta + \bar{\pi} - \tau) \circ (\bm{\delta} + 4\beta + 3\tau) \Psi + \text{c.c.}  \, , \label{h_nn} \\
		h_{n\bar{m}} &= - \frac{1}{2} \left\{ (\bm{\delta} + 4\beta - 2\bar{\pi} - \tau) \circ (\bm{D} + 3 \rho) + (\bm{D} + \bar{\rho} - \rho) \circ (\bm{\delta} + 4\beta + 3\tau) \right\} \Psi \, , \\
		h_{\bar{m}\bar{m}} &= - (\bm{D} - \rho)\circ(\bm{D} + 3 \rho) \Psi \, ,
	\end{align}
\end{subequations}
together with $h_{nm} = \bar{h}_{n\bar{m}}$ and $h_{mm} = \bar{h}_{\bar{m}\bar{m}}$. Here, $\bm{D} \equiv \ell^\alpha \partial_\alpha$ and $\bm{\delta} \equiv m^\alpha \partial_\alpha$, while the Greek letters are the usual spin coefficients of Kerr background in the Newman-Penrose formalism. In advanced Kerr coordinates, the relevant spin coefficients are explicitly given by~\cite{LoWh.02,YuGo.06}
\beq\label{spin_coeff}
		\rho = - \frac{1}{r - \ui a \cos{\theta}} \, , \quad \beta = - \bar{\rho} \, \frac{\cot{\theta}}{2\sqrt{2}} \, , \quad \pi = \ui a \rho^2 \, \frac{\sin{\theta}}{\sqrt{2}} \, , \quad  \tau = - \ui a \rho \bar{\rho} \, \frac{\sin{\theta}}{\sqrt{2}} \, .
\eeq
These expressions coincide with those obtained using the more conventional Boyer-Lindquist coordinates~\cite{BoLi.67} because the coordinate transformation from advanced Kerr coordinates to Boyer-Lindquist coordinates leaves $r$ and $\theta$ unchanged; see Eqs.~\eqref{Kerr->BL} below.

The directional derivative operators $\bm{D}$ and $\bm{\delta}$ in Eq.~\eqref{metric} involve the elements $\ell^\alpha$ and $m^\alpha$ of the Kinnersley tetrad, while the nonvanishing IRG metric perturbation components \eqref{h_IRG} depend on $\ell_\alpha$ and $m_\alpha$. In advanced Kerr coordinates $(v,r,\theta,\pAdv)$, their coordinate components read as\footnote{There are typos in Eqs.~(42) of Ref.~\cite{YuGo.06}, in which the covariant tetrad components $\ell_\alpha$, $n_\alpha$ and $m_\alpha$ are all off by an overall minus sign.}
\begin{subequations}\label{tetrad}
	\begin{align}
		\ell^\alpha &= \left( 2 \frac{r^2+a^2}{\Delta}, 1, 0, \frac{2a}{\Delta} \right) , \quad m^\alpha = - \frac{\bar{\rho}}{\sqrt{2}} \left( \ui a \sin{\theta}, 0, 1, \frac{\ui}{\sin{\theta}} \right) , \\
		\ell_\alpha &= \left( - 1, \frac{2\Sigma}{\Delta}, 0, a \sin^2{\theta} \right) , \quad m_\alpha = - \frac{\bar{\rho}}{\sqrt{2}} \Bigl( - \ui a \sin{\theta}, 0, \Sigma, \ui (r^2 + a^2) \sin{\theta} \Bigr) \, .
	\end{align}
\end{subequations}
We may now express the advanced Kerr coordinate components of the IRG metric perturbation \eqref{h_IRG} in terms of the nonvanishing tetrad components \eqref{metric}, or more precisely in terms of $h_{nn}$, $\bar{\rho} h_{n\bar{m}}$ and $\bar{\rho}^2 h_{\bar{m}\bar{m}}$, according to
\begin{subequations}\label{h_comp}
	\begin{align}
		h_{vv} &= h_{nn} - 2 \sqrt{2} \, a \sin{\theta} \; \Im{(\bar{\rho} h_{n\bar{m}})} - a^2 \sin^2{\theta} \; \Re{(\bar{\rho}^2 h_{\bar{m}\bar{m}})} \, , \label{h_vv} \\
		h_{vr} &= - \frac{2\Sigma}{\Delta} \, \bigl( h_{nn} - \sqrt{2} \, a \sin{\theta} \; \Im{(\bar{\rho} h_{n\bar{m}})} \bigr) \, , \\
		h_{v\theta} &= - \sqrt{2} \Sigma \, \Re{(\bar{\rho} h_{n\bar{m}})} + a \sin{\theta} \, \Sigma \, \Im{(\bar{\rho}^2 h_{\bar{m}\bar{m}})} \, , \\
		h_{v\pAdv} &= \sqrt{2} \, \sin{\theta} \, (\Sigma + 2a^2 \sin^2{\theta}) \, \Im{(\bar{\rho} h_{n\bar{m}})} - a \sin^2{\theta} \, \bigl( h_{nn} - (r^2 + a^2) \Re{(\bar{\rho}^2 h_{\bar{m}\bar{m}})} \bigr) \, , \\
		h_{rr} &= \frac{4\Sigma^2}{\Delta^2} \, h_{nn} \, , \\
		h_{r\theta} &= 2\sqrt{2} \, \frac{\Sigma^2}{\Delta} \, \Re{(\bar{\rho} h_{n\bar{m}})} \, , \\
		h_{r\pAdv} &= \frac{2\Sigma}{\Delta} \, a \sin^2{\theta} \, h_{nn} - 2\sqrt{2} \, \frac{\Sigma}{\Delta} (r^2 + a^2) \sin{\theta} \, \Im{(\bar{\rho} h_{n\bar{m}})} \, , \\
		h_{\theta\theta} &= \Sigma^2 \, \Re{(\bar{\rho}^2 h_{\bar{m}\bar{m}})} \, , \\
		h_{\theta\pAdv} &= \sqrt{2} \, a \sin^2{\theta} \, \Sigma \, \Re{(\bar{\rho} h_{n\bar{m}})} - \sin{\theta} \, \Sigma (r^2 + a^2) \, \Im{(\bar{\rho}^2 h_{\bar{m}\bar{m}})} \, , \label{h_thetaphi Kerr}\\
		h_{\pAdv\pAdv} &= a^2 \sin^4{\theta} \, h_{nn} - 2 \sqrt{2} \, a \sin^3{\theta} \, (r^2 + a^2) \, \Im{(\bar{\rho} h_{n\bar{m}})} - (r^2 + a^2)^2 \sin^2{\theta} \, \Re{(\bar{\rho}^2 h_{\bar{m}\bar{m}})} \, .
	\end{align}
\end{subequations}
The metric reconstruction detailed above determines the physical metric perturbation only up to a stationary and axisymmetric piece whose curvature scalar $\psi_0$ vanishes identically, and up to a gauge transformation. In the vacuum region $r_+ \leqslant r \ll \mathcal{R}$, this so-called completion piece reduces to a linear combination of mass and angular momentum perturbations~\cite{Wa.73,Me.al.16,vdM2.17}, which can always be absorbed into a redefinition of the background Kerr black hole mass and spin, and thus cannot affect the functional form of the higher-order Geroch-Hansen multipole moments of the perturbed Kerr geometry. Therefore, the completion piece of the full, physical metric perturbation will not affect our calculation of the TLNs of a Kerr black hole, to be performed in Sec.~\ref{sec:multipoles} below.  

\subsection{Tidally perturbed Schwarzschild black hole}\label{subsec:Schw}

We first consider the simple case of a nonspinning black hole, for which the linear response was shown to vanish identically [recall Eq.~\eqref{kappa=0}], so that $h_{\alpha\beta} = h_{\alpha\beta}^\text{tidal}$. Setting $a = 0$ in the expressions \eqref{h_comp} of the coordinate components of the metric perturbation  in advanced Kerr coordinates (which reduce to advanced Eddington-Finkelstein coordinates in that limit), we obtain the simple relationships
\begin{subequations}\label{h_comp_Schw}
	\begin{align}
		&h_{vv} = h_{nn} \, , \label{h_vv-h_nn} \\
		& h_{rr} = - 2 f^{-1} h_{vr} = 4 f^{-2} h_{nn} \, ,
		\label{h_rr-h_nn}
		\\
		&h_{r\pAdv} + \ui \sin{\theta} \, h_{r\theta} = -2f^{-1} (h_{v\pAdv} + \ui \sin{\theta} \, h_{v\theta}) = 
		- 2\sqrt{2} \, \ui f^{-1} r \sin{\theta} \, h_{n\bar{m}} \, , \label{h_rphi}\\
		&h_{\pAdv\pAdv} + \ui \sin{\theta} \, h_{\theta\pAdv} = - r^2 \sin^2{\theta} \, h_{\bar{m}\bar{m}} \, , \label{h_phiphi} \\
		&h_{\theta\theta} = - h_{\pAdv\pAdv} / \sin^2{\theta} \, ,
	\end{align}
\end{subequations}
where we introduced $f \equiv 1 - 2M/r$. Moreover, by setting $a = 0$ in the formulas \eqref{metric}--\eqref{tetrad}, we find $\rho = \bar{\rho} = - 1/r$ and $\pi = \tau = 0$. Neglecting as usual the small partial time derivatives, so that $\bm{D} \Psi = \partial_r \Psi$, the nonvanishing tetrad components \eqref{metric} now simply read
\begin{subequations}\label{tetrad_Schw}
	\begin{align}
		h_{nn} &= - \frac{1}{2r^2} \, \eth_{-1} \circ \eth_{-2} \Psi + \text{c.c.} \ , \label{truc} \\
		h_{n\bar{m}} &=  \frac{1}{\sqrt{2}r} \, (\partial_r - 2/r) \circ \eth_{-2} \Psi \, , \label{bidule} \\
		h_{\bar{m}\bar{m}} &= - (\partial_r - 2/r) \circ \partial_r \Psi \, , \label{bidule2}
	\end{align}
\end{subequations}
where we recognize in Eqs.~\eqref{truc} and \eqref{bidule} the spin lowering operators $\eth_{-1}$ and $\eth_{-2}$ acting on ${}_{2}\bar{Y}_{\ell m} \propto \exp{(-\ui m \pAdv)}$ [see Eq.~\eqref{eq:eth Ycc}], thanks to the mode decomposition \eqref{Psibar} of $\bar{\Psi}$. In the nonspinning limit, the radial part of the ($\ell,m$) mode of the (complex conjugated) Hertz potential is given by Eq.~\eqref{G_Schw}, so that the nonvanishing tetrad components \eqref{tetrad_Schw} become
\begin{subequations}\label{metric_bis}
	\begin{align}
		h_{nn} &= - \sum_{\ell = 2}^\infty C_\ell \; \mathcal{P}_\ell(r) \left( \sum_{m = -\ell}^\ell \bar{z}_{\ell m}(v) \, \bar{Y}_{\ell m}(\theta,\pAdv) + \text{c.c.} \right) , \label{h_nn_Schw} \\
		h_{n\bar{m}} &= - \sqrt{2} \sum_{\ell = 2}^\infty C_\ell \, \sqrt{\frac{(\ell-1)!}{(\ell+1)!}} \; r \mathcal{P}'_\ell(r) \sum_{m = -\ell}^\ell \bar{z}_{\ell m}(v) \, {}_{1}\bar{Y}_{\ell m}(\theta,\pAdv) \, , \label{bob} \\
		h_{\bar{m}\bar{m}} &= - 2\sum_{\ell = 2}^\infty C_\ell \, \sqrt{\frac{(\ell-2)!}{(\ell+2)!}} \left[ r^2 \mathcal{P}''_\ell(r) + 2r \mathcal{P}'_\ell(r) - 2\mathcal{P}_\ell(r) \right] \sum_{m = -\ell}^\ell \bar{z}_{\ell m}(v) \, {}_{2}\bar{Y}_{\ell m}(\theta,\pAdv) \, , \label{blip}
	\end{align}
\end{subequations}
where $\mathcal{P}_\ell(r) \equiv (r/M-2)^2 P_\ell''(r/M-1)$ is a real-valued polynomial of degree $\ell$ in the radial coordinate $r$, with a double zero at the location of the Schwarzschild black hole event horizon, and we introduced the $\ell$-dependant coefficients
\beq\label{C_l}
    C_\ell \equiv \frac{\ell!\left[(\ell-2)!\right]^{3/2}}{(2\ell)!\left[(\ell+2)!\right]^{1/2}} \; (2M)^\ell .
\eeq
Owing to the overall factor of $r^2$ in the radial part of $\Psi_{\ell m}$ in the complex conjugate of \eqref{G_Schw}, we notice that the nonvanishing tetrad components $h_{nn}$, $h_{n\bar{m}}$ and $h_{\bar{m}\bar{m}}$ are all polynomials (of degree $\ell$) in $r$, so that the metric perturbation coordinate components \eqref{h_comp_Schw} all share the polynomial behavior. Together with the fact that $\mathcal{P}_\ell(r)$ has at least a double zero at $r \!=\! 2M$, it follows that the IRG metric perturbation $h_{\alpha\beta}$ is smooth on the future event horizon in advanced Kerr coordinates; the only nonvanishing coordinate components there are $h_{\theta\theta}$, $h_{\theta\pAdv}$ and $h_{\pAdv\pAdv} = - \sin^2{\theta} \, h_{\theta\theta}$.

Recall that the slowly varying coefficients in Eq.~\eqref{metric_bis} are given by $\bar{z}_{\ell m} \!=\! \bar{\alpha}_{\ell m} - \ui \bar{\beta}_{\ell m}$, where $\alpha_{\ell m}$ and $\beta_{\ell m}$ are linear combinations of components of the tidal tensors $\mathcal{E}_L$ and $\mathcal{B}_L$ in \eqref{EL_BL_GR}. For $\ell = 2$ and $\ell = 3$, these coefficients were shown to obey the relations $\bar{\alpha}_{\ell m} = (-)^m \alpha_{\ell \, -m}$ and $\bar{\beta}_{\ell m} = (-)^m \beta_{\ell \, -m}$. In the Newtonian limit, the relation for $\alpha_{\ell m}$ was proven for all $(\ell,m)$ modes. If those two relations hold true for all $(\ell,m)$ modes in general relativity as well, then, by using $\bar{Y}_{\ell m} = (-)^m Y_{\ell \, -m}$, the term in round brackets in \eqref{h_nn_Schw} reduces to $2\sum_m \alpha_{\ell m} Y_{\ell m}$, so that the time-time component \eqref{h_vv-h_nn} of the metric perturbation is given by
\begin{align}\label{h_vv_Schw}
	h_{vv} = - 2 \sum_{\ell m} \frac{[(\ell-2)!]^2}{(2\ell)!} \, (2M)^\ell \, \mathcal{E}_{\ell m}(v) \, (r/M-2)^2 P''_\ell(r/M-1) \, Y_{\ell m}(\theta,\pAdv) \, ,
\end{align}
where we have also used the relation \eqref{alpha} between $\alpha_{\ell m}$ and $\mathcal{E}_{\ell m}$. Notice that $h_{vv}$ does not depend on the magnetic-type tidal tensors $\mathcal{B}_L$, as expected from parity arguments \cite{LaPo.15,Pa.al.15}. All the other metric perturbation coordinate components are similarly obtained by substituting for Eq.~\eqref{metric_bis} into \eqref{h_comp_Schw}. Recall that the resulting metric perturbation $h_{\alpha\beta}$ is entirely associated with the perturbing tidal field. Following an appropriate gauge transformation from the IRG to the light-cone gauge, the metric \eqref{h_comp_Schw} with \eqref{metric_bis} of a tidally perturbed Schwarzschild black hole could be compared to that derived in Ref.~\cite{BiPo.09}; see Eqs.~(1.3)--(1.7) therein, in which the electric-type and magnetic-type TLNs $k_\text{el}$ and $k_\text{mag}$ defined by the authors vanish for a nonspinning black hole.

Finally, in order to make contact with the Newtonian result \eqref{U_Newt_ter} established in Sec.~\ref{sec:Newton}, we introduce the effective Newtonian potential
\beq\label{Ueff}
    U_\text{eff} \equiv \frac{1}{2} \left( 1 + g_{vv} \right) = \frac{1}{2} \left( 1 + \mathring{g}_{vv} + h_{vv} \right) ,
\eeq
where $\mathring{g}_{vv} = - 1 + 2M/r$ from the Kerr metric \eqref{eq:metric}. We shall take the Newtonian limit of this effective potential, which corresponds to keeping only the leading order term as $r \to \infty$ in each multipole of $h_{vv}$. Using the Rodrigues formula for the Legendre polynomial appearing in Eq.~\eqref{h_vv_Schw}, namely $P_\ell(z) = \bigl( \ud^\ell \bigl[ (z^2-1)^\ell \bigr] / \ud z^\ell \bigr) / (2^\ell \ell!)$, the coefficient of the leading order term as $r \to \infty$ can easily be determined, and we find
\beq\label{eq:Ueff}
    U_\text{eff}(v,r,\theta,\pAdv) = \frac{M}{r} - \sum_{\ell m} \frac{(\ell-2)!}{\ell!} \, \mathcal{E}_{\ell m}(v) \left[ r^\ell + O(r^{\ell-1}) \right] Y_{\ell m}(\theta,\pAdv) \, .
\eeq
This result is compatible with the Newtonian formula \eqref{U_Newt_ter} in which $k_\ell = 0$. Given the exact vanishing of the linear response of a nonspinning black hole [Eq.~\eqref{kappa=0}], this is to be expected because the metric perturbation $h_{\alpha\beta} = h_{\alpha\beta}^\text{tidal}$ reconstructed from \eqref{metric_bis} is entirely associated with the external tidal field.

This calculation for a tidally perturbed, nonspinning black hole allowed us to test the IRG metric reconstruction by comparing the weak-field limit of \eqref{h_vv_Schw} to the Newtonian result \eqref{U_Newt_ter}. Our main interest, however, is in computing the nonvanishing TLNs of spinning black holes. Thereafter, we will restrict ourselves to \emph{linear} order in the spin, to obtain a metric perturbation in simple enough form so that the associated Geroch-Hansen multipole moments can be computed without any further approximation.

\subsection{Kerr black hole response to linear order in the spin}

We are finally in a position to compute the Kerr black hole response to the applied tidal field, at the level of the metric perturbation. In general, in order to compute the IRG metric perturbation $h_{\alpha\beta}^\text{resp}$ associated \emph{only} with the black hole response, one needs to substitute into Eqs.~\eqref{metric} and \eqref{h_comp} the part of the Hertz potential \eqref{Psibar} associated solely with the response, namely [recalling the split \eqref{G=Gtidal+Gresp}]
\beq\label{Psi_resp}
    \Psi^\text{resp} = 2\sum_{\ell m} \bar{\kappa}_{\ell m} \, \bar{z}_{\ell m}(v) \, \bar{G}_{\ell m}^\text{resp}(r) \, {}_{2}\bar{Y}_{\ell m}(\theta,\pAdv) \, ,
\eeq
with the explicit expressions \eqref{kappa_phys} and \eqref{G_resp} for the response coefficients $\kappa_{\ell m}$ and the radial part $G_{\ell m}^\text{resp}(r)$ of the black hole response contribution to the Hertz potential.

As we shall continue to perform the calculation explicitly to \emph{linear} order in the dimensionless black hole spin parameter $\ad$, it is convenient to introduce the symbol ``$\doteq$'' for an equality that holds to  that order. Hence all terms of $O(\ad^2)$ in the spin are discarded in such equations. While restricting the calculation to linear order in the black hole spin, a notable simplification occurs. In that case, Eq.~\eqref{kappa_small_a} shows that for small spin values $0 < \ad \ll 1$,
\beq\label{eq:k lin a}
    \bar{\kappa}_{\ell m} \doteq \ui m n_\ell \, \ad
    \, ,
\eeq
where $n_\ell \equiv (\ell-2)!(\ell+2)!(\ell!)^2/[4(2\ell)!(2\ell+1)!]$. Because the approximation \eqref{eq:k lin a} to $\bar{\kappa}_{\ell m}$ is linear in the black hole spin, it is sufficient to use the nonspinning form \eqref{G_resp_Schw} for $G_{\ell m}^\text{resp}(r)$ in Eq.~\eqref{Psi_resp}, which readily gives
\beq\label{Psi_resp_linear}
    \Psi^\text{resp} \doteq 2 \ui \ad \sum_{\ell m} m n_\ell q_\ell \,  \bar{z}_{\ell m}(v) \, r^2\mathcal{Q}_\ell(r) \, {}_{2}\bar{Y}_{\ell m}(\theta,\pAdv)
    \, ,
\eeq
where $q_\ell \equiv (2M)^\ell [4(2\ell+1)!(\ell-2)!]/(\ell![(\ell+2)!]^2)$ and $\mathcal{Q}_\ell(r) \equiv (r/M-2)^2 Q_\ell''(r/M-1)$ is a real-valued function of the radial coordinate $r$. Consequently, the nonvanishing tetrad components \eqref{tetrad_Schw} associated with the black hole response are given, to linear order in the spin, by
\begin{subequations}\label{metric_resp}
	\begin{align}
		h^\text{resp}_{nn} &\doteq - \ad \sum_{\ell = 2}^\infty C_\ell \, \mathcal{Q}_\ell(r) \left( \sum_{m = -\ell}^\ell \ui m \, \bar{z}_{\ell m}(v) \, \bar{Y}_{\ell m}(\theta,\pAdv) + \text{c.c.} \right)
		\, , \label{h_nn_resp} \\
		h^\text{resp}_{n\bar{m}} &\doteq - \sqrt{2} \ad \sum_{\ell = 2}^\infty C_\ell \, \sqrt{\frac{(\ell-1)!}{(\ell+1)!}} \; r \mathcal{Q}'_\ell(r) \sum_{m = -\ell}^\ell \ui m \, \bar{z}_{\ell m}(v) \, {}_{1}\bar{Y}_{\ell m}(\theta,\pAdv)
		\, , \label{h_nbarm_resp}\\
		h^\text{resp}_{\bar{m}\bar{m}} &\doteq - 2\ad \sum_{\ell = 2}^\infty C_\ell \, \sqrt{\frac{(\ell-2)!}{(\ell+2)!}} \left[ r^2 \mathcal{Q}''_\ell(r) + 2r \mathcal{Q}'_\ell(r) - 2\mathcal{Q}_\ell(r) \right] \sum_{m = -\ell}^\ell \ui m \, \bar{z}_{\ell m}(v) \, {}_{2}\bar{Y}_{\ell m}(\theta,\pAdv)
		\, .
	\end{align}
\end{subequations}
The IRG metric perturbation $h_{\alpha\beta}^\text{resp}$ associated with the Kerr black hole linear response to the applied tidal field is then obtained by substituting Eqs.~\eqref{metric_resp} into \eqref{h_comp_Schw}. This is another important result of this paper as it shows explicitly, at the level of the metric perturbation, and to linear order in the Kerr black hole spin $\ad$, that the response of a spinning black hole to a generic, nonaxisymmetric tidal perturbation does not vanish.\footnote{If there existed a gauge in which the metric perturbation $h^\text{resp}_{\alpha\beta}$ vanished identically, then the coordinate-invariant Geroch-Hansen multipole moments associated with the linear response would all vanish. As we shall see in the next Sec.~\ref{sec:multipoles}, the quadrupole moments associated with the linear response do not vanish, so that there is no gauge in which $h^\text{resp}_{\alpha\beta}$ vanishes.} Notice that (the leading orders in) Eqs.~\eqref{metric_resp} are identical to \eqref{metric_bis} with the formal substitution $\mathcal{P}_\ell(r) \to \ui m \ad \, \mathcal{Q}_\ell(r)$ in each ($\ell,m$) mode. Notice also that the IRG black hole response $h_{\alpha\beta}^\text{resp}$ in advanced Kerr coordinates is not smooth on the future event horizon in Kerr, because the radial function $\mathcal{Q}_\ell(r)$ is merely $C^1$ there.

Moreover, if we assume that $\bar{\alpha}_{\ell m} = (-)^m \alpha_{\ell \, -m}$ (see below \eqref{alpha} in the Newtonian limit) and $\bar{\beta}_{\ell m} = (-)^m \beta_{\ell \, -m}$, then, using the property $\bar{Y}_{\ell m} = (-)^m Y_{\ell -m}$, the term in round brackets in Eq.~\eqref{h_nn_resp} becomes $-2 \sum_m \ui m \, \alpha_{\ell m} Y_{\ell m}$, thus canceling out the dependence of $h_{nn}^\text{resp}$ on the magnetic-type tidal fields $\mathcal{B}_L$.  Consequently, for the advanced time-time component \eqref{h_vv-h_nn} of the metric perturbation we obtain
\begin{align}\label{h_vv_resp}
	h_{vv}^\text{resp} \doteq 2\ui \ad \sum_{\ell m} \frac{[(\ell-2)!]^2}{(2\ell)!} \; m \, \mathcal{E}_{\ell m}(v) \, (2M)^\ell \, (r/M-2)^2 Q''_\ell(r/M-1) \, Y_{\ell m}(\theta,\pAdv)
	\, ,
\end{align}
where we also used the relation \eqref{alpha} between $\alpha_{\ell m}$ and $\mathcal{E}_{\ell m}$. To compare this result with the formula \eqref{U_spinning} for the gravitational potential of an axisymmetric, spinning, Newtonian body, we must consider the weak-field limit of $\tfrac{1}{2} h_{vv}^\text{resp}$ [recall Eq.~\eqref{Ueff}]. By using the asymptotic formula $Q_\ell(z) \sim [\ell! / (2\ell+1)!!] \, z^{-\ell-1}$ as $|z| \to \infty$~\cite{Olv}, the asymptotic form of the black hole response  as $r \to \infty$ is given by
\beq\label{h_vv_resp_asym}
    \frac{1}{2} \, h_{vv}^\text{resp} \sim \sum_{\ell m} \mathcal{E}_{\ell m}(v) \, \ui m \ad \, \frac{(\ell+2)! \, \ell! \, [(\ell-2)!]^2}{2 (2\ell)! \, (2\ell+1)!} \, \frac{(2M)^{2\ell+1} }{r^{\ell+1}} \, Y_{\ell m}(\theta,\phi)+ o(\ad) \, .
\eeq
Comparing, through Eq.~\eqref{eq:g-U}, this expression with the exact Newtonian result \eqref{U_spinning} with $k_{\ell m, \ell' m'} = k_{\ell m} \, \delta_{\ell\ell'} \delta_{mm'}$, we find that the ``Newtonian TLNs" of a Kerr black hole of mass $M$ and spin $\chi$ are indeed given by the formula \eqref{k_phys} with \eqref{kappa_small_a}. In Sec.~\ref{sec:multipoles} below we shall compute the general relativistic TLNs \eqref{TLN_bis} from the coordinate-invariant Geroch-Hansen multipole moments, at quadrupolar order. We will find two families of nonvanishing TLNs that couple the electric-type (resp. magnetic-type) quadrupolar tidal fields to the mass-type (resp. current-type) quadrupole moments, whose amplitude match precisely the ``Newtonian TLNs" \eqref{k_phys} for $\ell=2$.

\section{Multipole moments and tidal Love numbers}\label{sec:multipoles}

In the previous Sec.~\ref{sec:metric}, we have computed the linear response $h_{\alpha\beta}^\text{resp}$ of a Kerr black hole to a weak and slowly varying, but otherwise arbitrary tidal environment, up to linear order in the black hole spin. (This calculation could have been extended to any order in the spin.) The resulting metric perturbation $h_{\alpha\beta}^\text{resp}$, given by Eqs.~\eqref{h_comp_Schw} with \eqref{metric_resp}, was constructed in the domain $r_+ \leqslant r \ll \mathcal{R}$. However, from now on we shall take it to be valid for all $r \geqslant r_+$~\cite{Pa.al.15,Pa.al2.15}, just like the body's response $U_\text{body}$ in the Newtonian gravitational potential \eqref{U_Newt}. The metric perturbation $h_{\alpha\beta}^\text{resp}$ is then an asymptotically flat solution of the linearized Einstein equation in vacuum. Physically, the metric
\beq\label{g_resp}
    g_{\alpha\beta} \equiv \mathring{g}_{\alpha\beta} + h_{\alpha\beta}^\text{resp}
\eeq
thus describes the vacuum spacetime of a Kerr black hole linearly perturbed by an external tidal field, but \emph{without} the contribution from the tidal field itself, which is not asymptotically flat. We emphasize that all of our results so far---including the proof that the Kerr TLNs for an axisymmetric tidal perturbation vanish---were valid for a slowly-varying tidal field, but not necessarily static, as it was allowed to possess a parametric dependence on time. From now on we neglect such parametric dependence.

By neglecting the slow parametric time evolution of the coefficients $z_{\ell m}(v)$ in $h_{\alpha\beta}^\text{resp}$, the metric \eqref{g_resp} is a stationary, asymptotically flat, source-free solution of the linearized Einstein equation. So the multipole moments \eqref{ML_SL_PL} associated with the metric \eqref{g_resp} are well defined. Since we are working to linear order in perturbation theory, we necessarily have
\begin{subequations}\label{ML_SL_linear}
    \begin{align}
        M_L &= \mathring{M}_L + \delta M_L \, , \\
        S_L &= \mathring{S}_L + \delta S_L \, ,
    \end{align}
\end{subequations}
where, as said earlier in Sec.~\ref{sec:GR}, $\mathring{M}_L$ and $\mathring{S}_L$ are the mass-type and current-type multipole moments associated with the background Kerr metric $\mathring{g}_{\alpha\beta}$ , respectively, while $\delta M_L$ and $\delta S_L$ are the corrections induced by the linear metric perturbation $h_{\alpha\beta}^\text{resp}$.

\subsection{Coordinate transformation}

We now proceed to compute the multipole moments \eqref{ML_SL_linear}. To do so it will prove especially convenient to first transform from advanced Kerr coordinates $(v,r,\theta,\pAdv)$ to Boyer-Lindquist (BL) coordinates $(t,r,\theta,\pBL)$, which are related by~\cite{BoLi.67}
\begin{subequations}\label{Kerr->BL}
    \begin{align}
        \ud v &\doteq \ud t + f^{-1} \, \ud r \, , \\
        \ud \pAdv &\doteq \ud \pBL + \frac{a}{r^2 f} \, \ud r \, , \label{phi-psi} 
    \end{align}
\end{subequations}
while the radial and polar coordinates $r$ and $\theta$ remain identical in both coordinate systems. Recall the notation $f = 1 - 2M/r$. Such a change in coordinates will not affect the values of the multipole moments \eqref{ML_SL_linear}, which are defined in a geometrical, coordinate-independent manner. In BL coordinates $(t,r,\theta,\pBL)$, the linear-in-$\ad$ background Kerr geometry \eqref{eq:metric} simply reads as
\begin{align}\label{Kerr_BL}
    \mathring{g}_{\alpha\beta} \, \ud x^\alpha \ud x^\beta \doteq - f \, \ud t^2 - \frac{4 M}{r^2} \, a \sin^2{\theta} \, \ud t \ud \pBL + f^{-1} \, \ud r^2 + r^2 \, \ud \theta^2 + r^2 \sin^2{\theta} \, \ud \pBL^2 \, . 
\end{align}
Having only one off-diagonal term, the form \eqref{Kerr_BL} of the Kerr metric will simplify the algebra of the forthcoming calculations. On the other hand, while transforming from advanced Kerr coordinates to BL coordinates, the coordinate components of the metric perturbation $h_{\alpha\beta}^\text{resp}$ associated with the black hole response are now given, still to linear order in the black hole spin $\ad$, by
\begin{subequations}\label{h_resp_BL}
    \begin{align}
        h^\text{resp}_{tt} &= h^\text{resp}_{vv} \doteq h^\text{resp}_{nn} \,
        , \\
        h^\text{resp}_{tr} &\doteq f^{-1} h^\text{resp}_{vv} + h^\text{resp}_{vr}
        \doteq - f^{-1} h^\text{resp}_{nn}\,
        , \\
        h^\text{resp}_{t\theta} &= h^\text{resp}_{v\theta} \doteq \sqrt{2}\, r\, \Re(h^\text{resp}_{n\bar m})\,
        , \\
        h^\text{resp}_{t\pBL} &= h^\text{resp}_{v\pAdv} \doteq
        -\sqrt{2} \, r \sin\theta\, \Im(h^\text{resp}_{n\bar m})\,
        , \\
        h^\text{resp}_{rr} &\doteq f^{-2} h^\text{resp}_{vv} + 2 f^{-1} h^\text{resp}_{vr} + h^\text{resp}_{rr}
        \doteq f^{-2} h^\text{resp}_{nn}
        \, , \\
        h^\text{resp}_{r\theta} &\doteq h^\text{resp}_{r\theta}+f^{-1}h^\text{resp}_{v\theta} 
        \doteq - \sqrt{2} \, r f^{-1} \Re(h^\text{resp}_{n\bar m}) \, , \\
        h^\text{resp}_{r\pBL} &\doteq h^\text{resp}_{r\pAdv} + f^{-1} h^\text{resp}_{v\pAdv} \doteq \sqrt{2}\, r f^{-1}\sin\theta\, \Im(h^\text{resp}_{n\bar m})
        \, , \\
        h^\text{resp}_{\theta\theta} &= h^\text{resp}_{\theta\theta} \doteq r^2\Re(h^\text{resp}_{\bar m\bar m}) \, , \\
        h^\text{resp}_{\theta\pBL} &= h^\text{resp}_{\theta\pAdv} \doteq -r^2\sin\theta\, \Im(h^\text{resp}_{\bar{m}\bar{m}})\,
        , \\
        h^\text{resp}_{\pBL\pBL} &= h^\text{resp}_{\pAdv\pAdv} \doteq -r^2\sin^2\theta\, \Re(h^\text{resp}_{\bar m\bar m})\, ,
    \end{align}
\end{subequations}
where we used the formulas \eqref{h_comp_Schw}. From Eq.~\eqref{phi-psi} it is clear that $\pAdv = \pBL + O(\ad)$. Since the metric perturbation \eqref{h_resp_BL} is $O(\ad)$, and because we systematically neglect all terms $O(\ad^2)$ or higher in $h_{\alpha\beta}^\text{resp}$, we can use either $\pAdv$ or $\pBL$ as the azimuthal angle in the spin-weighted spherical harmonics ${}_{s}Y_{\ell m}$ that appear in the tetrad components \eqref{metric_resp}.

\subsection{Perturbed scalar twist}

Recall from Sec.~\ref{sec:GR} that two key ingredients entering the definition of the Geroch-Hansen multipole moments of a stationary spacetime are the norm $\lambda$ and scalar twist $\omega$ associated with the timelike Killing field $\xi^\alpha$. Using BL coordinates ($t,r,\theta,\pBL$) and neglecting the slow variation of the perturbing tidal field, i.e.\ we now consider the coefficients $z_{\ell m}$ in \eqref{metric_resp} to be \emph{constants}, the metric \eqref{g_resp} does not depend on the coordinate time $t$. The timelike Killing field is thus $\xi^\alpha = (\partial_t)^\alpha$, whose norm is simply $\lambda = - g_{tt} \doteq f - h_{nn}^\text{resp}$, which explicitly reads
\beq\label{norm-lambda}
    \lambda \doteq 1 - \frac{2M}{r} + 2 \ad \sum_{\ell = 2}^\infty C_{\ell}\, \mathcal{Q}_\ell(r) \sum_{m = -\ell}^\ell  m \, \Im{\left[z_{\ell m} Y_{\ell m}(\theta,\pBL)\right]} \, ,
\eeq
where we recall that the constant $C_\ell$ is defined in \eqref{C_l} and $\mathcal{Q}_\ell(r) \equiv (r/M-2)^2 Q_\ell''(r/M-1)$. On the other hand, to linear order in $h^\text{resp}_{\alpha\beta} = O(\ad)$, the twist $\omega^\alpha = \varepsilon^{\alpha\beta\gamma\delta} \xi_\beta \nabla_\gamma \xi_\delta$ associated with the Killing field $\xi^\alpha$ reads
\beq\label{eq:twist-metric}
    \omega^\alpha \doteq \varepsilon^{\alpha\beta\gamma\delta} \left(\mathring{g}_{t\beta} \partial_\gamma \mathring{g}_{t\delta} + \mathring{g}_{t\beta} \partial_\gamma h^\text{resp}_{t\delta} + h^\text{resp}_{t\beta} \partial_\gamma \mathring{g}_{t\delta}\right) \, ,
\eeq
where $\varepsilon^{\alpha\beta\gamma\delta} \!=\! [\alpha\beta\gamma\delta]/\sqrt{|g|}$, with $g = \text{det}(g_{\alpha\beta})$ the determinant of the metric \eqref{g_resp} and $[\alpha\beta\gamma\delta]$ the totally antisymmetric permutation symbol, such that $[tr\theta\pBL] = 1$. Using Eqs.~\eqref{Kerr_BL} and \eqref{h_resp_BL}, the scalar twist $\omega$ associated with $\xi^\alpha$, which is such that $\omega_\alpha = \nabla_\alpha \omega$, can then be shown to obey the following set of first-order partial differential equations:
\begin{subequations}\label{eq:twist diff eqs}
    \begin{align}
        \partial_t \omega & \doteq 0 \, ,
        \label{eq:twist diff eqs t}
        \\
        \partial_r \omega &\doteq - \frac{4M}{r^3} \, a \cos{\theta} - \frac{\sqrt{2}}{r\sin\theta} \, \bigl[ \partial_{\pBL}\Re(h^\text{resp}_{n\bar{m}}) + \partial_{\theta}\left(\sin{\theta} \, \Im(h^\text{resp}_{n\bar{m}})\right) \bigr]
        \, , \\
        \partial_\theta \omega &\doteq - \frac{2M}{r^2} \, a \sin{\theta} - \frac{1}{\sin{\theta}} \, \partial_{\pBL} h^\text{resp}_{nn}
        + \sqrt{2} \, \bigl[ r f \partial_r \Im{(h^\text{resp}_{n\bar{m}})} + (1-4M/r) \, \Im(h^\text{resp}_{n\bar{m}}) \bigr] \, ,
        \\
        \partial_\pBL \omega &\doteq \sin{\theta} \, \partial_{\theta} h^\text{resp}_{nn}
        + \sqrt{2} \sin{\theta} \, \bigl[ r f \partial_r \Re{(h^\text{resp}_{n\bar{m}})} + (1-4M/r) \, \Re(h^\text{resp}_{n\bar{m}}) \bigr]
        \, .
    \end{align}
\end{subequations}
Here, the tetrad components $h^\text{resp}_{nn}$ and $h^\text{resp}_{n\bar{m}}$ do not depend on time $t$ and are expressed as functions of the BL azimuthal coordinate $\pBL$. As is proven in App.~\ref{app:twist}, the partial differential equations \eqref{eq:twist diff eqs} can be integrated. Imposing the boundary condition $\omega \to 0$ as $r \to \infty$, the scalar twist is found to be given, to linear order in black hole spin, by the simple formula
\beq\label{norm-twist}
    \omega \doteq \frac{2M^2}{r^2} \, \ad \cos{\theta} + 2 \ad \sum_{\ell = 2}^\infty C_{\ell}\, \mathcal{Q}_\ell(r) \sum_{m = -\ell}^\ell  m \, \Re{\left[z_{\ell m} Y_{\ell m}(\theta,\pBL)\right]}
    \, .
\eeq
Remarkably, the scalar twist perturbation in \eqref{norm-twist}, say $\delta\omega$, differs from $\delta \lambda$  in \eqref{norm-lambda} just in having the real part $\Re$ instead of the imaginary part $\Im$. Consequently, the perturbation in the complex combination $\epsilon \equiv \lambda + \ui \omega$, which is closely related to the Ernst potential~\cite{Er.68,Fo.al.89} for stationary-axisymmetric gravitational fields, simply reads
\beq\label{depsbar}
    \delta \epsilon \doteq 2 \ui \ad \sum_{\ell m} m \, C_{\ell} \, \bar{z}_{\ell m} \, \mathcal{Q}_\ell(r) \, \bar{Y}_{\ell m}(\theta,\pBL)
    \, .
\eeq

Since the scalar twist \eqref{norm-twist} is linear in the black hole spin, we can neglect the $O(\ad^2)$ term $\omega^2$ in the mass-type potential $\Phi_M$ in Eq.~\eqref{Phi}, and use the background (spin-independent) value $\mathring{\lambda} = f = 1 - 2M/r$ of the norm \eqref{norm-lambda} in the current-type potential $\Phi_S$. Hereafter, we shall thus consider the generating potential
\beq\label{potential}
    \Phi \doteq \frac{1}{4\lambda} - \frac{\lambda}{4} + \ui \,  \frac{\omega}{2f} \, .
\eeq

\subsection{Conformal transformation}

The next step is to determine the 3-metric $\gamma_{ab}$ on any spacelike hypersurface $V$ orthogonal to $\xi^\alpha$, and to perform a conformal transformation to compute the conformal metric $\tilde{\gamma}_{ab} = \Omega^2 \gamma_{ab}$ on $\tilde{V} = V \cup \Lambda$, with $\Omega$ the conformal factor. To linear order in the metric perturbation and the black hole spin, we readily find from \eqref{Kerr_BL}--\eqref{norm-lambda}
that the BL coordinate components of the induced 3-metric \eqref{gamma} on $V$ read
\begin{subequations}\label{gamma_com}
    \begin{align}
        \gamma_{rr} &\doteq 1
        \, , \\
        \gamma_{r\theta} &\doteq - \sqrt{2} \, r \, \Re(h^\text{resp}_{n\bar{m}})
        \, , \\
        \gamma_{r\pBL} &\doteq \sqrt{2} \, r \sin{\theta} \, \Im(h^\text{resp}_{n\bar{m}})
        \, , \\
        \gamma_{\theta\theta} &\doteq r^2 f \left[1 + \Re(h^\text{resp}_{\bar{m}\bar{m}}) - f^{-1} h^\text{resp}_{nn} \right] , 
        \\
        \gamma_{\theta\pBL} &\doteq - r^2 f \sin{\theta} \, \Im(h^\text{resp}_{\bar{m}\bar{m}})
        \, , \\
        \gamma_{\pBL\pBL} &= \sin^2{\theta} \, \gamma_{\theta\theta}
        \, ,
    \end{align}
\end{subequations}
where we recall that the tetrad components $h_{nn}^\text{resp}$, $h^\text{resp}_{n\bar{m}}$ and $h^\text{resp}_{\bar{m}\bar{m}}$ are given by Eqs.~\eqref{metric_resp}. In order to identify an appropriate conformal factor, we first introduce a new radial coordinate $R(r)$ through
\beq\label{r(R)}
    r \equiv \frac{1}{R} \left( 1 + \frac{M}{2} R \right)^2 ,
\eeq
such that the point $\Lambda$ (spatial infinity) corresponds to $R = 0$. For a nonspinning black hole, the radial coordinate $R(r)$ is related in a simple manner to the isotropic Schwarzschild radial coordinate. Performing the change of coordinates $(r,\theta,\pBL) \to (R,\theta,\pBL)$, the metric \eqref{gamma_com} on $V$ can be written as $\gamma_{ab} = \Omega^{-2} \, \tilde{\gamma}_{ab}$, where the conformal factor is given by
\beq\label{Omega}
    \Omega = R^2 \left( 1 - \frac{M^2}{4} R^2 \right)^{-1} ,
\eeq
and the components of the conformal metric $\tilde{\gamma}_{ab}$ on $\tilde{V}$ with respect to the spherical polar coordinates $(R,\theta,\pBL)$ read
\begin{subequations}\label{gamma_tilde}
    \begin{align}
        \tilde{\gamma}_{RR} &\doteq 1
        \, , \\
        \tilde{\gamma}_{R\theta} &\doteq \sqrt{2} R \, \frac{1+\tfrac{1}{2}MR}{1-\tfrac{1}{2}MR} \, \Re(h^\text{resp}_{n\bar{m}})
        \, , \\
        \tilde{\gamma}_{R\pBL} &\doteq - \sqrt{2} R \, \frac{1+\tfrac{1}{2}MR}{1-\tfrac{1}{2}MR} \, \sin{\theta} \, \Im(h^\text{resp}_{n\bar{m}})
        \, , \\
        \tilde{\gamma}_{\theta\theta} &\doteq R^2 \Biggl[1 + \Re(h^\text{resp}_{\bar{m}\bar{m}}) - \biggl( \frac{1+\tfrac{1}{2}MR}{1-\tfrac{1}{2}MR} \biggr)^{\!2} \, h^\text{resp}_{nn} \Biggr]
        \, , \\
        \tilde{\gamma}_{\theta\pBL} &\doteq - R^2 \sin{\theta} \, \Im(h^\text{resp}_{\bar{m}\bar{m}}) 
        \, , \\
        \tilde{\gamma}_{\pBL\pBL} &= \sin^2{\theta} \, \tilde{\gamma}_{\theta\theta}
        \, .
    \end{align}
\end{subequations}
It can be checked that the conditions $\Omega = \tilde{D}_a \Omega = 0$ and $\tilde{D}_a \tilde{D}_b \Omega = 2 \tilde{\gamma}_{ab}$ at $\Lambda$ are all satisfied. In particular, to verify the last equality we use the fact that, to leading order in $R$, the tetrad components $h^\text{resp}_{nn}$, $h^\text{resp}_{n\bar{m}}$ and $h^\text{resp}_{\bar{m}\bar{m}}$ in \eqref{gamma_tilde} scale as $O(R^3)$, corresponding to the quadrupolar contributions in \eqref{metric_resp}, so that the only nonvanishing component of the conformal metric at spatial infinity is $\tilde{\gamma}_{RR}|_\Lambda \doteq 1$.

We notice that the conformal metric \eqref{gamma_tilde} is flat in the background, so that the conformal Ricci tensor $\tilde{R}_{ab}$ and the conformal scalar curvature $\tilde{R}$ are both of $O(h^\text{resp})$. Thereafter, we shall merely need $\tilde{R}$ and the coordinate component $\tilde{R}_{RR}$ of $\tilde{R}_{ab}$ to leading order in a Taylor series expansion in the radial coordinate $R$. An explicit calculation using the formulas \eqref{metric_resp} and \eqref{gamma_tilde} reveals that
\begin{subequations}\label{Ricci}
    \begin{align}
        \tilde{R}_{RR} &\doteq - \frac{32}{15} \, a M^4 R \,\, \Im \sum_{m=-2}^2 m \, \bar{z}_{2m}\,{}_2\bar{Y}_{2m} + O(R^2) \, , \\
        \tilde{R} &\doteq -\frac{16}{45} \, a M^4 R\,\,\Im \sum _{m=-2}^2 m \, \bar{z}_{2m} \, \bigl(13 + 4\cot^2\theta + 4\ui\csc\theta\cot\theta\,\partial_\pBL \nonumber \\ &\qquad\qquad\qquad\qquad\qquad\qquad\qquad\! - \csc^2\theta\,\partial^2_\pBL + \cot\theta\,\partial_\theta \bigr)\,{}_2\bar{Y}_{2m} + O(R^2) \, .
    \end{align}
\end{subequations}

\subsection{Monopole and dipole moments}\label{subsec:mono-dipo}

We are finally ready to compute from the conformally rescaled scalar potential $\tilde{\Phi} \equiv \Phi / \sqrt{\Omega}$ the multipole moments \eqref{ML_SL_PL} associated with the perturbed Kerr geometry \eqref{g_resp}. First, by combining Eqs.~\eqref{norm-lambda}, \eqref{norm-twist}, \eqref{potential}, \eqref{r(R)} and \eqref{Omega}, we readily find
\begin{align}\label{tildePhi}
    \tilde{\Phi} &\doteq M + \ui a M R \cos{\theta} + \frac{5}{8} M^3 R^2 + \frac{4}{15} \, \ui a M^4 R^2 \! \sum_{m=-2}^2 \! m (\mathcal{E}_{2m} + \ui \mathcal{B}_{2m}) Y_{2m}(\theta,\pBL) + O(R^3) \, .
\end{align}
This accuracy will be sufficient to compute the complex monopole $P$, dipole $P_a$ and quadrupole $P_{ab}$, all to be evaluated at $\Lambda$. The Taylor series expansion in the coordinate radius $R$ would have to be carried out to higher orders in order to compute higher-order multipole moments. We perform a last coordinate transformation on $\tilde{V}$ from the spherical-type coordinates $(R,\theta,\pBL)$ to the associated Cartesian-type coordinates $(x^i) \equiv (x,y,z)$. Using the chain rule for partial derivatives, we obtain for the monopole and dipole moments:
\begin{subequations}
    \begin{align}
        P|_\Lambda &= \lim_{R \to 0} \tilde{\Phi} = M \, , \\
        P_i|_\Lambda &= \lim_{R \to 0} \partial_i \tilde{\Phi} = \ui M a \, s^i \, , \label{dipole}
    \end{align}
\end{subequations}
where $s^i$ is the unit $z$-direction vector at the point $\Lambda$, so that $s^i = (0,0,1)$. The reconstructed metric perturbation $h_{\alpha\beta}^\text{resp}$ does not contribute to the monopole, the mass dipole moment nor to the angular momentum of the perturbed spacetime \eqref{g_resp}, namely $\delta M = \delta M_i = \delta S_i = 0$. Such contributions would come from the completion piece of the metric perturbation~\cite{Wa.73,Me.al.16,vdM2.17}. In particular, the mass dipole moment vanishes and the angular momentum is equal to the spin of the background Kerr black hole: $M_i = \Re{(P_i|_\Lambda)} = 0$ and $S_i = \Im{(P_i|_\Lambda)} = J s^i$.

\subsection{Quadrupole moments}\label{subsec:quad}

From the general definition \eqref{P's}--\eqref{ML_SL_PL}, the complex-valued quadrupole moment is given by
\beq\label{Pab}
    M_{ab} + \ui S_{ab} = \frac{1}{3} P_{ab}|_\Lambda = \frac{1}{3} \lim_{R \to 0} \left( \tilde{D}_{\langle a} \tilde{D}_{b \rangle} \tilde{\Phi} - \frac{1}{2} \tilde{R}_{\langle ab \rangle} \tilde{\Phi} \right) .
\eeq
At the quadrupolar order, we find it more convenient to work with the modes $M_{2 m}$ and $S_{2m}$ of the quadrupole moments $M_{ab} = \Re{(P_{ab}|_\Lambda)}/3$ and $S_{ab} = \Im{(P_{ab}|_\Lambda)}/3$, for all $|m| \leqslant 2$. Those are defined in Eq.~\eqref{MlmSlm}, where $n^a$ is the unit radial direction that lives in $\tilde{V}$, so that
\begin{align}\label{quad_modes} 
    M_{2m} + \ui S_{2m} &= \frac{1}{3} \oint_{\mathbb{S}^2} n^a n^b P_{ab}|_\Lambda \, \bar{Y}_{2 m} \, \ud \Omega \nonumber \\ &= \frac{1}{3} \oint_{\mathbb{S}^2} \lim_{R \to 0} \left( n^a n^b \tilde{D}_a \tilde{D}_b \tilde{\Phi} - \frac{1}{3} \tilde{\Delta} \tilde{\Phi} - \frac{1}{2} n^a n^b \tilde{R}_{ab} \, \tilde{\Phi} + \frac{1}{6} \tilde{R} \tilde{\Phi} \right) \bar{Y}_{2 m} \, \ud \Omega \, ,
\end{align}
where $\tilde{\Delta}$ is the Laplace operator associated with $\tilde{\gamma}_{ab}$. Since the integrand in \eqref{quad_modes} is covariant, we may conveniently evaluate it while using the spherical polar coordinates $(R,\theta,\pBL)$, with respect to which $n^i|_\Lambda = (1,0,0)$. Using Eqs.~\eqref{Ricci} and \eqref{tildePhi}, as well as the Levi-Civita connection $\tilde{\Gamma}^a_{\phantom{a}bc}$ associated with the conformal metric $\tilde{\gamma}_{ab}$, the four terms in the integrand of Eq.~\eqref{quad_modes} individually read as
\begin{subequations}\label{terms}
    \begin{align}
        n^a n^b \tilde{D}_a \tilde{D}_b \tilde{\Phi} &\doteq \frac{5}{4} M^3 + \frac{8}{15} M^4 \, \ui a \! \sum_{m'=-2}^2 \! m' (\mathcal{E}_{2m'} + \ui \mathcal{B}_{2m'}) Y_{2m'} + O(R) \, , \label{nnDDPhi} \\
        \tilde{\Delta} \tilde{\Phi} &\doteq \frac{15}{4} M^3 + O(R) \, , \label{Laplace} \\
        n^a n^b \tilde{R}_{ab} &\doteq O(R) \, , \\
        \tilde{R} \tilde{\Phi} &\doteq O(R) \, ,
    \end{align}
\end{subequations}
where we used the fact that $\tilde{\Delta} [R^2 \, Y_{2m}(\theta,\pBL)]|_\Lambda = 0$ in \eqref{Laplace}. The second term in the right-hand side of \eqref{nnDDPhi} is the only contribution coming from the perturbing tidal field, and it is closely related to the quadrupolar ($\ell=2$) contribution in \eqref{depsbar}, defined out of the norm and scalar twist of the perturbed spacetime. By substituting  \eqref{terms} into \eqref{quad_modes}, and using the orthonormality \eqref{ortho} of the spherical harmonics, we obtain the simple formula
\beq\label{M2+iS2}
    M_{2m} + \ui S_{2m} \doteq \frac{8}{45} \, \ui ma M^4 (\mathcal{E}_{2m} + \ui \mathcal{B}_{2m})
    \, .
\eeq
To linear order in the spin, the selection rules for $\ell$-mode couplings would in principle have allowed for the octupolar tidal fields $\mathcal{E}_{3m}$ and $\mathcal{B}_{3m}$ to couple to the quadrupole moments $S_{2m}$ and $M_{2m}$, respectively~\cite{Pa.al.15,Pa.al2.15}. Interestingly, those contributions do not appear in Eq.~\eqref{M2+iS2}. We also find no contribution from the Kerr background in Eq.~\eqref{M2+iS2}, in agreement with the Hansen formula \eqref{M_S}, since $\mathring{M}_{2m} \propto a^2 \delta_{m0} \doteq 0$ and $\mathring{S}_{2m} = 0$.

Now, as a consequence of the definition \eqref{MlmSlm}, the modes $M_{\ell m}$ of the mass-type multipole moments $M_L$ obey the general property $M_{\ell\,-m} = (-)^m \bar{M}_{\ell m}$, and similarly for $S_{\ell m}$, $\mathcal{E}_{\ell m}$, $\mathcal{B}_{\ell m}$. Consequently, Eq.~\eqref{M2+iS2} implies
\beq\label{M2bar+iS2bar}
    \bar{M}_{2m} + \ui \bar{S}_{2m} \doteq - \frac{8}{45} \, \ui ma M^4 (\bar{\mathcal{E}}_{2m} + \ui \bar{\mathcal{B}}_{2m})
    \, .
\eeq
By taking the complex conjugate of Eq.~\eqref{M2bar+iS2bar} and by considering separately the sum and the difference of the result with Eq.~\eqref{M2+iS2}, we readily obtain the spherical-harmonic modes of the mass-type and current-type quadrupole moments as
\begin{subequations}\label{M2S2}
    \begin{align}
        M_{2m} &\doteq \delta M_{2m} \doteq \frac{8}{45} \, \ui ma M^4 \mathcal{E}_{2m}
        \, , \label{M2} \\
        S_{2m} &\doteq \delta S_{2m} \doteq \frac{8}{45} \, \ui ma M^4 \mathcal{B}_{2m}
        \, .
    \end{align}
\end{subequations}
Interestingly, to linear order in the black hole spin, we find that the modes of the mass-type (resp. current-type) quadrupole moment are sourced \emph{only} by the modes of the electric-type (resp. magnetic-type) quadrupolar tidal field. The coupling \eqref{M2S2} between $(\mathcal{E}_{2m},\mathcal{B}_{2m})$ and $(M_{2m},S_{2m})$ is akin to a Zeeman-like splitting proportional to the azimuthal number $m$~\cite{Pa.al2.15}. Moreover, the factor of ``$\ui$" in that coupling suggests that the black hole tidal bulge is rotated by $45^\circ$ with respect to the quadrupolar tidal perturbation~\cite{LeCa.21}.

We would also like to relate the perturbations in the STF quadrupole moments $M_{ij}$ and $S_{ij}$ to the perturbing quadrupolar tidal fields $\mathcal{E}_{ij}$ and $\mathcal{B}_{ij}$ themselves. Now, by summing the products $ Y_{2m} \, \delta M_{2m}$ and $Y_{2m} \, \delta S_{2m}$ over the $m$ modes, the formulas \eqref{M2S2} imply
\begin{subequations}\label{MabSab}
    \begin{align}
    n^i n^j \delta M_{ij} &\doteq \frac{16}{45} \, a M^4 \, n^i \partial_\varphi n^j \mathcal{E}_{ij} 
    \, , \\
    n^i n^j \delta S_{ij} &\doteq \frac{16}{45} \, a M^4 \, n^i \partial_\varphi n^j \mathcal{B}_{ij}
    \, .
    \end{align}
\end{subequations}
Notice that $\partial_\varphi n^i = \sin{\theta} \, e_\varphi^i$ is orthogonal to the unit radial direction $n^i$. Since $n^i n^j \delta M_{ij}$ and $n^i n^j \delta S_{ij}$ are not proportional to $n^i n^j \mathcal{E}_{ij}$ and $n^i n^j \mathcal{B}_{ij}$, respectively, the perturbed quadrupole moments cannot satisfy simple proportionality relations akin to Eq.~\eqref{lambda}. Rather they must obey more general {\it tensorial} relations of the form
\beq\label{M=lambdaE}
    \delta M_{ij} \doteq \lambda_{ijkl} \, \mathcal{E}^{kl} \quad \text{and} \quad \delta S_{ij} \doteq \lambda_{ijkl} \, \mathcal{B}^{kl} \, ,
\eeq
where the constant tensor $\lambda_{ijkl} = O(\ad)$ is the quadrupolar TLT of a Kerr black hole. Such a complication with respect to the nonspinning case [recall Eq.~\eqref{lambda}] stems from the fact that the nonzero black hole spin breaks the spherical symmetry of the background spacetime. Recall that the notion of TLT was previously introduced, in Sec.~\ref{subsec:U-psi0_spin} above, in a Newtonian context. Because $\delta M_{ij}$ and $\mathcal{E}_{ij}$ are both STF tensors, it follows that $\lambda_{ijkl} = \lambda_{\langle ij \rangle kl}$ and that  it is sufficient to determine $\lambda_{ij \langle kl \rangle}$.

Since the perturbed quadrupole moments $\delta M_{ij}$ and $\delta S_{ij}$  in \eqref{M=lambdaE} obey identical relations with respect to the electric- and magnetic-type tidal tensors,\footnote{In other words, there is no parity mixing and the quadrupolar mass-electric TLT $\lambda_{ijkl}^{M\mathcal{E}}$ and current-magnetic TLT $\lambda_{ijkl}^{S\mathcal{B}}$ are equal: $\lambda_{ijkl}^{M\mathcal{E}}=\lambda_{ijkl}^{S\mathcal{B}}=\lambda_{ijkl}$. Together with the property we have found that there is no $\ell$-mode coupling in the $\ell=2$ result. It follows that the quadrupolar TLTs and TLNs of Kerr continue to be related via \eqref{plop}  under $\lambda_{2 m,\ell' m'} \to \lambda_{2 \ell' m}^{M \mathcal{E}} \, \delta_{\ell' 2} \, \delta_{m'm} = \lambda_{2 \ell' m}^{S \mathcal{B}} \, \delta_{\ell' 2} \, \delta_{m'm}$.} hereafter we shall restrict our attention to the perturbed mass-type quadrupole $\delta M_{ij}$. This STF tensor can be expressed in terms of its spherical-harmonic modes $\delta M_{2m}$ according to~\cite{PoWi}
\beq\label{Mij}
    \delta M_{ij} = \sum_{m=-2}^2 \bar{\mathscr{Y}}_{ij}^{2m} \, \delta M_{2m} \doteq \frac{8}{45} \, a M^4 \sum_{m=-2}^2 \ui m \, \bar{\mathscr{Y}}_{ij}^{2m} \mathcal{E}_{2m} \, ,
\eeq
where we used Eq.~\eqref{M2} in the second equality. The constant STF tensors $\mathscr{Y}_L^{\ell m}$ are defined in App.~\ref{subapp:STF}. Combined with Eqs.~\eqref{M=lambdaE} and \eqref{F^L}, the formula \eqref{Mij} implies
\beq\label{confinement}
    \left( \lambda_{ijkl} - \frac{64\pi}{675} \, a M^4 \! \sum_{m=-2}^2 \ui m \,  \bar{\mathscr{Y}}_{ij}^{2m} \mathscr{Y}_{kl}^{2m} \right) \mathcal{E}^{kl} \doteq 0 \, .
\eeq
This equation must be verified for any quadrupolar tidal tensor $\mathcal{E}_{kl}$, so the difference between parentheses must vanish, up to a term antisymmetric in $(k,l)$ and a term proportional to $\delta_{kl}$. By taking the STF part with respect to the indices $(k,l)$ and using the symmetry property $\mathscr{Y}_{ij}^{2 \, -m} \!=\! (-)^m \bar{\mathscr{Y}}_{ij}^{2m}$, we obtain
\beq\label{eq:TLT}
    \lambda_{ij \langle kl \rangle} \doteq - \frac{128\pi}{675} \, a M^4 \sum_{m=1}^2 m \, \Im{\left( \bar{\mathscr{Y}}_{ij}^{2m} \mathscr{Y}_{kl}^{2m} \right)} \doteq - \frac{8}{45} \, \chi M^5 I_{ijkl}^{(1)} \, ,
\eeq
which verifies the antisymmetry property $\lambda_{ij \langle kl \rangle} = - \lambda_{kl \langle ij \rangle}$ by exchange $ij \leftrightarrow kl$ of the pairs of indices. The TLT \eqref{eq:TLT} fits the generic pattern \eqref{TLT_exp} identified for a tidally-perturbed, spinning, axisymmetric, Newtonian body which displays no mode coupling. For a Kerr black hole we find, in particular, $\lambda_2^{(0)} = 0$, $\lambda_2^{(1)}(\chi=0) = (2M)^5 / 180$ and $\lambda_2^{(n)}(\chi=0) =0$ for $n \geqslant 2$, or equivalently $k_2^{(0)} = 0$, $k_2^{(1)}(\chi=0) = - 1/120$ and $k_2^{(n)}(\chi=0) = 0$ for $n \geqslant 2$.

The rank four tensor in Eq.~\eqref{eq:TLT} is explicitly given by Eq.~\eqref{I1}, so that the quadrupolar TLT \eqref{eq:TLT} can be written in the simple, compact form
\beq\label{TLT}
    \lambda_{ij \langle kl \rangle} \doteq - \frac{16}{45} \, M^3 \, \delta_{(i|\langle k} \, \varepsilon_{l \rangle|j)q} S^q \, ,
\eeq
where we recall that $S^i = J s^i$ is the Kerr black hole spin vector pointing in the unit direction $s^i$, and that $\delta_{ij}$ is the Kronecker symbol, while $\varepsilon_{ijk}$ is the totally antisymmetric Levi-Civita symbol with $\varepsilon_{123} = +1$. Consequently, the tidally-induced quadrupole moments \eqref{M=lambdaE} of a Kerr black hole explicitly read
\beq\label{youpi}
    \delta M_{ij} \doteq \frac{16}{45} \, M^3 \, \mathcal{E}^k_{\phantom{k}(i} \, \varepsilon_{j)kl} S^l \quad \text{and} \quad \delta S_{ij} \doteq \frac{16}{45} \, M^3 
    \, \mathcal{B}^k_{\phantom{k}(i} \, \varepsilon_{j)kl} S^l \, .
\eeq

\subsection{Quadrupolar tidal Love numbers}

Going back to the general definition \eqref{TLN} of the TLNs of a compact object, we conclude from the expressions \eqref{M2S2} that the quadrupolar TLNs of a spinning black hole are given, to linear order in the black hole spin, by
\begin{subequations}\label{TLN_quad_octu}
    \begin{align}
        \lambda_{2m}^{M\mathcal{E}} &\doteq \lambda_{2m}^{S\mathcal{B}} \doteq \frac{\ui m \ad}{180} \, (2M)^5 \, ,
        \\
        \lambda_{2m}^{M\mathcal{B}} &\doteq \lambda_{2m}^{S\mathcal{E}} \doteq 0 \, .
    \end{align}
\end{subequations}
This is another key result of this paper, as we have proven that a spinning black hole becomes tidally deformed under the effect of a weak and nonaxisymmetric tidal field. Remarkably, the quadrupolar TLNs \eqref{TLN_quad_octu} of a Kerr black hole are in full agreement with the formulas \eqref{TLN_conjecture} that were conjectured earlier, for an arbitrary multipolar index $\ell \in \mathbb{L}$, based on the large-$r$ behavior of the perturbed curvature scalar $\psi_0$. This supports the validity of the conjecture put forward in Sec.~\ref{subsec:TLN}. It would be interesting to compare the formulas \eqref{TLN_conjecture} to a direct calculation of the TLNs of a Kerr black hole for a multipolar index $\ell \geqslant 3$.

By analogy with the Newtonian TLNs $k_{\ell m}$ introduced in Sec.~\ref{sec:Newton}, we define some \emph{dimensionless} relativistic TLNs $k_{\ell m}^{M\mathcal{E}}$ according to
\beq\label{pouet-pouet}
    \lambda_{\ell m}^{M\mathcal{E}} \equiv - \frac{2(\ell-2)!}{(2\ell-1)!!} \, k_{\ell m}^{M\mathcal{E}} \, (2M)^{2\ell+1} \, ,
\eeq
and similarly for the $M\mathcal{B}$, $S\mathcal{E}$ and $S\mathcal{B}$ couplings, since the gravitational radius of a nonspinning Kerr black hole is equal to $2M$. The results \eqref{TLN_quad_octu} then imply
\begin{subequations}\label{TLN_quad_octu_dim}
    \begin{align}
        k_{2m}^{M\mathcal{E}} &\doteq k_{2m}^{S\mathcal{B}} \doteq - \frac{\ui m \ad}{120} \, , \label{k2m} \\
        k_{2m}^{M\mathcal{B}} &\doteq k_{2m}^{S\mathcal{E}} \doteq 0 \, ,
    \end{align}
\end{subequations}
which is consistent with the Newtonian TLNs \eqref{k_phys} at the quadrupolar order. The nonzero TLNs $k_{\ell m}^{M\mathcal{E}}$ and $k_{\ell m}^{S\mathcal{B}}$ generalize to a rotating background the dimensionless gravito-electric and gravito-magnetic TLNs $k_\ell^\text{el}$ (resp. $k_\ell$) and $k_\ell^\text{mag}$ (resp. $j_\ell$) that have been introduced in Ref.~\cite{BiPo.09} (resp. Ref.~\cite{DaNa2.09}) for a tidally perturbed, nonspinning compact object. Indeed, the rotation of the central compact object breaks the spherically symmetry of the background spacetime, so that the TLNs now depend on the azimuthal number $m$, in addition to the multipolar index $\ell$, just like in Newtonian gravity (recall Sec.~\ref{subsec:U-psi0_spin}).

For a black hole spin $\ad \sim 0.12$, the results \eqref{k2m} give $|k_{2,\pm 2}^{M\mathcal{E}}| = |k_{2,\pm2}^{S\mathcal{B}}| \sim 0.002$. This small number can be compared, for instance, to the values $k_2^\text{el} \sim 0.05-0.15$ and $|k_2^\text{mag}| \lesssim 6 \times 10^{-4}$ of the gravito-electric and gravito-magnetic quadrupolar TLNs of a \textit{non}spinning neutron star, depending on the equation of state~\cite{DaNa2.09,BiPo.09}. Moreover, the large-$\ell$ asymptotics \eqref{kappa_large-l} of the coefficients $k_{\ell m}$ in \eqref{k_phys} show that the linear response of a Kerr black hole to a perturbing tidal field decays quickly with increasing $\ell$. Therefore, although spinning black holes become tidally deformed like all other self-gravitating bodies, this physical effect is numerically very small. In this precise sense, black holes are especially ``rigid'' compact objects, in accordance with their extreme compactness.

The results \eqref{TLN_quad_octu}--\eqref{TLN_quad_octu_dim} are complementary to those previously derived in Refs.~\cite{Pa.al.15,LaPo.15}. In particular, Pani et al.~\cite{Pa.al.15,Pa.al2.15} used the same definition as us for the TLNs of a tidally perturbed compact object, and found zero black hole TLNs for an \emph{axisymmetric} tidal field of electric type, up to quadratic order in the spin. The formulas \eqref{TLN_quad_octu} with $m = 0$ agree with that conclusion. However, Landry and Poisson~\cite{LaPo.15} introduced an alternative definition of the TLNs of a slowly spinning compact object, which appear as integration constants attached to the decaying solutions of the homogeneous perturbation equations, and parameterize the near-zone metric of the tidally perturbed compact body, expressed in light-cone coordinates. While those near-zone TLNs have been shown to be invariant under infinitesimal coordinate (i.e. perturbative gauge) transformations that preserve the geometrical meaning of the light-cone coordinates~\cite{BiPo.09,LaPo.15}, they lack the fully coordinate-invariant nature of the TLNs defined in Eq.~\eqref{TLN}. The authors considered a generic quadrupolar tidal perturbation of a slowly rotating compact object and introduced six TLNs, denoted $K_2^\text{el}$, $K_2^\text{mag}$, $\mathfrak{E}^\mathrm{q}$, $\mathfrak{F}^\mathrm{o}$, $\mathfrak{B}^\mathrm{q}$ and $\mathfrak{K}^\mathrm{o}$, which \emph{all vanish} for a Kerr black hole, to first order in spin, as a consequence of imposing the response solution to be smooth on the horizon. The first two of those six constants are proportional to the TLNs $k_2^\text{el}$ and $k_2^\text{mag}$ introduced above for nonspinning compact objects. The last four are new  ``rotational-type'' TLNs. Two of those correspond to the $\ell$-mode couplings (here quadrupole-octupole) that are allowed at linear order in spin~\cite{Pa.al2.15}, while the remaining two are analogous to our TLNs $k^{M\mathcal{E}}_{2m}$ and $k^{S\mathcal{B}}_{2m}$. While this difference in definitions of TLNs could superficially explain why our results \eqref{k2m} appear to enter in conflict with the results of Ref.~\cite{LaPo.15}, we argue in App.~\ref{app:splits} that the vanishing TLNs found there is directly related to their choice of tidal/response split.

\section{Tidal torquing of spinning black holes}\label{sec:torquing}

In this section, we establish that the tidally-induced quadrupole moments \eqref{youpi} of a Kerr black hole, as derived in Sec.~\ref{sec:multipoles}, are closely related to the well-known physical phenomenon of tidal torquing of a spinning body interacting with a tidal gravitational environment.\footnote{This possibility was first suggested to us (and to the audience) by Walter Goldberger during the online workshop ``Rethinking the Relativistic Two-Body Problem" organized by the Albert Einstein Institute (August 24--28, 2020).}

In a seminal paper Thorne and Hartle~\cite{ThHa.85} showed that an arbitrary spinning body that interacts with a weak and slowly varying tidal environment suffers a tidal torque that drives the evolution of the body's spin angular momentum $S^i$. Let $\mathcal{E}_{ij}$ and $\mathcal{B}_{ij}$ denote the electric-type and magnetic-type quadrupolar tidal fields that characterize the tidal environment with characteristic lengthscale of variation $\mathcal{R}$. Let $M_{ij}$ and $S_{ij}$ denote the mass- and current-type quadrupole moments of the spinning body. If $M / \mathcal{R} \ll 1$, then the average rate of change of the body's spin magnitude $S$ is given by\footnote{\label{ftn:3/4}As shown in Ref.~\cite{Gu.83}, the Geroch-Hansen current quadrupole moment $S_{ij}$ is equal to $4/3$ times the Thorne current quadrupole moment used in~\cite{ThHa.85,Po2.04}. This explains the absence of a factor of $4/3$ in the second term on the right-hand side of \eqref{torque} in relation to these references.}~\cite{ThHa.85,Po2.04}
\beq\label{torque}
    \langle \dot{S} \rangle = - \varepsilon^{ijk} s_i \, \langle M_{jl} \mathcal{E}^l_{\phantom{l}k} + S_{jl} \mathcal{B}^l_{\phantom{l}k} \rangle \, . 
\eeq

We shall now apply the general law \eqref{torque} to the particular case of a Kerr black hole whose quadrupole moments are given, to linear in the spin, by the formulas \eqref{youpi}. Substituting  those expressions into Eq.~\eqref{torque}, we readily obtain
\beq\label{torque_Kerr}
    \langle \dot{S} \rangle \doteq - \varepsilon^{ijk} s_i \, \langle \delta M_{jl} \mathcal{E}^l_{\phantom{l}k} + \delta S_{jl} \mathcal{B}^l_{\phantom{l}k} \rangle \doteq - \frac{8}{45} \, M^5 \ad \left[ 2 \langle E_1 + B_1 \rangle - 3 \langle E_2 + B_2 \rangle \right] ,
\eeq
where we introduced the tidal invariants $E_1 \equiv \mathcal{E}_{ij}\mathcal{E}^{ij}$ and $E_2 \equiv \mathcal{E}_{ij} s^j \mathcal{E}^{ik} s_k$, and similarly for the magnetic-type quadrupolar tidal fields. This formula can be compared to the results of Poisson~\cite{Po2.04}, who precisely computed the tidal torquing of a Kerr black hole to leading order in a multipole expansion of the tidal environment. The result, given in Eq.~(9.39) therein, is valid for a \textit{generic} Kerr black hole spin $\ad$, and reads
\beq\label{torque_Kerr_Poisson}
    \langle \dot{S} \rangle = - \frac{2}{45} \, M^5 \ad \left[8(1+3\ad^2) \langle E_1 + B_1 \rangle - 3(4+17\ad^2) \langle E_2 + B_2 \rangle + 15 \ad^2 \langle E_3 + B_3 \rangle \right] ,
\eeq
which involves the additional tidal invariants $E_3 \equiv (\mathcal{E}_{ij} s^i s^j)^2$ and $B_3 \equiv (\mathcal{B}_{ij} s^i s^j)^2$. Clearly, the restriction to $O(\ad)$ of the formula \eqref{torque_Kerr_Poisson} is in perfect agreement with the formula \eqref{torque_Kerr}. This consistency test provides strong support to the validity of the long and involved calculation that yielded the tidally-induced quadrupole moments \eqref{youpi}.

\subsection{Tidally-induced quadrupole moments}

Interestingly, in Eqs.~(9.44)--(9.45) of~\cite{Po2.04}, Poisson gave expressions for the tidally-induced quadrupole moments of a Kerr black hole. Those expressions were \textit{inferred} from a comparison of the general tidal-torquing formula \eqref{torque} to the specific Kerr black hole result \eqref{torque_Kerr_Poisson}. Those tidally-induced quadrupole moments read (recall footnote \ref{ftn:3/4})
\begin{subequations}\label{youpi_Poisson}
    \begin{align}
        \! M^\text{torque}_{ij} &= \frac{2}{45} \, M^5 \ad \left[ 8(1+3\ad^2) \mathcal{E}^k_{\phantom{k}(i} \varepsilon_{j)kl} s^l + 30 \ad^2 s_{(i} \varepsilon_{j)kl} s^k \mathcal{E}^{lq} s_q - 3 \lambda(\ad) s_{\langle i} s_{j \rangle} \, \mathcal{E}_{kl} s^k s^l \right] , \label{youpi_deltaM} \\
        \! S^\text{torque}_{ij} &= \frac{2}{45} \, M^5 \ad \left[ 8(1+3\ad^2) \mathcal{B}^k_{\phantom{k}(i} \varepsilon_{j)kl} s^l + 30 \ad^2 s_{(i} \varepsilon_{j)kl} s^k \mathcal{B}^{lq} s_q - 3 \mu(\ad) s_{\langle i} s_{j \rangle} \, \mathcal{B}_{kl} s^k s^l \right] ,
    \end{align}
\end{subequations}
where $\lambda(\ad)$ and $\mu(\ad)$ are two undetermined functions of the dimensionless Kerr parameter. Up to the terms involving those functions, we notice that the quadrupole moments \eqref{youpi_Poisson} only involve terms linear and cubic in the black hole spin $\ad$. Moreover the formulas \eqref{youpi_Poisson} display no parity mixing nor mode coupling. It can easily be checked that the common TLT associated with the couplings \eqref{youpi_Poisson} simply reads
\beq\label{TLT_Poisson}
    \lambda_{ij\langle kl \rangle}^\text{torque} = \frac{2}{45} \, M^5 \ad \left[ - 8 (1+3\ad^2) \, \delta_{(i|\langle k} \, \varepsilon_{l \rangle|j)q} s^q + 30 \ad^2 \, s_{(i} \varepsilon_{j) q \langle k} s_{l \rangle} s^q - 3 \lambda(\ad) \, s_{\langle i} s_{j \rangle} s_{\langle k} s_{l \rangle} \right] ,
\eeq
and similarly for the current-type quadrupole with the substitution $\lambda(\ad) \to \mu(\ad)$.

Clearly, by requiring agreement between Eqs.~\eqref{youpi} and \eqref{youpi_Poisson} to linear order in the spin, or equivalently between Eqs.~\eqref{TLT} and \eqref{TLT_Poisson}, the functions $\lambda(\ad)$ and $\mu(\ad)$ must satisfy the constraints
\beq
    \lambda(0) = \mu(0) = 0 \, .
\eeq
One can go even further by appealing to the general analysis performed in Sec.~\ref{subsec:U-psi0_spin} above; see specifically Eqs.~\eqref{TLT_exp}--\eqref{eq:symm TLN} there. Since the quadrupolar TLT \eqref{TLT_Poisson} displays no mode coupling nor parity mixing, its tensorial structure is necessarily encoded into the discrete set of STF tensors \eqref{STF_tensors}. If $\lambda(\ad) \neq 0$, then the tensor $s_{\langle i} s_{j \rangle} s_{\langle k} s_{l \rangle}$ must necessarily be a linear combination of the STF tensors \eqref{TLT_matrix_Re}, because it is symmetric by exchange of the pair of indices $ij$ and $kl$. However this is impossible because the matrix associated with the tensor $s_{\langle i} s_{j \rangle} s_{\langle k} s_{l \rangle}$ is diagonal with nonvanishing components in the $(3,3)$ block, while this block vanishes identically for all the tensors \eqref{TLT_matrix_Re}. We therefore conclude that the undetermined functions in Eqs.~\eqref{youpi_Poisson} and \eqref{TLT_Poisson} vanish identically:
\beq\label{lambda-mu}
    \lambda(\ad) = \mu(\ad) = 0 \, .
\eeq

The previous discussion tentatively suggests that the tidally-induced quadrupole moments that we have computed (to linear order in spin), out of the Geroch-Hansen definition, agree in general with those inferred in~\cite{Po2.04}, out of a completely independent tidal torquing calculation. We thus conjecture that the \textit{exact} expressions (i.e., valid to all orders in spin) for the tidally-induced quadrupole moments are simply given by
\begin{subequations}\label{quad_mom_conj}
    \begin{align}
        \delta M_{ij} &= \frac{4}{45} \, M^5 \ad \left[ 4(1+3\ad^2) \mathcal{E}^k_{\phantom{k}(i} \, \varepsilon_{j)kl} s^l + 15 \ad^2 s_{(i} \, \varepsilon_{j)kl} s^k \mathcal{E}^{lq} s_q \right] , \\
        \delta S_{ij} &= \frac{4}{45} \, M^5 \ad \left[ 4(1+3\ad^2) \mathcal{B}^k_{\phantom{k}(i} \, \varepsilon_{j)kl} s^l + 15 \ad^2 s_{(i} \, \varepsilon_{j)kl} s^k \mathcal{B}^{lq} s_q \right] .
    \end{align}
\end{subequations}
Remarkably, such expressions would \textit{truncate} at $O(\ad^3)$. This would be consistent with the fact that the ``Newtonian TLNs'' \eqref{k_phys} of a Kerr black hole were found to truncate as well. This conjecture could be explored by pushing up to $O(\ad^3)$ the calculation of the quadrupole moments performed in Sec.~\ref{sec:multipoles}; in particular it predicts no $O(\ad^2)$ contribution.

We shall now prove that the expressions \eqref{quad_mom_conj} can easily be recovered out of the postulated formulas \eqref{TLN_conjecture}, according to which the nonvanishing quadrupolar TLNs of a Kerr black hole should exactly read\footnote{Interestingly, comparing the expression \eqref{TLNs_l=2} with the formula between Eqs.~(9.32) and (9.33) of Ref.~\cite{Po2.04}, we notice that the leading-order tidal torquing is simply given by $\langle \dot{S} \rangle = \sum_{m \neq 0} (\ui\lambda_{2m})/(2m) \, \langle |z_{2m}|^2 \rangle$. This expression is likely a consequence of the general formula \eqref{torque}, where the tidally-induced quadrupoles are given in terms of the quadrupolar tidal fields via the TLT which are associated with the TLNs in \eqref{TLNs_l=2}.}
\beq\label{TLNs_l=2}
    \lambda_{2m} \equiv \lambda_{2m}^{M\mathcal{E}} = \lambda_{2m}^{S\mathcal{B}} = \frac{2M^5}{45} \, \ui m\chi \left[ (1-\ad^2) + m^2 \ad^2 \right] \! \left[ 4 (1-\ad^2) + m^2 \ad^2 \right] .
\eeq
Applying the general result \eqref{TLT_exp} to the quadrupolar case ($\ell = 2$), with $L=ij$ and $L'=kl$, together with the explicit formula \eqref{TLNs_l=2} for the TLNs, the common (mass- and current-type) quadrupolar TLT should then be given by
\begin{align}\label{TLT_quad}
    \lambda_{ij \langle kl \rangle} &= \frac{8\pi}{15} \sum_{m=-2}^2 \lambda_{2m} \bar{\mathscr{Y}}_{ij}^{2m} \mathscr{Y}_{kl}^{2m} \nonumber \\ &= - \frac{2}{45} M^5 \chi \, \Bigl[ 4(1-\ad^2)^2 \, I^{(1)}_{ijkl} + 5\chi^2(1-\ad^2) \, I^{(3)}_{ijkl} + \chi^4 \, I^{(5)}_{ijkl} \Bigr] \, .
\end{align}
This formula fits the general tensorial pattern \eqref{TLT_exp}, with the only nonvanishing functions $\lambda_2^{(1)}(\chi) = (1-\chi^2)^2 (2M)^5 / 180$, $\lambda_2^{(3)}(\chi) = - \chi^2(1-\chi^2) (2M)^5 / 144$ and $\lambda_2^{(5)}(\chi) = \chi^4 (2M)^5 / 720$. The tensors $I^{(1)}_{ijkl}$, $I^{(3)}_{ijkl}$ and $I^{(5)}_{ijkl}$ appearing in \eqref{TLT_quad} are explicitly given by Eqs.~\eqref{I1}, \eqref{I3} and \eqref{I5}. Most remarkably, while substituting those expressions, the three contributions of $O(\chi^5)$ cancel out exactly in the final expression, leaving a quadrupolar TLT that truncates at $O(\chi^3)$. More precisely, we find
\beq\label{TLT_quad_bis}
    \lambda_{ij\langle kl \rangle} = \frac{4}{45} \, M^5 \ad \left[ - 4 (1+3\ad^2) \, \delta_{(i|\langle k} \, \varepsilon_{l \rangle|j)q} s^q + 15 \ad^2 \, s_{(i} \varepsilon_{j) q \langle k} s_{l \rangle} s^q \right] ,
\eeq
which fully agrees with Eq.~\eqref{TLT_Poisson} with $\lambda(\chi)=0$. The associated tidally-induced quadrupole moments $\delta M_{ij} = \lambda_{ij \langle kl \rangle} \, \mathcal{E}^{kl}$ and $\delta S_{ij} = \lambda_{ij \langle kl \rangle} \, \mathcal{B}^{kl}$ are indeed given by the formulas \eqref{quad_mom_conj}.

In summary, the tidal torquing calculation performed in Ref.~\cite{Po2.04} could be used to infer the exact tidally-induced quadrupole moments of a Kerr black hole, given in closed form in \eqref{youpi_Poisson}, up to two undetermined functions that we have shown must vanish, Eq.~\eqref{lambda-mu}. Remarkably, by using the postulated formulas \eqref{TLN_conjecture} for the TLNs of a Kerr black hole, we could readily reproduce the expressions \eqref{quad_mom_conj} for those quadrupole moments. This observation gives much support to the conjecture that the TLNs of a Kerr black holes are given, for any multipolar index and to all orders in the spin, by the simple algebraic formulas \eqref{TLN_conjecture}.

\subsection{Tidally-induced octupole moments}

The tidal torquing calculation of~\cite{Po2.04} was later extended to next-to-leading order in the tidal coupling~\cite{Ch.al.13}. The generalization of \eqref{torque_Kerr} to next-to-leading order involves additional terms that depend on the first time derivatives of the slowly evolving quadrupolar tidal fields, but it does not involve the octupolar tidal moments. Hence the clever ``trick'' used to infer the quadrupole moments in \eqref{youpi_Poisson} cannot be extended to constrain the octupole moments. Rather, given the newly confidence gained in the validity of the formula \eqref{TLN_conjecture}, we may use it for $\ell=3$, yielding
\beq\label{TLNs_l=3}
    \lambda_{3m}^{M\mathcal{E}} = \frac{3}{4} \lambda_{3m}^{S\mathcal{B}} = \frac{2M^7}{14175} \, \ui m\chi \left[ (1-\ad^2) + m^2 \ad^2 \right] \! \left[ 4 (1-\ad^2) + m^2 \ad^2 \right] \! \left[ 9 (1-\ad^2) + m^2 \ad^2 \right] .
\eeq
Applying the general result \eqref{TLT_exp} to the octupolar case ($\ell = 3$), with $L=ijk$ and $L'=lpq$, together with the Kerr formula \eqref{TLNs_l=3} for the TLNs, the mass-electric and current-magnetic octupolar TLTs should then be given by
\begin{align}\label{TLT_octu}
    \lambda^{M\mathcal{E}}_{ijk \langle lpq \rangle} = \frac{3}{4} \lambda^{S\mathcal{B}}_{ijk \langle lpq \rangle} &= \frac{24\pi}{105} \sum_{m=-3}^3 \lambda^{M\mathcal{E}}_{3m} \bar{\mathscr{Y}}_{ijk}^{3m} \mathscr{Y}_{lpq}^{3m} \nonumber \\ &= - \frac{2M^7}{14175} \, \chi \, \Bigl[ 36(1-\ad^2)^3 \, I^{(1)}_{ijklpq} + 49 \chi^2(1-\chi^2)^2 \, I^{(3)}_{ijklpq} \nonumber \\ &\qquad\qquad\qquad + 14 \chi^4(1-\chi^2) \, I^{(5)}_{ijklpq} + \chi^6 \, I^{(7)}_{ijklpq} \Bigr] \, .
\end{align}
Remarkably, by using the general definition \eqref{I's} of the tensors $I^{(2n-1)}_{LL'}$ for $\ell=3$, we find that the coefficient of the term of $O(\chi^7)$ in \eqref{TLT_octu} vanishes identically, so that the octupolar TLTs \eqref{TLT_octu} truncate at $O(\chi^5)$. This prediction could be checked by a future calculation of the tidally-induced Geroch-Hansen octupole moments of a Kerr black hole.

\section{Summary and prospects}

In this paper we addressed the issue of deformability of black holes under the effect of a weak, slowly-varying perturbing tidal field. Such deformability can be measured by means of TLNs, defined in Eq.~\eqref{TLN} as the rates of change of the multipole moments of the perturbed black hole geometry under small variations of the perturbing tidal field. In previous works it was shown that the static TLNs vanish for nonspinning (i.e. Schwarzschild) black holes~\cite{BiPo.09,DaNa2.09,KoSm.12,Ch.al2.13,Gu.15}, as well as for spinning (i.e. Kerr) black holes~\cite{LaPo.15,Pa.al.15}, but only to first order in the spin, for a generic quadrupolar tidal field, and to quadratic order in the spin, for an axisymmetric and quadrupolar tidal field of electric type.

The results established in Ref.~\cite{Pa.al.15} were derived by solving the metric perturbation equations directly. Those equations, however, cannot be solved in closed form for an arbitrary multipolar index $\ell$, and so the authors of~\cite{Pa.al.15} could not unambiguously disentangle the part of the metric perturbation that corresponds to the perturbing tidal field from the part that corresponds to the linear response of the compact body. On the other hand, the vanishing TLNs in Ref.~\cite{LaPo.15} appear as integration constants attached to the decaying solutions of the homogeneous perturbation equations, and thus need not coincide with the TLNs that we have defined here, out of the coordinate-invariant Geroch-Hansen multipole moments. Moreover, the tidal/response split of the full physical solution used in~\cite{LaPo.15} is different from that performed in this work.

In this paper, we tackled the case of a rotating black hole for \emph{arbitrary} spin and (weak, slowly varying) tidal field. What enabled us to carry out the investigation is that we solved the mode frequency $\omega=0$  Teukolsky equation for the Weyl scalar $\psi_0$ (instead of the metric perturbation equations), as this equation separates by variables and the radial equation admits, in the static case, two simple linearly independent  solutions known in closed form. The fact that we have obtained the solutions for a generic multipolar number $\ell$  allowed us to unambiguously identify one solution as the external tidal field and the other one as the corresponding black hole linear response. This linear response vanishes identically for a Schwarzschild black hole and for an axisymmetric perturbation of a Kerr black hole, in full agreement with the literature. For a nonaxisymmetric perturbation of a spinning black hole, however, the linear response does \emph{not} vanish.

Although this is at the level of the curvature scalar, by analyzing the operations that take the Weyl scalar to the metric perturbation, those results continue to hold at the level of the metric perturbation. We have thus shown that the TLNs of rotating black holes, for \emph{arbitrary} values of the spin and multipolar number, are generically nonzero. As a first application, we computed the nonzero Kerr black hole TLNs that couple the mass-type (resp. current-type) quadrupole moments to the electric-type (resp. magnetic-type) quadrupolar tidal fields, to linear order in the black hole spin. In addition, we conjectured that the TLNs of a Kerr black hole display no mode coupling nor parity mixing in general, and only involve odd powers of the black hole spin.

In the context of four-dimensional general relativity, all previous works on the tidal deformability of black holes suggested that these compact bodies behave in a peculiar manner, as they were found not to deform under the effect of a static tidal perturbation~\cite{BiPo.09,DaNa2.09,KoSm.12,Ch.al2.13,Gu.15,LaPo.15,Pa.al.15}. Our work shows, on the contrary, that a generic spinning black hole embedded in a generic, static tidal environment will in fact become tidally deformed, like any other self-gravitating body. In particular, while in a binary system, a spinning black hole develops multipole moments induced by the tidal field of its companion. We further showed that, at quadrupolar order, the tidally-induced multipole moments of a Kerr black hole reproduce the known tidal torquing phenomenon resulting from the interaction of a spinning body with a tidal gravitational environment~\cite{ThHa.85,Po2.04}.

This unexpected result has important consequences for the ongoing effort to model the dynamics and gravitational-wave emission from binary systems of spinning black holes~\cite{BuSa.15}. In particular, it shows that in the context of the post-Newtonian approximation~\cite{Bl.14} and the effective one-body model~\cite{DaNa}, the effective description of spinning black holes must \textit{a priori} account for the tidally-induced quadrupole (and higher order) moments of these compact objects. Such finite-size effects can be accounted for through the multipolar gravitational skeleton framework~\cite{StPu.10,RaLe.21} or by using effective-field-theory techniques~\cite{Po.16,Le.20}. Crucially, one should assess carefully if and how, while in a binary system, the tidally-induced multipole moments of a spinning black hole affect the gravitational-wave signal generated by the orbital motion, beyond the known effects of tidal heating and tidal torquing~\cite{Al.01,Po.08}.

Future work could be built on the present analysis in order to compute the TLNs of a Kerr black hole at higher orders in the spin, to higher multipolar orders, or to second order in the perturbing tidal field. One may also attempt to relate the gravitational TLNs defined and computed here to the ``shape'' Love numbers~\cite{DaLe.09,DaNa2.09,LaPo.14} of a spinning black hole, and to the intrinsic curvature of the perturbed event horizon~\cite{Ve.al.11}, in particular when the tidal perturbation is sourced by a small orbiting companion~\cite{OsHu.14,OsHu.16,Pe.al.17}. Finally, it would also be interesting to assess whether or not the black hole tidal polarizability effect uncovered here could be seen in the context of numerical relativity simulations of compact binary systems involving at least one spinning black hole.

\textit{Note added:} following the submission of this paper for publication, Refs.~\cite{Go.al.20,Ch.al.21} appeared on the arXiv preprint server. These two papers established, in the context of effective field theory, that the Kerr black hole tidal deformation uncovered here (see also \cite{LeCa.21}) gives rise to purely dissipative effects. Reference \cite{Ch.al.21} also vindicated the analytic continuation of $\ell \in \mathbb{R}$ advocated here and showed, moreover, that it is motivated from the effective-field-theory point of view. Physically, Refs.~\cite{Ch2.20,Go.al.20,Ch.al.21} are all in agreement with our result of nonzero and---as they show---purely dissipative  deformation of a Kerr black hole under a static tidal field. However, Refs.~\cite{Ch2.20,Go.al.20,Ch.al.21} disagree on the \textit{nomenclature} for ``tidal Love numbers'': whereas they describe this result  as vanishing tidal Love numbers, we instead prefer to describe it as nonzero tidal Love numbers, in keeping with the nomenclature followed until now within the literature in Newtonian gravity~\cite{PoWi,Og.14} and in General Relativity~\cite{Ch.al2.13,Pa.al.15,Pa.al2.15}.

\begin{acknowledgments}

We are grateful to Eric Poisson, Rafael Porto, Adrian Otewill and Sam Gralla for helpful correspondence. ALT acknowledges the financial support of the Action Fédératrice PhyFOG and of the Scientific Council of the Paris Observatory. All authors acknowledge partial financial support by CNPq (Brazil), process nos.\ 310200/2017-2 (MC), 302397/2019-1 (EF), 300328/2020-6 (EF), 301088/2020-9 (EF) and 312917/2019-8 (ALT).
\end{acknowledgments}

\appendix

\section{Spin-weighted spherical harmonics and applications}\label{app:SWSH}

In this appendix we briefly recall the definition of the spin-weighted spherical harmonics ${}_sY_{\ell m}$~\cite{NePe.66}, and summarize some of their main properties, since those angular functions play an important role in our black hole perturbative analysis.

\subsection{Spin-weighted spherical harmonics}

First, the ordinary scalar spherical harmonics are defined as
\beq\label{eq:sphcal}
    Y_{\ell m}(\theta,\pAdv) \equiv \sqrt{\frac{(2\ell+1)(\ell-m)!}{4\pi(\ell+m)!}} \,
P_\ell^m(\cos{\theta}) \, e^{\ui m \pAdv}=(-)^m \, \bar{Y}_{\ell \, -m}(\theta,\pAdv) \, ,
\eeq
where $P_\ell^m(z)$ is the associated Legendre polynomial of degree $\ell$ and of order $m$. Although a standard angular mode decomposition corresponds to integers $(\ell,m) \in \mathbb{N} \times \mathbb{Z}$ with $|m| \leqslant \ell$, in this work we consider the analytic continuation to $\ell \in \mathbb{R}$.

We now introduce the spin-weighted spherical harmonics. For any integer $s \!\in\! \mathbb{N}$, we define the spin-$s$ raising operator
\begin{equation}\label{eq:eth}
    \eth_s \equiv -\left(\partial_\theta + \ui \csc{\theta} \, \partial_\pAdv - s \cot{\theta}\right),
\end{equation}
out of which the spin-weighted spherical harmonics ${}_{s}Y_{\ell m}$ are built recursively from the scalar spherical harmonic ${}_{0}Y_{\ell m} \equiv Y_{\ell m}$, for any $\ell \geqslant |s|$, according to
\begin{equation}\label{eq:eth Y}
    {}_{s+1}Y_{\ell m} \equiv \left[(\ell-s)(\ell+s+1)\right]^{-1/2} \, \eth_s({}_sY_{\ell m}) \, .
\end{equation}
The spin-weighted spherical harmonics ${}_{s}Y_{\ell m}(\theta,\phi)$ provide an orthonormal basis of functions on the unit 2-sphere $\mathbb{S}^2$, in the sense that
\beq\label{ortho}
    \oint_{\mathbb{S}^2} {}_{s}Y_{\ell m} \, {}_{s}\bar{Y}_{\ell'm'} \, \ud\Omega = \delta_{\ell\ell'} \, \delta_{mm'} \, .
\eeq

An identity of particular interest that follows from the iterative definition \eqref{eq:eth Y}, and which was used in Eq.~\eqref{psi0_Newt mode} above, is
\beq\label{eq:Y2 via eth}
    \eth_1 \eth_0 Y_{\ell m} = \sqrt{\frac{(\ell+2)!}{(\ell-2)!}} \; {}_{2}Y_{\ell m} \, .
\eeq
From Eqs.~\eqref{eq:sphcal} and \eqref{eq:eth Y}, the following property follows:
\begin{equation}\label{eq:sphcal cc}
    {}_s\bar{Y}_{\ell m}=(-)^{s+m}{}_{-s}Y_{\ell \, -m} \, .
\end{equation}
Finally, from Eqs.~\eqref{eq:sphcal cc} and \eqref{eq:eth Y}, the operator $\eth_s$ acts as a spin-lowering operator on ${}_{-s}\bar{Y}_{\ell m}$ according to
\begin{equation}\label{eq:eth Ycc}
    {}_{-s-1}\bar{Y}_{\ell m} = - \left[(\ell-s)(\ell+s+1)\right]^{-1/2} \, \eth_s({}_{-s}\bar{Y}_{\ell m}) \, .
\end{equation}

\subsection{Spherical harmonics and symmetric trace-free tensors}\label{subapp:STF}

The scalar spherical harmonics \eqref{eq:sphcal} provide a basis of functions on the unit 2-sphere $\mathbb{S}^2$. Alternatively, it can prove convenient to use the basis made of STF products $n^{\langle L \rangle}$ of $\ell$ copies of the unit radial vector $n^i = (\sin\theta \cos\phi, \sin\theta \sin\phi, \cos\theta)$. The relationship between the two basis is provided by $Y_{\ell m} = \bar{\mathscr{Y}}^L_{\ell m} n_{\langle L \rangle}$, or equivalently by $n^{\langle L \rangle} = 4\pi \ell! / (2\ell+1)!! \sum_{m=-\ell}^\ell \mathscr{Y}^L_{\ell m} Y_{\ell m}$, where we introduced the constant STF tensors~\cite{Th.80,BlDa.86,PoWi}
\beq\label{eq:YL}
     \mathscr{Y}^L_{\ell m} \equiv \frac{(2\ell+1)!!}{4\pi \ell!} \oint_{\mathbb{S}^2} n^{\langle L \rangle} \bar{Y}_{\ell m}(\theta,\phi) \, \ud\Omega \, ,
\eeq
which verify the symmetry property $\mathscr{Y}^L_{\ell \, -m} =  (-)^m \bar{\mathscr{Y}}^L_{\ell m}$. Any STF tensor $F^L$ can be expanded over its spherical-harmonic modes $F_{\ell m}$ according to
\beq\label{F^L}
    F^L = \sum_{m=-\ell}^\ell \bar{\mathscr{Y}}^L_{\ell m} F_{\ell m} \, , \quad \text{where} \quad F_{\ell m} = \frac{4\pi \ell!}{(2\ell+1)!!} \, \mathscr{Y}^L_{\ell m} F_L \, .
\eeq
Contracting this mode expansion with $n^L$ yields the formula \eqref{nLEL} introduced in Sec.~\ref{sec:Newton}.

\section{Regge-Wheeler-Zerilli formalism and Bardeen-Press equation}\label{app:NS_check}

Perturbations of the Schwarzschild black hole are often studied within the Regge-Wheeler-Zerilli formalism~\cite{ReWh.57,Ze3.70}. The background metric is perturbed to linear order, and the metric perturbation $h_{\alpha\beta}$ is decomposed in spherical harmonics and separated according to its even or odd parity. In Schwarzschild coordinates $(t,r,\theta,\pAdv)$ and in the Regge-Wheeler gauge~\cite{ReWh.57},
\begin{subequations}
\begin{align}
\label{heven}
h_{\alpha\beta}^{\text{even}} &= \sum_{\ell m}
\left(\begin{array}{cccc}
f H_0^{\ell m} & H_1^{\ell m} & 0 & 0 \\
 H_1^{\ell m} & f^{-1}H_2^{\ell m} & 0 &0\\
 0 &0 & r^2 K^{\ell m}& 0\\ 
0&0 &0 & r^2\sin^2{\theta} K^{\ell m}
\end{array}\right)Y_{\ell m} \,, \\
\label{hodd}
h_{\alpha\beta}^{\text{odd}}
&= \sum_{\ell m}
\left(\begin{array}{cccc}
0 & 0 & - h_0^{\ell m} \csc\theta\,\partial_\pAdv & h_0^{\ell m} \sin\theta\,\partial_\theta \\
0 &0 & - h_1^{\ell m} \csc\theta\,\partial_\pAdv & h_1^{\ell m} \sin\theta\,\partial_\theta \\
- h_0^{\ell m} \csc\theta\,\partial_\pAdv & - h_1^{\ell m} \csc\theta\,\partial_\pAdv &0 &0\\
h_0^{\ell m} \sin\theta\,\partial_\theta & h_1^{\ell m} \sin\theta\,\partial_\theta & 0 &0 \end{array}
\right)Y_{\ell m} \, ,
\end{align}
\end{subequations}
where $f(r) \equiv 1-2M/r$, while $H_{0,1,2}^{\ell m}=H_{0,1,2}^{\ell m}(t,r)$, $K^{\ell m}=K^{\ell m}(t,r)$ and $h_{0,1}^{\ell m}=h_{0,1}^{\ell m}(t,r)$ are functions to be determined by solving the field equations. Because of the spherical symmetry of the background spacetime, the perturbation equations cannot mix terms that belong to  a different $\ell$ and parity, and $m$ can be set equal to zero. Therefore, in the following we drop these indices. Moreover, one typically assumes harmonic time dependence.

The Einstein field equations impose $H_0=H_2\equiv H$. In the static limit, they also impose $h_1=0$. The equations governing the radial part of the perturbations can be combined into wave equations for the Cunningham-Price-Moncrief~\cite{Cu.al.78} and Zerilli functions $\Psi_\text{CPM}$ and $\Psi_\text{Z}$, which are given combinations of the metric perturbation functions and their derivatives.\footnote{The Regge-Wheeler function $\Psi_\text{RW}$ and the Cunningham-Price-Moncrief function $\Psi_\text{CPM}$ formally satisfy the same equation. However, the Regge-Wheeler function is defined as $\Psi_\text{RW}(x) = x\,h_1(x)/(x+1)^2$, which in the static limit we are considering, vanishes identically.\label{footnote:RWCPM}} In the static limit, written in terms of the advanced Eddington-Finkelstein {dimensionless radial} coordinate $x \equiv r/(2M)-1$, the Cunningham-Price-Moncrief (or Regge-Wheeler) and Zerilli equations are
\begin{subequations}
\begin{align}\label{eq:ODE RW}
0 &= x\left(x+1\right) \Psi''_\text{CPM} + \Psi'_\text{CPM} - \frac{2n (x+1) + 2x - 1}{x+1}\,\Psi_\text{CPM} \,,\\
0 &= x\left(x+1\right) \Psi''_\text{Z} + \Psi'_\text{Z} - \frac{8n^3 (x+1)^3 + 4n^2 (2x+5) (x+1)^2 + 18n (x+1) + 9}{(x+1) [2n (x+1) + 3]^2}\,\Psi_\text{Z} \,,
\end{align}
\end{subequations}
where $n \equiv (\ell-1)(\ell+2)/2$. The initial metric perturbation functions can be obtained from linear integro-differential operators acting on the Cunningham-Price-Moncrief and Zerilli functions. For our purposes we only need expressions for the following quantities:
\begin{subequations}
\begin{align}
h_0(x) &= (x+1)^2 \int^x \ud u\,\frac{\Psi_\text{CPM}(u)}{(u+1)^3},\label{h0fromRW}\\
H(x) &= \frac{2 n (x+1) (2 n (x+1) (2 n (x+1)+2 x+5)+9)+9}{2 (x+1)^2 (2 n (x+1)+3)^2}\,\Psi_\text{Z}(x)\nonumber\\
&\phantom{=}+\frac{2 n \left(2 x^2+x-1\right)-3}{2 (x+1) (2 n (x+1)+3)}\,\Psi'_\text{Z}(x)\,.\label{HfromZ}
\end{align}
\end{subequations}

Alternatively, all the physical results about static perturbations of a Schwarzschild black hole can be derived from the static Bardeen-Press equation~\cite{BaPr.73} (i.e.\ Eq.~\eqref{Teuk} with $a=0$). Following Chandrasekhar~\cite{Ch2.75}, we give here the explicit relations between the solutions to the Cunningham-Price-Moncrief and Zerilli equations and the Bardeen-Press equation:
\begin{subequations}
    \begin{align}
        \Psi_\text{CPM}(x) &= \frac{(x+1)^2 (1 + (n-1) x)}{2 n (n+1)}\,R_{\ell m}(x) - \frac{x (x+1)^2 (2 x-1)}{4 n (n+1)}\,R'_{\ell m}(x)\,,\label{RWfromBP}\\
        \Psi_\text{Z}(x) &= \frac{(x+1) (n (x+1) \left[x (2 n (x+1)-2 x+3)+2\right]+6 x+3)}{2n (n+1) [2n (x+1) + 3]}\,R_{\ell m}(x)\nonumber\\ &\phantom{=} - \frac{x (x+1)^2 \left(2 n \left(2 x^2+x-1\right)-3\right)}{4n (n+1) [2n (x+1) + 3]}\,R'_{\ell m}(x)\,.\label{ZefromBP}
    \end{align}
\end{subequations}

For the even sector, substituting for Eq.~\eqref{ZefromBP} into \eqref{HfromZ} and making use of the relationship $(1-x^2) {P_\ell^2}'(x) = (\ell+1) x P_\ell^2(x)-(\ell-1) P_{\ell+1}^2(x)$ and the following recurrence relation for the associated Legendre functions $P_{\ell}^2$ (which are valid also for $Q_{\ell}^2$),
\beq\label{recurrenceP}
\ell P_{\ell+2}^2(2x+1) - (2 \ell+3) (2 x+1) P_{\ell+1}^2(2 x+1) + (\ell+3) P_{\ell}^2(2 x+1) = 0\,,
\eeq
it can be verified that Eq.~\eqref{sol_NS} is mapped to
\beq
H(x) = a_P P_{\ell}^2(2 x+1) + a_Q Q_{\ell}^2(2 x+1)\,,
\eeq
where $a_P\equiv a_\ell/2$ and $a_Q\equiv b_\ell/2$. This result is in agreement with Eq.~(36) of Ref.~\cite{DaNa2.09}.\footnote{Notice that in Ref.~\cite{DaNa2.09} the authors define $x=r/M-1$, whereas we are using $x=r/(2M)-1$.
Therefore, in order to compare our results with theirs, we need first to perform the transformation $x\to(x-1)/2$.}
It is clear that the condition \eqref{b_l} is equivalent to $a_Q=0$.

For the odd sector, the Einstein field equations imply the following differential equation for $h_0$:
\beq
h_0'' - \frac{\ell(\ell+1)(x+1) - 2}{x (x+1)^2}\,h_0 = 0\,.
\eeq
Its general solution is
\begin{align}\label{solh0}
    h_0(x) &= b_P\,x(x+1)^2\,F\left(-\ell+2,\ell+3,2;-x\right) \nonumber \\ &+ b_Q\,(x+1)^{-\ell}\,F\bigl(\ell-1, \ell+2, 2\ell+2; (x+1)^{-1}\bigr) \, ,
\end{align}
where $b_P$ and $b_Q$ are integration constants. The first term in Eq.~\eqref{solh0} is in agreement with Eq.~(38) in Ref.~\cite{ReWh.57} and Eq.~(55) in Ref.~\cite{DaNa2.09} when $\ell=2$; the second term in Eq.~\eqref{solh0} is in agreement (up to numerical factors) with Eq.~(56) in Ref.~\cite{DaNa2.09} when $\ell=2$.

We now show that the Cunningham-Price-Moncrief function computed with Eqs.~\eqref{RWfromBP} and \eqref{sol_NS} is proportional to that obtained by inverting Eq.~\eqref{h0fromRW} together with \eqref{solh0}. First, we need the hypergeometric representation of the associated Legendre functions,
\begin{subequations}
    \begin{align}
        P_\ell^2(1+2x) &= \frac{\Gamma(\ell+3)}{2\Gamma(\ell-1)} \, x (x+1) \, F(\ell+3,-\ell+2,3;-x)\,,\label{PfromF21}\\
        Q_\ell^2(1+2x) &= \frac{\Gamma(\ell+1)\Gamma(\ell+3)}{2\Gamma(2\ell+2)} \, \frac{x}{(x+1)^{\ell+2}} \, F\bigl(\ell+3,\ell+1,2\ell+2;(x+1)^{-1}\bigr) \,,\label{QfromF21}
    \end{align}
\end{subequations}
where Eq.~\eqref{QfromF21} has been obtained from Eq.~(14.3.7) in Ref.~\cite{NIST:DLMF}, together with a quadratic transformation (see Eq.~(15.8.13) in Ref.~\cite{NIST:DLMF}) and the duplication formula $\Gamma(z) \Gamma(z+1/2) = \sqrt{\pi}\,2^{1-2z}\Gamma(2z)$. Using Eqs.~\eqref{2F1_ids} below, it can be verified that the term proportional to $a_\ell$ in Eq.~\eqref{sol_NS} becomes mapped, via \eqref{RWfromBP}, to the term proportional to $b_P$ in Eq.~\eqref{solh0} with $a_\ell=b_P$;
similarly, the term proportional to $b_\ell$ is mapped to the term proportional to $b_Q$ with $b_\ell = \beta_\ell\,b_Q$, where
\beq
\beta_\ell \equiv -\frac{2(\ell-1)\ell(\ell+2)\Gamma(2\ell+2)}{\Gamma(\ell+1)\Gamma(\ell+3)}\,.
\eeq
We conclude that the condition \eqref{b_l} is also equivalent to $b_Q=0$.

We finish this appendix with some identities involving hypergeometric functions:
\begin{subequations}\label{2F1_ids}
\begin{align}
z \frac{\ud F}{\ud z} = (c-1)\left(F(c-)-F\right),\label{2F1_ids_a}\\
(c-1) c (1-z) F(c-) - c \left[z (a+b-2 c+1)+c-1\right] F = z (a-c) (b-c) F(c+)\,,\label{2F1_ids_b}\\
(a-c) F(a-)+\left[z (a-b)-2 a+c\right] F = a (z-1) F(a+)\,,\label{2F1_ids_c}\\
(a-c) F(a-) - (a+b-c) F = b (z-1) F(b+)\,,\label{2F1_ids_d}\\
\left[z (c-a)-b\right] F(b+,c+) + (b-c) F(c+) = c (z-1) F(b+)\,,\label{2F1_ids_e}\\
z (c-a) (b-c) F(c+) + c \left[z (c-a)-b\right] F = b c (z-1) F(b+)\,,\label{2F1_ids_f}
\end{align}
\end{subequations}
where we use the shortcuts $F\equiv F(a,b,c;z)$, $F(a\pm)\equiv F(a\pm1,b,c;z)$ and so on.

\section{Tidal field, linear response and physical solution}\label{app:continuity}

The crucial expression that we use to split the radial factor \eqref{sol_Kerr} of the full perturbed curvature scalar into its tidal and response parts \eqref{tidal-resp} is taken from Eq.~(15.8.2) in Ref.~\cite{NIST:DLMF}, namely
\begin{align}\label{Eq15.8.2}
    \frac{\sin{(\pi(b-a))}}{\pi} \, \mathbf{F}(a,b,c;-x) &= \frac{x^{-a}}{\Gamma(b)\Gamma(c-a)} \, \mathbf{F}(a,a-c+1,a-b+1;-1/x) \nonumber \\ &- \frac{x^{-b}}{\Gamma(a)\Gamma(c-b)} \, \mathbf{F}(b,b-c+1,b-a+1;-1/x) \, ,
\end{align}
where we recall that the regularized hypergeometric function is defined by the Gauss series
\beq\label{Gauss}
    \mathbf{F}(a,b,c;z) = \sum_{k=0}^{\infty}\frac{(a)_k (b)_k}{\Gamma(c+k)}\frac{z^k}{k!} \, ,
\eeq
the Pochhammer symbol being defined as $(a)_k \equiv \Gamma(a+k)/\Gamma(a) = a(a+1)(a+2)\cdots(a+k-1)$. As emphasized in Ref.~\cite{Olv}, the formulas \eqref{Eq15.8.2} and \eqref{Gauss} are valid for any $(a,b,c) \in \mathbb{C}^3$. We are interested in applying the identity \eqref{Eq15.8.2} in the particular case where $a=-\ell-2$, $b=\ell-1$ and $c \!=\! -1+2\ui m \gamma$, namely
\begin{align}\label{Eq15.8.2bis}
    &\frac{\sin{(\pi(2\ell+1))}}{\pi} \, \mathbf{F}(-\ell-2,\ell-1,-1+2\ui m\gamma;-x) \nonumber \\
    = \; &\frac{x^{\ell+2}}{\Gamma(\ell-1)\Gamma(\ell+1+2\ui m \gamma)} \, \mathbf{F}(-\ell-2,-\ell-2\ui m\gamma,-2\ell;-1/x) \nonumber \\
    - \, &\frac{x^{-\ell+1}}{\Gamma(-\ell-2)\Gamma(-\ell+2\ui m\gamma)} \, \mathbf{F}(\ell-1,\ell+1-2\ui m\gamma,2\ell+2;-1/x) \, .
\end{align}

We shall perform a local analysis in a neighbourhood of an integer value for the multipole index $\ell$. Hence we write $\ell = \ell_0 + \varepsilon$, where $\ell_0 \in \mathbb{L}$ and $0 < \varepsilon \ll 1$. It can easily be checked that all three terms in \eqref{Eq15.8.2bis} vanish like $\varepsilon$ as $\ell \to \ell_0$. In particular, Euler's reflection formula implies
\beq
    \frac{\sin{(\pi(2\ell+1))}}{\pi} = \frac{1}{\Gamma(2\ell+1)\Gamma(-2\ell)} = -2\varepsilon + O(\varepsilon^2) \, ,
\eeq
where the last equality follows from the Laurent series expansion of Euler's Gamma function near one of its poles (i.e., any non-positive integer). Similarly, the second term in the right-hand side of \eqref{Eq15.8.2bis} vanishes like
\beq
    \frac{1}{\Gamma(-\ell-2)} = (-)^{\ell_0+1} \, \Gamma(\ell_0+3) \, \varepsilon +  O(\varepsilon^2) \, ,
\eeq
while the associated regularized hypergeometric function has a nonzero, finite limit as $\varepsilon \to 0$. The behavior as $\varepsilon \to 0$ of the regularized hypergeometric function in the first term on the right-hand side of \eqref{Eq15.8.2bis} is controlled by the coefficients of the Gauss series \eqref{Gauss}, and more precisely by
\beq\label{coeffs}
    \frac{(-\ell-2)_k}{\Gamma(-2\ell+k)} \sim
    \begin{cases}
    (-)^{\ell_0+1} \, \frac{\Gamma\left(\ell_0+3\right)\Gamma\left(k-\ell_0-2\right)}{\Gamma\left(k-2\ell_0\right)} \, \varepsilon & \mbox{if } k \geqslant 2\ell_0+1 \, , \\
    -2\frac{\Gamma\left(\ell_0+3\right)\Gamma\left(2\ell_0+1-k\right)}{\Gamma\left(\ell_0+3-k\right)} \, \varepsilon & \mbox{if } 0 \leqslant k \leqslant \ell_0 + 2  \, , \\
    O(\varepsilon^2) & \mbox{otherwise} \, .
    \end{cases}
\eeq

Given that the simple zeros of each term in Eq.~\eqref{Eq15.8.2bis} are under control, we may multiply this identity by $\Gamma(2\ell+1) \Gamma(-2\ell) \sim 1/(-2\varepsilon)$, to obtain the decomposition
\begin{align}\label{Eq15.8.2ter}
    &\mathbf{F}(-\ell-2,\ell-1,-1+2\ui m\gamma;-x) \nonumber \\
    = \; &\frac{\Gamma(2\ell+1)}{\Gamma(\ell-1)\Gamma(\ell+1+2\ui m \gamma)} \; x^{\ell+2} \, F(-\ell-2,-\ell-2\ui m\gamma,-2\ell;-1/x) \nonumber \\
    - \, &\frac{\Gamma(2\ell+1) \Gamma(-2\ell)}{\Gamma(-\ell-2)\Gamma(-\ell+2\ui m\gamma)\Gamma(2\ell+2)} \; x^{-\ell+1} \, F(\ell-1,\ell+1-2\ui m\gamma,2\ell+2;-1/x) \, ,
\end{align}
where it should be understood that the hypergeometric function $F(a,b,c;z) = \Gamma(c) \, \mathbf{F}(a,b,c;z)$ in the first term on the right-hand side of \eqref{Eq15.8.2ter} is defined by taking the limit as $\varepsilon \to 0$ \textit{simultaneously} in its first and third arguments, as per \eqref{coeffs}. This coincides with the interpretation (15.2.6) in Ref.~\cite{NIST:DLMF}.\footnote{As opposed to using (15.2.5)~\cite{NIST:DLMF}, which would instead just yield a polynomial in $x$ (via (15.2.4)~\cite{NIST:DLMF}) for $x^{\ell+2}$ times the hypergeometric function  in the first term on the right-hand side of \eqref{Eq15.8.2ter}. Furthermore, it can be checked that one must use the simultaneous limit of (15.2.6)~\cite{NIST:DLMF} (and not (15.2.5)~\cite{NIST:DLMF}) so that (15.8.2)~\cite{NIST:DLMF} (which is the equivalent of \eqref{Eq15.8.2} for the unregularized hypergeometric function)
continuously goes to (15.8.6)~\cite{NIST:DLMF} when using our values of $a$, $b$ and $c$ given above and taking $\ell\in\mathbb{L}$.} As can be seen from the Gauss series \eqref{Gauss} and \eqref{coeffs}, such a hypergeometric function times $x^{\ell+2}$ then contains a polynomial of order $\ell+2$ in $x$ plus an infinite series in $1/x$.

The regularized hypergeometric function in the left-hand side of \eqref{Eq15.8.2ter} coincides, up to an overall constant factor, with the physical solution \eqref{sol_Kerr} times $x^2(1+x)^2$. Up to that same factor, the first term in the right-hand side of \eqref{Eq15.8.2ter} coincides with the tidal solution \eqref{R_tidal}, while the second term is proportional to the response solution \eqref{R_resp}. Hence the identity \eqref{Eq15.8.2ter} is equivalent to the decomposition \eqref{eq:tidal,resp comb} with the constant \eqref{kappa}.

For convenience, let us denote by $T_a$ and $T_b$ the first and second terms, respectively, on the right-hand side of the decomposition \eqref{Eq15.8.2ter}. As mentioned above, the term $T_a$ contains a polynomial in $x$ plus an infinite series in $1/x$. We shall now prove that, in the limit where $\ell \in \mathbb{L}$, this series exactly cancels out with $T_b$, leaving a polynomial that coincides with the left-hand side of \eqref{Eq15.8.2ter}. Taking the limit as $\ell\to\ell_0$ we have in terms of the Gauss series \eqref{Gauss}:
\begin{subequations}
    \begin{align}
        T_a &= \lim_{\ell\to\ell_0} \, \frac{\Gamma(2\ell+1)}{\Gamma(\ell-1)\Gamma(\ell+1+2\ui m \gamma)} \, x^{\ell+2} \sum_{k=0}^\infty \frac{(-\ell-2)_k (-\ell-2\ui m\gamma)_k}{(-2\ell)_k}\frac{(-1/x)^k}{k!}\,,\label{T_a} \\
        T_b &= - \lim_{\ell\to\ell_0} \, \frac{\Gamma(2\ell+1)\Gamma(-2\ell)}{\Gamma(2\ell+2)\Gamma(-\ell-2)\Gamma(-\ell+2\ui m\gamma)} \, x^{-\ell+1} \sum_{k=0}^\infty \frac{(\ell-1)_k (\ell+1-2\ui m\gamma)_k}{(2\ell+2)_k}\frac{(-1/x)^k}{k!}\,.\label{T_b}
    \end{align}
\end{subequations}
According to \eqref{coeffs}, the sum in Eq.~\eqref{T_a} splits into two sums: one from $k=0$ to $k=\ell_0+2$, which corresponds to the polynomial part, i.e., to Eq.~(15.8.6)~\cite{NIST:DLMF};
and one from $k=2\ell_0+1$ to infinity, which is the infinite series in $1/x$, say
\beq\label{T_a^np}
    T_a^\text{np} \equiv \lim_{\ell\to\ell_0} \, \frac{\Gamma(2\ell+1)}{\Gamma(\ell-1)\Gamma(\ell+1+2\ui m \gamma)} \; x^{\ell+2} \! \sum_{k=2\ell_0+1}^\infty \frac{(-\ell-2)_k (-\ell-2\ui m\gamma)_k}{(-2\ell)_k}\frac{(-1/x)^k}{k!}\,.
\eeq
Indeed, for $k\in\{\ell_0+3,\ldots,2\ell_0\}$ the asymptotics \eqref{coeffs} implies $\lim_{\ell\to\ell_0} (-\ell-2)_k / (-2\ell)_k = 0$.

Now, by changing the index $k$ to $n \equiv k - 2\ell_0 - 1$ in the series in \eqref{T_a^np}, using the identity $(a)_{n+2\ell_0+1} \!=\! (a+2\ell_0+1)_n (a)_{2\ell_0+1}$, the relationship between the Pochhammer symbol and the Gamma function and $\Gamma(n+1)=n!$,
Eq.~\eqref{T_a^np} becomes
\beq\label{eq:Tanp}
    T_a^\text{np} = \lim_{\ell\to\ell_0} \,
    \frac{\Gamma(\ell+1-2\ui m\gamma)\Gamma(-\ell+2\ui m\gamma)}{\Gamma(\ell+1+2\ui m\gamma)\Gamma(-\ell-2\ui m\gamma)}\,T_b\,.
\eeq
Using Euler's reflection formula $\Gamma(z)\Gamma(1-z) = \pi/\sin(\pi z)$, it is easy to verify that the factor in front of $T_b$ in \eqref{eq:Tanp} is equal to $-1$, and hence $T_a^\text{np}+ T_b = 0$, as claimed above. Analogous manipulations on the polynomial part $T_a - T_a^\text{np}$ of the first term in the right-hand side of \eqref{Eq15.8.2ter} show that it coincides with the left-hand side of \eqref{Eq15.8.2ter} as $\ell \in \mathbb{L}$.

Similarly, the split \eqref{G=Gtidal+Gresp} of the radial factor \eqref{G} of the full Hertz potential into its tidal and response parts \eqref{Gtidal-Gresp} is taken by applying the identity \eqref{Eq15.8.2} to the particular case where $a=-\ell+2$, $b=\ell+3$ and $c=3+2\ui m\gamma$. Following an analysis analogous to that detailed above for the curvature scalar, we readily obtain the decomposition
\begin{align}\label{Eq15.8.2quad}
    &\mathbf{F}(-\ell+2,\ell+3,3+2\ui m\gamma;-x) \nonumber \\
    = \; &\frac{\Gamma(2\ell+1)}{\Gamma(\ell+3)\Gamma(\ell+1+2\ui m \gamma)} \; x^{\ell-2} \, F(-\ell+2,-\ell-2\ui m\gamma,-2\ell;-1/x) \nonumber \\
    - \, &\frac{\Gamma(2\ell+1) \Gamma(-2\ell)}{\Gamma(-\ell+2)\Gamma(-\ell+2\ui m\gamma)\Gamma(2\ell+2)} \; x^{-\ell-3} \, F(\ell+3,\ell+1-2\ui m\gamma,2\ell+2;-1/x) \, .
\end{align}
The regularized hypergeometric function in the left-hand side of \eqref{Eq15.8.2quad} coincides, up to an overall constant factor, with the physical solution \eqref{G} over $x^2(1+x)^2$. Up to that same factor, the first term in the right-hand side of \eqref{Eq15.8.2quad} coincides with the tidal solution \eqref{G_tidal}, while the second term is proportional to the response solution \eqref{G_resp}. Hence the identity \eqref{Eq15.8.2quad} is equivalent to the decomposition \eqref{G=Gtidal+Gresp} with the constant \eqref{kappa}.

\section{An alternative tidal/response split}\label{app:splits}

In this appendix, we discuss an alternative prescription to split the full physical solution \eqref{sol_Kerr} into tidal and response contributions, as used in~\cite{LaPo.15} to define and compute the TLNs of a slowly spinning compact object, and which does not rely on the analytic continuation of the multipolar number $\ell \in \mathbb{R}$ that we advocate in this work. 

As pointed out in Ref.~\cite{Gr.18} and briefly discussed in Sec.~\ref{subsec:split} above, for a given harmonic index $\ell \in \mathbb{L}$, the choice of a radial coordinate has an influence on the numerical value of the TLNs associated with the compact body's linear response to the $2^\ell$-polar tidal perturbation. A calculation for general $\ell$, as we do in this work, however, resolves this coordinate ambiguity. Now, given a choice of radial coordinate $r$, there is a unique solution $R^\text{decay}_{\ell m}(r)$ of the static radial Teukolsky equation that has the decaying behavior $\sim r^{-(\ell+3)}$ at large radii, and which is then identified as the linear response contribution of a Kerr black hole. For instance, by using advanced Kerr coordinates $(v,r,\theta,\phi)$, the static radial Teukolsky equation is given by Eq.~\eqref{Teuk} and the associated decaying solution by Eq.~\eqref{R_resp}. The full physical solution can then be written as
\beq\label{split1}
    R_{\ell m}(r) = R^\text{grow}_{\ell m}(r) + 2\kappa_{\ell m} R^\text{decay}_{\ell m}(r) \, ,
\eeq
where $R^\text{grow}_{\ell m}(r)$ is a growing solution that behaves as $\sim r^{\ell-2}$ at large radii and is associated with the perturbing tidal field, while $\kappa_{\ell m}$ is a constant that is closely related to the Kerr black hole TLNs. Clearly, the split \eqref{split1} between a growing/dominant solution $R^\text{grow}_{\ell m}(r)$ and the decaying/subdominant solution $R^\text{decay}_{\ell m}(r)$ is not unique, because one can always shift a multiple of the decaying solution to a growing solution, yielding an alternative growing/decaying (or tidal/response) decomposition:
\beq\label{split2}
    R_{\ell m}(r) = \tilde{R}^\text{grow}_{\ell m}(r) + 2\tilde{\kappa}_{\ell m} R^\text{decay}_{\ell m}(r) \, .
\eeq
Here, we introduced the alternative growing solution
\beq
    \tilde{R}^\text{grow}_{\ell m}(r) \equiv R^\text{grow}_{\ell m}(r) + 2\alpha_{\ell m} R^\text{decay}_{\ell m}(r)
\eeq
for some nonzero constant $\alpha_{\ell m}$, so that the coefficient $\tilde{\kappa}_{\ell m} \equiv \kappa_{\ell m} - \alpha_{\ell m}$ that multiplies the decaying solution in the decomposition \eqref{split2} has been shifted, thus ultimately affecting the numerical values of the black hole TLNs.

Without imposing any further condition, the split of the physical solution into tidal and response solutions appears ambiguous. To lift this ambiguity, Ref.~\cite{LaPo.15} (see also~\cite{Gr.18}) adopted the prescription of imposing smoothness of the growing solution everywhere, including on the event horizon. In the problem of interest of a weakly tidally perturbed Kerr black hole, the growing solution \eqref{R_tidal} is not smooth on the event horizon, and the linear combination \eqref{eq:tidal,resp comb} with \eqref{kappa_phys} is the only (up to a normalization) smooth, growing solution. Under the prescription of Ref.~\cite{LaPo.15}, the physical solution \eqref{eq:tidal,resp comb} would then be identified as the tidal contribution, yielding zero response contribution and thus vanishing Kerr TLNs.

However, we emphasize that the tidal and response solutions of the linearized Einstein equation need not be separately smooth on the event horizon, because neither of them bears a physical meaning by itself, in the sense that a local observer can only probe the \textit{full} field $R_{\ell m}$ in \eqref{eq:tidal,resp comb}, which is the only field that needs to be---and is---smooth everywhere in a regular tetrad. Therefore, the prescription adopted in~\cite{LaPo.15,Gr.18} imposes a physical condition (namely, smoothness) on a nonphysical field (namely, the tidal field). Furthermore, such prescription yields a tidal/response split which, in the nonrelativistic limit, does not correctly reproduce the Newtonian tidal/response  split in Eq.~\eqref{psi0_Newt_spin}, as we observe when taking $\ell\in\mathbb{R}$. Finally, this prescription implies vanishing tidally-induced multipole moments for a Kerr black hole, which appears to be in contradiction with the tidal torquing calculation of~\cite{Po2.04}; see Sec.~\ref{sec:torquing} above.

\section{Calculation of the scalar twist perturbation}\label{app:twist}

In this appendix we detail the calculation of the perturbative part $\delta \omega$ of the scalar twist $\omega = \mathring{\omega} + \delta \omega$ that entered the computation of the mass-type and current-type quadrupole moments in Sec.~\ref{sec:multipoles}. Subtracting the contribution coming from the background scalar twist $\mathring{\omega}$, the first-order partial differential equations \eqref{eq:twist diff eqs} can be rewritten as
\begin{subequations}
    \begin{align}
        \partial_r \delta\omega &\doteq - \frac{\sqrt{2}}{r\sin\theta} \, \bigl[ \partial_{\pBL}\Re(h^\text{resp}_{n\bar{m}}) + \partial_{\theta}\left(\sin{\theta} \, \Im(h^\text{resp}_{n\bar{m}})\right) \bigr] \, ,\label{eq:twist diff eqs r} \\
        \partial_\theta \delta\omega &\doteq - \frac{1}{\sin{\theta}} \, \partial_{\pBL} h^\text{resp}_{nn}
        + \sqrt{2} \, \bigl[ r f \partial_r \Im{(h^\text{resp}_{n\bar{m}})} + (1-4M/r) \, \Im(h^\text{resp}_{n\bar{m}}) \bigr]
        \, , 
        \label{eq:twist diff eqs theta}
        \\
        \partial_\pBL \delta\omega &\doteq \sin{\theta} \, \partial_{\theta} h^\text{resp}_{nn}
        + \sqrt{2} \sin{\theta} \, \bigl[ r f \partial_r \Re{(h^\text{resp}_{n\bar{m}})} + (1-4M/r) \, \Re(h^\text{resp}_{n\bar{m}}) \bigr] \, ,
        \label{eq:twist diff eqs phi}
    \end{align}
\end{subequations}
where we recall that $f = 1 - 2M/r$. Let us first integrate Eq.~\eqref{eq:twist diff eqs r}. To do so, we note that it can be rewritten as
\begin{equation}\label{drdeltaomega}
        \partial_r \delta\omega \doteq 
        \frac{\sqrt{2}}{r} \, \Im\left(\eth_{-1} h^\text{resp}_{n\bar{m}}\right) \doteq 2 \ad \sum_{\ell = 2}^\infty C_{\ell} \,  \mathcal{Q}'_\ell(r) \sum_{m = -\ell}^\ell  m \, \Re{\left[\bar{z}_{\ell m} \bar{Y}_{\ell m}(\theta,\pBL)\right]} \, ,
\end{equation}
where we used the expression \eqref{h_nbarm_resp} for $h^\text{resp}_{n\bar{m}}$, as well as the properties \eqref{eq:sphcal cc} and \eqref{eq:eth Ycc} with spin $s=-1$. We can now readily integrate Eq.~\eqref{drdeltaomega} as
\begin{equation}\label{domega,int-r}
		\delta\omega\doteq
		2 \ad \sum_{\ell = 2}^\infty C_{\ell} \,  \mathcal{Q}_\ell(r) \sum_{m = -\ell}^\ell  m \, \Re{\left[\bar{z}_{\ell m} \bar{Y}_{\ell m}(\theta,\pBL)\right]} + F_1(\theta,\pBL) \, ,
\end{equation}
for some function $F_1$ of the angular coordinates $(\theta,\varphi)$.

The integration of Eq.~\eqref{eq:twist diff eqs phi} with respect to $\pBL$ is trivial: the result is the same operator that is acting on $h^\text{resp}_{nn}$ and $h^\text{resp}_{n\bar{m}}$, but instead acting on the $\pBL$-integrals $H_{nn}$ and $H_{n\bar{m}}$ of, respectively, $h^\text{resp}_{nn}$ and $h^\text{resp}_{n\bar{m}}$, plus some function $F_2(\theta,r)$. That is,
\begin{equation}\label{domega,int-phi}
       \delta\omega \doteq \sin{\theta} \, \partial_{\theta} H_{nn} + \sqrt{2} \sin{\theta} \, \bigl[ r f \partial_r \Re{(H_{n\bar{m}})} + (1-4M/r) \, \Re(H_{n\bar{m}}) \bigr] + F_2(\theta,r) \, ,
\end{equation}
where, from Eqs.~\eqref{h_nn_resp} and \eqref{h_nbarm_resp},
\begin{subequations}\label{soubis}
    \begin{align}
        H_{nn} &\equiv \int^\pBL \ud\pBL' \, h^\text{resp}_{nn}(\theta,\pBL')\doteq \ad \sum_{\ell = 2}^\infty C_{\ell} \, \mathcal{Q}_\ell(r) \left( \sum_{m = -\ell}^\ell \, \bar{z}_{\ell m} \bar{Y}_{\ell m}(\theta,\pBL) + \text{c.c.} \right) \, , \\
        H_{n\bar{m}} &\equiv \int^\pBL \ud\pBL' \, h^\text{resp}_{n\bar{m}}(\theta,\pBL') \doteq \sqrt{2} \ad \sum_{\ell = 2}^\infty C_\ell \, \frac{r\mathcal{Q}'_\ell(r)}{\sqrt{\ell(\ell+1)}} \, \sum_{m = -\ell}^\ell \bar{z}_{\ell m} \, {}_{1}\bar{Y}_{\ell m}(\theta,\pBL)  \, .
    \end{align}
\end{subequations}
Substituting for Eqs.~\eqref{soubis} into \eqref{domega,int-phi} gives the twist perturbation explicitly as a mode-sum that reads
\begin{align}\label{domega,int-phi mode}
   \delta\omega \doteq 2\ad \sin{\theta} \sum_{\ell = 2}^\infty C_{\ell} \, \Biggl( &\frac{r f \partial_r \bigl(r\mathcal{Q}'_\ell(r)\bigr) + (r-4M) \mathcal{Q}'_\ell(r)}{\sqrt{\ell(\ell+1)}} \sum_{m = -\ell}^\ell \Re{\left[z_{\ell m} \, {}_1Y_{\ell m}(\theta,\pBL)\right]} \nonumber \\ &+ \mathcal{Q}_\ell(r) \sum_{m = -\ell}^\ell \Re{\left[z_{\ell m} \, \partial_{\theta}Y_{\ell m}(\theta,\pBL)\right]} \Biggr) + F_2(\theta,r)  \, .
\end{align}
Now, using the Legendre equation, it can be shown that the radial factor in the first term in Eq.~\eqref{domega,int-phi mode} can in fact be simplified as
\begin{equation}\label{eq:ODE Ql}
       r f \partial_r \bigl(r\mathcal{Q}'_\ell(r)\bigr) + (r-4M) \mathcal{Q}'_\ell(r) =
\ell(\ell+1)\mathcal{Q}_\ell(r) \, .
\end{equation}
Using this equality, together with Eq.~\eqref{eq:eth Y}, the twist perturbation \eqref{domega,int-phi mode} simplifies into
\begin{equation}\label{domega,int-phi like r}
	\delta\omega\doteq 2 \ad \sum_{\ell = 2}^\infty C_\ell \, \mathcal{Q}_\ell(r) \sum_{m = -\ell}^\ell  m \, \Re{\left[ z_{\ell m} Y_{\ell m}(\theta,\pBL) \right]} + F_2(\theta,r) \, .
\end{equation}
By comparing the formula \eqref{domega,int-phi like r} with Eq.~\eqref{domega,int-r} it follows that $F_1(\theta,\pBL) = F_2(\theta,r) \equiv F(\theta)$.

Finally, we impose the remaining Eq.~\eqref{eq:twist diff eqs theta}. First, differentiating Eq.~\eqref{domega,int-r} or \eqref{domega,int-phi like r} with respect to $\theta$, we obtain
\begin{equation}\label{dthetadomega}
	\partial_{\theta}\delta\omega \doteq 2 \ad \sum_{\ell = 2}^\infty C_\ell \, \mathcal{Q}_\ell(r) \sum_{m = -\ell}^\ell  m \, \Re{\left[z_{\ell m} \, \partial_{\theta}Y_{\ell m}(\theta,\pBL)\right]} + F'(\theta)  \, .
\end{equation}
In its turn, Eq.~\eqref{eq:twist diff eqs theta} can be written,  using \eqref{metric_resp}, in terms of mode sums as
\begin{align}\label{dthetadomega mid}
	\partial_{\theta}\delta\omega \doteq 2\ad \sum_{\ell = 2}^\infty C_{\ell} \,\Biggl( &- \frac{r f \partial_r \bigl(r\mathcal{Q}'_\ell(r)\bigr) + (r-4M) \mathcal{Q}'_\ell(r)}{\sqrt{\ell(\ell+1)}} \sum_{m = -\ell}^\ell m \, \Re{\left[ z_{\ell m} \, {}_1Y_{\ell m}(\theta,\pBL) \right]} \nonumber \\ &+ \frac{\mathcal{Q}_\ell(r)}{\sin\theta} \sum_{m = -\ell}^\ell  m^2 \, \Re{\left[z_{\ell m} Y_{\ell m}(\theta,\pBL)\right]} \Biggr)  \, .
\end{align}
Using again Eqs.~\eqref{eq:ODE Ql} and \eqref{eq:eth Y}, we obtain that $\partial_{\theta}\delta\omega$ in \eqref{dthetadomega mid} is equal to the first term on the right hand side of \eqref{dthetadomega}. It thus follows that $F(\theta) = F = \const$. Imposing $\delta\omega \to 0$ as $r \to \infty$ on \eqref{domega,int-phi like r} then yields $F=0$, where we used that $\lim_{r\to\infty}\mathcal{Q}_\ell(r)=0$. Thus, the final expression for the twist perturbation is
\begin{equation}
	\delta\omega \doteq 2 \ad \sum_{\ell = 2}^\infty C_{\ell} \, \mathcal{Q}_\ell(r) \sum_{m = -\ell}^\ell  m \, \Re{\left[ z_{\ell m} Y_{\ell m}(\theta,\pBL) \right]} \, .
\end{equation}

\bibliography{}

\begin{thebibliography}{108}%
\makeatletter
\providecommand \@ifxundefined [1]{%
 \@ifx{#1\undefined}
}%
\providecommand \@ifnum [1]{%
 \ifnum #1\expandafter \@firstoftwo
 \else \expandafter \@secondoftwo
 \fi
}%
\providecommand \@ifx [1]{%
 \ifx #1\expandafter \@firstoftwo
 \else \expandafter \@secondoftwo
 \fi
}%
\providecommand \natexlab [1]{#1}%
\providecommand \enquote  [1]{``#1''}%
\providecommand \bibnamefont  [1]{#1}%
\providecommand \bibfnamefont [1]{#1}%
\providecommand \citenamefont [1]{#1}%
\providecommand \href@noop [0]{\@secondoftwo}%
\providecommand \href [0]{\begingroup \@sanitize@url \@href}%
\providecommand \@href[1]{\@@startlink{#1}\@@href}%
\providecommand \@@href[1]{\endgroup#1\@@endlink}%
\providecommand \@sanitize@url [0]{\catcode `\\12\catcode `\$12\catcode
  `\&12\catcode `\#12\catcode `\^12\catcode `\_12\catcode `\%12\relax}%
\providecommand \@@startlink[1]{}%
\providecommand \@@endlink[0]{}%
\providecommand \url  [0]{\begingroup\@sanitize@url \@url }%
\providecommand \@url [1]{\endgroup\@href {#1}{\urlprefix }}%
\providecommand \urlprefix  [0]{URL }%
\providecommand \Eprint [0]{\href }%
\providecommand \doibase [0]{http://dx.doi.org/}%
\providecommand \selectlanguage [0]{\@gobble}%
\providecommand \bibinfo  [0]{\@secondoftwo}%
\providecommand \bibfield  [0]{\@secondoftwo}%
\providecommand \translation [1]{[#1]}%
\providecommand \BibitemOpen [0]{}%
\providecommand \bibitemStop [0]{}%
\providecommand \bibitemNoStop [0]{.\EOS\space}%
\providecommand \EOS [0]{\spacefactor3000\relax}%
\providecommand \BibitemShut  [1]{\csname bibitem#1\endcsname}%
\let\auto@bib@innerbib\@empty
\bibitem [{\citenamefont {Kerr}(1963)}]{Ke.63}%
  \BibitemOpen
  \bibfield  {author} {\bibinfo {author} {\bibfnamefont {R.~P.}\ \bibnamefont
  {Kerr}},\ }\href@noop {} {\bibfield  {journal} {\bibinfo  {journal} {Phys.
  Rev. Lett.}\ }\textbf {\bibinfo {volume} {11}},\ \bibinfo {pages} {237}
  (\bibinfo {year} {1963})}\BibitemShut {NoStop}%
\bibitem [{\citenamefont {Israel}(1967)}]{Is.67}%
  \BibitemOpen
  \bibfield  {author} {\bibinfo {author} {\bibfnamefont {W.}~\bibnamefont
  {Israel}},\ }\href@noop {} {\bibfield  {journal} {\bibinfo  {journal} {Phys.
  Rev.}\ }\textbf {\bibinfo {volume} {164}},\ \bibinfo {pages} {1776} (\bibinfo
  {year} {1967})}\BibitemShut {NoStop}%
\bibitem [{\citenamefont {Carter}(1971)}]{Ca.71}%
  \BibitemOpen
  \bibfield  {author} {\bibinfo {author} {\bibfnamefont {B.}~\bibnamefont
  {Carter}},\ }\href@noop {} {\bibfield  {journal} {\bibinfo  {journal} {Phys.
  Rev. Lett.}\ }\textbf {\bibinfo {volume} {26}},\ \bibinfo {pages} {331}
  (\bibinfo {year} {1971})}\BibitemShut {NoStop}%
\bibitem [{\citenamefont {Robinson}(1975)}]{Ro.75}%
  \BibitemOpen
  \bibfield  {author} {\bibinfo {author} {\bibfnamefont {D.}~\bibnamefont
  {Robinson}},\ }\href@noop {} {\bibfield  {journal} {\bibinfo  {journal}
  {Phys. Rev. Lett.}\ }\textbf {\bibinfo {volume} {34}},\ \bibinfo {pages}
  {905} (\bibinfo {year} {1975})}\BibitemShut {NoStop}%
\bibitem [{\citenamefont {Geroch}(1970)}]{Ge.70}%
  \BibitemOpen
  \bibfield  {author} {\bibinfo {author} {\bibfnamefont {R.~P.}\ \bibnamefont
  {Geroch}},\ }\href@noop {} {\bibfield  {journal} {\bibinfo  {journal} {J.
  Math. Phys.}\ }\textbf {\bibinfo {volume} {11}},\ \bibinfo {pages} {2580}
  (\bibinfo {year} {1970})}\BibitemShut {NoStop}%
\bibitem [{\citenamefont {Hansen}(1974)}]{Ha.74}%
  \BibitemOpen
  \bibfield  {author} {\bibinfo {author} {\bibfnamefont {R.~O.}\ \bibnamefont
  {Hansen}},\ }\href@noop {} {\bibfield  {journal} {\bibinfo  {journal} {J.
  Math. Phys.}\ }\textbf {\bibinfo {volume} {15}},\ \bibinfo {pages} {46}
  (\bibinfo {year} {1974})}\BibitemShut {NoStop}%
\bibitem [{\citenamefont {Love}(1911)}]{Lov}%
  \BibitemOpen
  \bibfield  {author} {\bibinfo {author} {\bibfnamefont {A.~E.~H.}\
  \bibnamefont {Love}},\ }\href@noop {} {\emph {\bibinfo {title} {Some problems
  of geodynamics}}}\ (\bibinfo  {publisher} {Cornell University Library},\
  \bibinfo {address} {Ithaca},\ \bibinfo {year} {1911})\BibitemShut {NoStop}%
\bibitem [{\citenamefont {Poisson}\ and\ \citenamefont {Will}(2014)}]{PoWi}%
  \BibitemOpen
  \bibfield  {author} {\bibinfo {author} {\bibfnamefont {E.}~\bibnamefont
  {Poisson}}\ and\ \bibinfo {author} {\bibfnamefont {C.~M.}\ \bibnamefont
  {Will}},\ }\href@noop {} {\emph {\bibinfo {title} {Gravity: Newtonian,
  post-{N}ewtonian, relativistic}}}\ (\bibinfo  {publisher} {Cambridge
  University Press},\ \bibinfo {address} {Cambridge},\ \bibinfo {year}
  {2014})\BibitemShut {NoStop}%
\bibitem [{\citenamefont {Flanagan}\ and\ \citenamefont
  {Hinderer}(2008)}]{FlHi.08}%
  \BibitemOpen
  \bibfield  {author} {\bibinfo {author} {\bibfnamefont {{\'E}.~{\'E}.}\
  \bibnamefont {Flanagan}}\ and\ \bibinfo {author} {\bibfnamefont
  {T.}~\bibnamefont {Hinderer}},\ }\href@noop {} {\bibfield  {journal}
  {\bibinfo  {journal} {Phys. Rev. D}\ }\textbf {\bibinfo {volume} {77}},\
  \bibinfo {pages} {021502(R)} (\bibinfo {year} {2008})},\ \Eprint
  {http://arxiv.org/abs/0709.1915} {arXiv:0709.1915 [astro-ph]} \BibitemShut
  {NoStop}%
\bibitem [{\citenamefont {Hinderer}(2008)}]{Hi.08}%
  \BibitemOpen
  \bibfield  {author} {\bibinfo {author} {\bibfnamefont {T.}~\bibnamefont
  {Hinderer}},\ }\href@noop {} {\bibfield  {journal} {\bibinfo  {journal}
  {Astrophys. J.}\ }\textbf {\bibinfo {volume} {677}},\ \bibinfo {pages} {1216}
  (\bibinfo {year} {2008})},\ \bibinfo {note} {\textit{{E}rratum:} Astrophys.
  J. \textbf{697}, 964 (2009)},\ \Eprint {http://arxiv.org/abs/0711.2420}
  {arXiv:0711.2420 [astro-ph]} \BibitemShut {NoStop}%
\bibitem [{\citenamefont {Hinderer}\ \emph {et~al.}(2010)\citenamefont
  {Hinderer}, \citenamefont {Lackey}, \citenamefont {Lang},\ and\ \citenamefont
  {Read}}]{Hi.al2.10}%
  \BibitemOpen
  \bibfield  {author} {\bibinfo {author} {\bibfnamefont {T.}~\bibnamefont
  {Hinderer}}, \bibinfo {author} {\bibfnamefont {B.~D.}\ \bibnamefont
  {Lackey}}, \bibinfo {author} {\bibfnamefont {R.~N.}\ \bibnamefont {Lang}}, \
  and\ \bibinfo {author} {\bibfnamefont {J.~S.}\ \bibnamefont {Read}},\
  }\href@noop {} {\bibfield  {journal} {\bibinfo  {journal} {Phys. Rev. D}\
  }\textbf {\bibinfo {volume} {81}},\ \bibinfo {pages} {123016} (\bibinfo
  {year} {2010})},\ \Eprint {http://arxiv.org/abs/0911.3535} {arXiv:0911.3535
  [astro-ph.HE]} \BibitemShut {NoStop}%
\bibitem [{\citenamefont {Cardoso}\ \emph {et~al.}(2017)\citenamefont
  {Cardoso}, \citenamefont {Franzin}, \citenamefont {Maselli}, \citenamefont
  {Pani},\ and\ \citenamefont {Raposo}}]{Ca.al.17}%
  \BibitemOpen
  \bibfield  {author} {\bibinfo {author} {\bibfnamefont {V.}~\bibnamefont
  {Cardoso}}, \bibinfo {author} {\bibfnamefont {E.}~\bibnamefont {Franzin}},
  \bibinfo {author} {\bibfnamefont {A.}~\bibnamefont {Maselli}}, \bibinfo
  {author} {\bibfnamefont {P.}~\bibnamefont {Pani}}, \ and\ \bibinfo {author}
  {\bibfnamefont {G.}~\bibnamefont {Raposo}},\ }\href@noop {} {\bibfield
  {journal} {\bibinfo  {journal} {Phys. Rev. D}\ }\textbf {\bibinfo {volume}
  {95}},\ \bibinfo {pages} {084014} (\bibinfo {year} {2017})},\ \Eprint
  {http://arxiv.org/abs/1701.01116} {arXiv:1701.01116 [gr-qc]} \BibitemShut
  {NoStop}%
\bibitem [{\citenamefont {{B.~P.~Abbott et al. (LIGO Scientific Collaboration
  and Virgo Collaboration)}}(2017)}]{Ab.al3.17}%
  \BibitemOpen
  \bibfield  {author} {\bibinfo {author} {\bibnamefont {{B.~P.~Abbott et al.
  (LIGO Scientific Collaboration and Virgo Collaboration)}}},\ }\href@noop {}
  {\bibfield  {journal} {\bibinfo  {journal} {Phys. Rev. Lett.}\ }\textbf
  {\bibinfo {volume} {119}},\ \bibinfo {pages} {161101} (\bibinfo {year}
  {2017})},\ \Eprint {http://arxiv.org/abs/1710.05832} {arXiv:1710.05832
  [gr-qc]} \BibitemShut {NoStop}%
\bibitem [{\citenamefont {Most}\ \emph {et~al.}(2018)\citenamefont {Most},
  \citenamefont {Weih}, \citenamefont {Rezzolla},\ and\ \citenamefont
  {{Schaffner-Bielich}}}]{Mo.al.18}%
  \BibitemOpen
  \bibfield  {author} {\bibinfo {author} {\bibfnamefont {E.~R.}\ \bibnamefont
  {Most}}, \bibinfo {author} {\bibfnamefont {L.~R.}\ \bibnamefont {Weih}},
  \bibinfo {author} {\bibfnamefont {L.}~\bibnamefont {Rezzolla}}, \ and\
  \bibinfo {author} {\bibfnamefont {J.}~\bibnamefont {{Schaffner-Bielich}}},\
  }\href@noop {} {\bibfield  {journal} {\bibinfo  {journal} {Phys. Rev. Lett.}\
  }\textbf {\bibinfo {volume} {120}},\ \bibinfo {pages} {261103} (\bibinfo
  {year} {2018})},\ \Eprint {http://arxiv.org/abs/1803.00549} {arXiv:1803.00549
  [gr-qc]} \BibitemShut {NoStop}%
\bibitem [{\citenamefont {De}\ \emph {et~al.}(2018)\citenamefont {De} \emph
  {et~al.}}]{De.al.18}%
  \BibitemOpen
  \bibfield  {author} {\bibinfo {author} {\bibfnamefont {S.}~\bibnamefont {De}}
  \emph {et~al.},\ }\href@noop {} {\bibfield  {journal} {\bibinfo  {journal}
  {Phys. Rev. Lett.}\ }\textbf {\bibinfo {volume} {121}},\ \bibinfo {pages}
  {091102} (\bibinfo {year} {2018})},\ \Eprint
  {http://arxiv.org/abs/1804.08583} {arXiv:1804.08583 [astro-ph.HE]}
  \BibitemShut {NoStop}%
\bibitem [{\citenamefont {{B.~P.~Abbott et al. (LIGO Scientific Collaboration
  and Virgo Collaboration)}}(2018)}]{Ab.al2.18}%
  \BibitemOpen
  \bibfield  {author} {\bibinfo {author} {\bibnamefont {{B.~P.~Abbott et al.
  (LIGO Scientific Collaboration and Virgo Collaboration)}}},\ }\href@noop {}
  {\bibfield  {journal} {\bibinfo  {journal} {Phys. Rev. Lett.}\ }\textbf
  {\bibinfo {volume} {121}},\ \bibinfo {pages} {161101} (\bibinfo {year}
  {2018})},\ \Eprint {http://arxiv.org/abs/1805.11581} {arXiv:1805.11581
  [gr-qc]} \BibitemShut {NoStop}%
\bibitem [{\citenamefont {{B.~P.~Abbott et al. (LIGO Scientific Collaboration
  and Virgo Collaboration)}}(2020)}]{Ab.al.20}%
  \BibitemOpen
  \bibfield  {author} {\bibinfo {author} {\bibnamefont {{B.~P.~Abbott et al.
  (LIGO Scientific Collaboration and Virgo Collaboration)}}},\ }\href@noop {}
  {\bibfield  {journal} {\bibinfo  {journal} {Astrophys. J. Lett.}\ }\textbf
  {\bibinfo {volume} {892}},\ \bibinfo {pages} {L3} (\bibinfo {year} {2020})},\
  \Eprint {http://arxiv.org/abs/2001.01761} {arXiv:2001.01761 [astro-ph.HE]}
  \BibitemShut {NoStop}%
\bibitem [{\citenamefont {{Amaro-Seoane}}\ \emph {et~al.}(2017)\citenamefont
  {{Amaro-Seoane}} \emph {et~al.}}]{Am.al.17}%
  \BibitemOpen
  \bibfield  {author} {\bibinfo {author} {\bibfnamefont {P.}~\bibnamefont
  {{Amaro-Seoane}}} \emph {et~al.},\ }\href@noop {} {\enquote {\bibinfo {title}
  {Laser {I}nterferometer {S}pace {A}ntenna},}\ } (\bibinfo {year} {2017}),\
  \Eprint {http://arxiv.org/abs/1702.00786} {arXiv:1702.00786 [astro-ph.IM]}
  \BibitemShut {NoStop}%
\bibitem [{\citenamefont {Pani}\ and\ \citenamefont {Maselli}(2019)}]{PaMa.19}%
  \BibitemOpen
  \bibfield  {author} {\bibinfo {author} {\bibfnamefont {P.}~\bibnamefont
  {Pani}}\ and\ \bibinfo {author} {\bibfnamefont {A.}~\bibnamefont {Maselli}},\
  }\href@noop {} {\bibfield  {journal} {\bibinfo  {journal} {Int. J. Mod. Phys.
  D}\ }\textbf {\bibinfo {volume} {28}},\ \bibinfo {pages} {1944001} (\bibinfo
  {year} {2019})},\ \Eprint {http://arxiv.org/abs/1905.03947} {arXiv:1905.03947
  [gr-qc]} \BibitemShut {NoStop}%
\bibitem [{\citenamefont {Cardoso}\ \emph {et~al.}(2018)\citenamefont
  {Cardoso}, \citenamefont {Kimura}, \citenamefont {Maselli},\ and\
  \citenamefont {Senatore}}]{Ca.al.18}%
  \BibitemOpen
  \bibfield  {author} {\bibinfo {author} {\bibfnamefont {V.}~\bibnamefont
  {Cardoso}}, \bibinfo {author} {\bibfnamefont {M.}~\bibnamefont {Kimura}},
  \bibinfo {author} {\bibfnamefont {A.}~\bibnamefont {Maselli}}, \ and\
  \bibinfo {author} {\bibfnamefont {L.}~\bibnamefont {Senatore}},\ }\href@noop
  {} {\bibfield  {journal} {\bibinfo  {journal} {Phys. Rev. Lett.}\ }\textbf
  {\bibinfo {volume} {121}},\ \bibinfo {pages} {251105} (\bibinfo {year}
  {2018})},\ \Eprint {http://arxiv.org/abs/1808.08962} {arXiv:1808.08962
  [gr-qc]} \BibitemShut {NoStop}%
\bibitem [{\citenamefont {Cai}\ and\ \citenamefont {Wang}(2019)}]{CaWa.19}%
  \BibitemOpen
  \bibfield  {author} {\bibinfo {author} {\bibfnamefont {S.}~\bibnamefont
  {Cai}}\ and\ \bibinfo {author} {\bibfnamefont {K.-D.}\ \bibnamefont {Wang}},\
  }\href@noop {} {\  (\bibinfo {year} {2019})},\ \Eprint
  {http://arxiv.org/abs/1906.06850} {arXiv:1906.06850 [hep-th]} \BibitemShut
  {NoStop}%
\bibitem [{\citenamefont {Chakravarti}\ \emph {et~al.}(2019)\citenamefont
  {Chakravarti}, \citenamefont {Chakraborty}, \citenamefont {Bose},\ and\
  \citenamefont {{SenGupta}}}]{Ch.al.19}%
  \BibitemOpen
  \bibfield  {author} {\bibinfo {author} {\bibfnamefont {K.}~\bibnamefont
  {Chakravarti}}, \bibinfo {author} {\bibfnamefont {S.}~\bibnamefont
  {Chakraborty}}, \bibinfo {author} {\bibfnamefont {S.}~\bibnamefont {Bose}}, \
  and\ \bibinfo {author} {\bibfnamefont {S.}~\bibnamefont {{SenGupta}}},\
  }\href@noop {} {\bibfield  {journal} {\bibinfo  {journal} {Phys. Rev. D}\
  }\textbf {\bibinfo {volume} {99}},\ \bibinfo {pages} {024036} (\bibinfo
  {year} {2019})},\ \Eprint {http://arxiv.org/abs/1811.11364} {arXiv:1811.11364
  [gr-qc]} \BibitemShut {NoStop}%
\bibitem [{\citenamefont {Cardoso}\ \emph {et~al.}(2019)\citenamefont
  {Cardoso}, \citenamefont {Gualtieri},\ and\ \citenamefont
  {Moore}}]{Ca.al.19}%
  \BibitemOpen
  \bibfield  {author} {\bibinfo {author} {\bibfnamefont {V.}~\bibnamefont
  {Cardoso}}, \bibinfo {author} {\bibfnamefont {L.}~\bibnamefont {Gualtieri}},
  \ and\ \bibinfo {author} {\bibfnamefont {C.~J.}\ \bibnamefont {Moore}},\
  }\href@noop {} {\bibfield  {journal} {\bibinfo  {journal} {Phys. Rev. D}\
  }\textbf {\bibinfo {volume} {100}},\ \bibinfo {pages} {124037} (\bibinfo
  {year} {2019})},\ \Eprint {http://arxiv.org/abs/1910.09557} {arXiv:1910.09557
  [gr-qc]} \BibitemShut {NoStop}%
\bibitem [{\citenamefont {Tan}(2020)}]{Ta.20}%
  \BibitemOpen
  \bibfield  {author} {\bibinfo {author} {\bibfnamefont {H.~S.}\ \bibnamefont
  {Tan}},\ }\href@noop {} {\bibfield  {journal} {\bibinfo  {journal} {Phys.
  Rev. D}\ }\textbf {\bibinfo {volume} {102}},\ \bibinfo {pages} {044061}
  (\bibinfo {year} {2020})},\ \Eprint {http://arxiv.org/abs/2001.00403}
  {arXiv:2001.00403 [gr-qc]} \BibitemShut {NoStop}%
\bibitem [{\citenamefont {Addazi}\ \emph {et~al.}(2019)\citenamefont {Addazi},
  \citenamefont {Marciano},\ and\ \citenamefont {Yunes}}]{Ad.al.19}%
  \BibitemOpen
  \bibfield  {author} {\bibinfo {author} {\bibfnamefont {A.}~\bibnamefont
  {Addazi}}, \bibinfo {author} {\bibfnamefont {A.}~\bibnamefont {Marciano}}, \
  and\ \bibinfo {author} {\bibfnamefont {N.}~\bibnamefont {Yunes}},\
  }\href@noop {} {\bibfield  {journal} {\bibinfo  {journal} {Phys. Rev. Lett.}\
  }\textbf {\bibinfo {volume} {122}},\ \bibinfo {pages} {081301} (\bibinfo
  {year} {2019})},\ \Eprint {http://arxiv.org/abs/1810.10417} {arXiv:1810.10417
  [gr-qc]} \BibitemShut {NoStop}%
\bibitem [{\citenamefont {Maselli}\ \emph {et~al.}(2019)\citenamefont {Maselli}
  \emph {et~al.}}]{Ma.al.19}%
  \BibitemOpen
  \bibfield  {author} {\bibinfo {author} {\bibfnamefont {A.}~\bibnamefont
  {Maselli}} \emph {et~al.},\ }\href@noop {} {\bibfield  {journal} {\bibinfo
  {journal} {Class. Quant. Grav.}\ }\textbf {\bibinfo {volume} {36}},\ \bibinfo
  {pages} {167001} (\bibinfo {year} {2019})},\ \Eprint
  {http://arxiv.org/abs/1811.03689} {arXiv:1811.03689 [gr-qc]} \BibitemShut
  {NoStop}%
\bibitem [{\citenamefont {Cardoso}\ and\ \citenamefont
  {Duque}(2020)}]{CaDu.20}%
  \BibitemOpen
  \bibfield  {author} {\bibinfo {author} {\bibfnamefont {V.}~\bibnamefont
  {Cardoso}}\ and\ \bibinfo {author} {\bibfnamefont {F.}~\bibnamefont
  {Duque}},\ }\href@noop {} {\bibfield  {journal} {\bibinfo  {journal} {Phys.
  Rev. D}\ }\textbf {\bibinfo {volume} {101}},\ \bibinfo {pages} {064028}
  (\bibinfo {year} {2020})},\ \Eprint {http://arxiv.org/abs/1912.07616}
  {arXiv:1912.07616 [gr-qc]} \BibitemShut {NoStop}%
\bibitem [{\citenamefont {Chirenti}\ \emph {et~al.}(2020)\citenamefont
  {Chirenti}, \citenamefont {Posada},\ and\ \citenamefont {Guedes}}]{Ch.al.20}%
  \BibitemOpen
  \bibfield  {author} {\bibinfo {author} {\bibfnamefont {C.}~\bibnamefont
  {Chirenti}}, \bibinfo {author} {\bibfnamefont {C.}~\bibnamefont {Posada}}, \
  and\ \bibinfo {author} {\bibfnamefont {V.}~\bibnamefont {Guedes}},\
  }\href@noop {} {\bibfield  {journal} {\bibinfo  {journal} {Class. Quantum
  Grav.}\ }\textbf {\bibinfo {volume} {37}},\ \bibinfo {pages} {195017}
  (\bibinfo {year} {2020})},\ \Eprint {http://arxiv.org/abs/2005.10794}
  {arXiv:2005.10794 [gr-qc]} \BibitemShut {NoStop}%
\bibitem [{\citenamefont {Poisson}(2009)}]{Po.09}%
  \BibitemOpen
  \bibfield  {author} {\bibinfo {author} {\bibfnamefont {E.}~\bibnamefont
  {Poisson}},\ }\href@noop {} {\bibfield  {journal} {\bibinfo  {journal} {Phys.
  Rev. D}\ }\textbf {\bibinfo {volume} {80}},\ \bibinfo {pages} {064029}
  (\bibinfo {year} {2009})},\ \Eprint {http://arxiv.org/abs/0907.0874}
  {arXiv:0907.0874 [gr-qc]} \BibitemShut {NoStop}%
\bibitem [{\citenamefont {Penna}(2018)}]{Pe.18}%
  \BibitemOpen
  \bibfield  {author} {\bibinfo {author} {\bibfnamefont {R.~F.}\ \bibnamefont
  {Penna}},\ }\href@noop {} {\  (\bibinfo {year} {2018})},\ \Eprint
  {http://arxiv.org/abs/1812.05643} {arXiv:1812.05643 [gr-qc]} \BibitemShut
  {NoStop}%
\bibitem [{\citenamefont {Binnington}\ and\ \citenamefont
  {Poisson}(2009)}]{BiPo.09}%
  \BibitemOpen
  \bibfield  {author} {\bibinfo {author} {\bibfnamefont {T.}~\bibnamefont
  {Binnington}}\ and\ \bibinfo {author} {\bibfnamefont {E.}~\bibnamefont
  {Poisson}},\ }\href@noop {} {\bibfield  {journal} {\bibinfo  {journal} {Phys.
  Rev. D}\ }\textbf {\bibinfo {volume} {80}},\ \bibinfo {pages} {084018}
  (\bibinfo {year} {2009})},\ \Eprint {http://arxiv.org/abs/0906.1366}
  {arXiv:0906.1366 [gr-qc]} \BibitemShut {NoStop}%
\bibitem [{\citenamefont {Damour}\ and\ \citenamefont
  {Nagar}(2009)}]{DaNa2.09}%
  \BibitemOpen
  \bibfield  {author} {\bibinfo {author} {\bibfnamefont {T.}~\bibnamefont
  {Damour}}\ and\ \bibinfo {author} {\bibfnamefont {A.}~\bibnamefont {Nagar}},\
  }\href@noop {} {\bibfield  {journal} {\bibinfo  {journal} {Phys. Rev. D}\
  }\textbf {\bibinfo {volume} {80}},\ \bibinfo {pages} {084035} (\bibinfo
  {year} {2009})},\ \Eprint {http://arxiv.org/abs/0906.0096} {arXiv:0906.0096
  [gr-qc]} \BibitemShut {NoStop}%
\bibitem [{\citenamefont {Kol}\ and\ \citenamefont {Smolkin}(2012)}]{KoSm.12}%
  \BibitemOpen
  \bibfield  {author} {\bibinfo {author} {\bibfnamefont {B.}~\bibnamefont
  {Kol}}\ and\ \bibinfo {author} {\bibfnamefont {M.}~\bibnamefont {Smolkin}},\
  }\href@noop {} {\bibfield  {journal} {\bibinfo  {journal} {JHEP}\ }\textbf
  {\bibinfo {volume} {1202}},\ \bibinfo {pages} {010} (\bibinfo {year}
  {2012})},\ \Eprint {http://arxiv.org/abs/1110.3764} {arXiv:1110.3764
  [hep-th]} \BibitemShut {NoStop}%
\bibitem [{\citenamefont {Chakrabarti}\ \emph {et~al.}(2013)\citenamefont
  {Chakrabarti}, \citenamefont {Delsate},\ and\ \citenamefont
  {Steinhoff}}]{Ch.al2.13}%
  \BibitemOpen
  \bibfield  {author} {\bibinfo {author} {\bibfnamefont {S.}~\bibnamefont
  {Chakrabarti}}, \bibinfo {author} {\bibfnamefont {T.}~\bibnamefont
  {Delsate}}, \ and\ \bibinfo {author} {\bibfnamefont {J.}~\bibnamefont
  {Steinhoff}},\ }\href@noop {} {\  (\bibinfo {year} {2013})},\ \Eprint
  {http://arxiv.org/abs/1304.2228} {arXiv:1304.2228 [gr-qc]} \BibitemShut
  {NoStop}%
\bibitem [{\citenamefont {G{\"u}rlebeck}(2015)}]{Gu.15}%
  \BibitemOpen
  \bibfield  {author} {\bibinfo {author} {\bibfnamefont {N.}~\bibnamefont
  {G{\"u}rlebeck}},\ }\href@noop {} {\bibfield  {journal} {\bibinfo  {journal}
  {Phys. Rev. Lett.}\ }\textbf {\bibinfo {volume} {114}},\ \bibinfo {pages}
  {151102} (\bibinfo {year} {2015})},\ \Eprint
  {http://arxiv.org/abs/1503.03240} {arXiv:1503.03240 [gr-qc]} \BibitemShut
  {NoStop}%
\bibitem [{\citenamefont {Landry}\ and\ \citenamefont
  {Poisson}(2015)}]{LaPo.15}%
  \BibitemOpen
  \bibfield  {author} {\bibinfo {author} {\bibfnamefont {P.}~\bibnamefont
  {Landry}}\ and\ \bibinfo {author} {\bibfnamefont {E.}~\bibnamefont
  {Poisson}},\ }\href@noop {} {\bibfield  {journal} {\bibinfo  {journal} {Phys.
  Rev. D}\ }\textbf {\bibinfo {volume} {91}},\ \bibinfo {pages} {104018}
  (\bibinfo {year} {2015})},\ \Eprint {http://arxiv.org/abs/1503.07366}
  {arXiv:1503.07366 [gr-qc]} \BibitemShut {NoStop}%
\bibitem [{\citenamefont {Pani}\ \emph
  {et~al.}(2015{\natexlab{a}})\citenamefont {Pani}, \citenamefont {Gualtieri},
  \citenamefont {Maselli},\ and\ \citenamefont {Ferrari}}]{Pa.al.15}%
  \BibitemOpen
  \bibfield  {author} {\bibinfo {author} {\bibfnamefont {P.}~\bibnamefont
  {Pani}}, \bibinfo {author} {\bibfnamefont {L.}~\bibnamefont {Gualtieri}},
  \bibinfo {author} {\bibfnamefont {A.}~\bibnamefont {Maselli}}, \ and\
  \bibinfo {author} {\bibfnamefont {V.}~\bibnamefont {Ferrari}},\ }\href@noop
  {} {\bibfield  {journal} {\bibinfo  {journal} {Phys. Rev. D}\ }\textbf
  {\bibinfo {volume} {92}},\ \bibinfo {pages} {024010} (\bibinfo {year}
  {2015}{\natexlab{a}})},\ \Eprint {http://arxiv.org/abs/1503.07365}
  {arXiv:1503.07365 [gr-qc]} \BibitemShut {NoStop}%
\bibitem [{\citenamefont {{Le Tiec}}\ and\ \citenamefont
  {Casals}(2021)}]{LeCa.21}%
  \BibitemOpen
  \bibfield  {author} {\bibinfo {author} {\bibfnamefont {A.}~\bibnamefont {{Le
  Tiec}}}\ and\ \bibinfo {author} {\bibfnamefont {M.}~\bibnamefont {Casals}},\
  }\href@noop {} {\bibfield  {journal} {\bibinfo  {journal} {Phys. Rev. Lett.}\
  }\textbf {\bibinfo {volume} {126}},\ \bibinfo {pages} {131102} (\bibinfo
  {year} {2021})},\ \Eprint {http://arxiv.org/abs/2007.00214} {arXiv:2007.00214
  [gr-qc]} \BibitemShut {NoStop}%
\bibitem [{\citenamefont {Teukolsky}(1973)}]{Te.73}%
  \BibitemOpen
  \bibfield  {author} {\bibinfo {author} {\bibfnamefont {S.~A.}\ \bibnamefont
  {Teukolsky}},\ }\href@noop {} {\bibfield  {journal} {\bibinfo  {journal}
  {Astrophys. J.}\ }\textbf {\bibinfo {volume} {185}},\ \bibinfo {pages} {635}
  (\bibinfo {year} {1973})}\BibitemShut {NoStop}%
\bibitem [{\citenamefont {Teukolsky}\ and\ \citenamefont
  {Press}(1974)}]{TePr.74}%
  \BibitemOpen
  \bibfield  {author} {\bibinfo {author} {\bibfnamefont {S.~A.}\ \bibnamefont
  {Teukolsky}}\ and\ \bibinfo {author} {\bibfnamefont {W.~H.}\ \bibnamefont
  {Press}},\ }\href@noop {} {\bibfield  {journal} {\bibinfo  {journal}
  {Astrophys. J.}\ }\textbf {\bibinfo {volume} {193}},\ \bibinfo {pages} {443}
  (\bibinfo {year} {1974})}\BibitemShut {NoStop}%
\bibitem [{\citenamefont {Cohen}\ and\ \citenamefont
  {Kegeles}(1974)}]{CoKe.74}%
  \BibitemOpen
  \bibfield  {author} {\bibinfo {author} {\bibfnamefont {J.~M.}\ \bibnamefont
  {Cohen}}\ and\ \bibinfo {author} {\bibfnamefont {L.~S.}\ \bibnamefont
  {Kegeles}},\ }\href@noop {} {\bibfield  {journal} {\bibinfo  {journal} {Phys.
  Rev. D}\ }\textbf {\bibinfo {volume} {10}},\ \bibinfo {pages} {1070}
  (\bibinfo {year} {1974})}\BibitemShut {NoStop}%
\bibitem [{\citenamefont {Chrzanowski}(1975)}]{Ch.75}%
  \BibitemOpen
  \bibfield  {author} {\bibinfo {author} {\bibfnamefont {P.~L.}\ \bibnamefont
  {Chrzanowski}},\ }\href@noop {} {\bibfield  {journal} {\bibinfo  {journal}
  {Phys. Rev. D}\ }\textbf {\bibinfo {volume} {11}},\ \bibinfo {pages} {2042}
  (\bibinfo {year} {1975})}\BibitemShut {NoStop}%
\bibitem [{\citenamefont {Kegeles}\ and\ \citenamefont
  {Cohen}(1979)}]{KeCo.79}%
  \BibitemOpen
  \bibfield  {author} {\bibinfo {author} {\bibfnamefont {L.~S.}\ \bibnamefont
  {Kegeles}}\ and\ \bibinfo {author} {\bibfnamefont {J.~M.}\ \bibnamefont
  {Cohen}},\ }\href@noop {} {\bibfield  {journal} {\bibinfo  {journal} {Phys.
  Rev. D}\ }\textbf {\bibinfo {volume} {19}},\ \bibinfo {pages} {1641}
  (\bibinfo {year} {1979})}\BibitemShut {NoStop}%
\bibitem [{\citenamefont {Chia}(2020)}]{Ch2.20}%
  \BibitemOpen
  \bibfield  {author} {\bibinfo {author} {\bibfnamefont {H.~S.}\ \bibnamefont
  {Chia}},\ }\href@noop {} {\  (\bibinfo {year} {2020})},\ \Eprint
  {http://arxiv.org/abs/2010.07300} {arXiv:2010.07300 [gr-qc]} \BibitemShut
  {NoStop}%
\bibitem [{\citenamefont {Pani}\ \emph
  {et~al.}(2015{\natexlab{b}})\citenamefont {Pani}, \citenamefont {Gualtieri},\
  and\ \citenamefont {Ferrari}}]{Pa.al2.15}%
  \BibitemOpen
  \bibfield  {author} {\bibinfo {author} {\bibfnamefont {P.}~\bibnamefont
  {Pani}}, \bibinfo {author} {\bibfnamefont {L.}~\bibnamefont {Gualtieri}}, \
  and\ \bibinfo {author} {\bibfnamefont {V.}~\bibnamefont {Ferrari}},\
  }\href@noop {} {\bibfield  {journal} {\bibinfo  {journal} {Phys. Rev. D}\
  }\textbf {\bibinfo {volume} {92}},\ \bibinfo {pages} {124003} (\bibinfo
  {year} {2015}{\natexlab{b}})},\ \Eprint {http://arxiv.org/abs/1509.02171}
  {arXiv:1509.02171 [gr-qc]} \BibitemShut {NoStop}%
\bibitem [{\citenamefont {Bardeen}\ and\ \citenamefont
  {Press}(1973)}]{BaPr.73}%
  \BibitemOpen
  \bibfield  {author} {\bibinfo {author} {\bibfnamefont {J.~M.}\ \bibnamefont
  {Bardeen}}\ and\ \bibinfo {author} {\bibfnamefont {W.~H.}\ \bibnamefont
  {Press}},\ }\href@noop {} {\bibfield  {journal} {\bibinfo  {journal} {J.
  Math. Phys.}\ }\textbf {\bibinfo {volume} {14}},\ \bibinfo {pages} {7}
  (\bibinfo {year} {1973})}\BibitemShut {NoStop}%
\bibitem [{\citenamefont {Wald}(1984)}]{Wal}%
  \BibitemOpen
  \bibfield  {author} {\bibinfo {author} {\bibfnamefont {R.~M.}\ \bibnamefont
  {Wald}},\ }\href@noop {} {\emph {\bibinfo {title} {General relativity}}}\
  (\bibinfo  {publisher} {University of Chicago Press},\ \bibinfo {address}
  {Chicago},\ \bibinfo {year} {1984})\BibitemShut {NoStop}%
\bibitem [{\citenamefont {Kinnersley}(1969)}]{Ki.69}%
  \BibitemOpen
  \bibfield  {author} {\bibinfo {author} {\bibfnamefont {W.}~\bibnamefont
  {Kinnersley}},\ }\href@noop {} {\bibfield  {journal} {\bibinfo  {journal} {J.
  Math. Phys.}\ }\textbf {\bibinfo {volume} {10}},\ \bibinfo {pages} {1195}
  (\bibinfo {year} {1969})}\BibitemShut {NoStop}%
\bibitem [{\citenamefont {Newman}\ and\ \citenamefont
  {Penrose}(1962)}]{NePe.62}%
  \BibitemOpen
  \bibfield  {author} {\bibinfo {author} {\bibfnamefont {E.}~\bibnamefont
  {Newman}}\ and\ \bibinfo {author} {\bibfnamefont {R.}~\bibnamefont
  {Penrose}},\ }\href@noop {} {\bibfield  {journal} {\bibinfo  {journal} {J.
  Math. Phys.}\ }\textbf {\bibinfo {volume} {3}},\ \bibinfo {pages} {566}
  (\bibinfo {year} {1962})},\ \bibinfo {note} {\textit{{E}rratum:} J. Math.
  Phys. \textbf{4}, 998 (1963)}\BibitemShut {NoStop}%
\bibitem [{\citenamefont {Wahl}\ \emph {et~al.}(2017)\citenamefont {Wahl},
  \citenamefont {Hubbard},\ and\ \citenamefont {Militzerac}}]{Wa.al.17}%
  \BibitemOpen
  \bibfield  {author} {\bibinfo {author} {\bibfnamefont {S.~A.}\ \bibnamefont
  {Wahl}}, \bibinfo {author} {\bibfnamefont {W.~B.}\ \bibnamefont {Hubbard}}, \
  and\ \bibinfo {author} {\bibfnamefont {B.}~\bibnamefont {Militzerac}},\
  }\href@noop {} {\bibfield  {journal} {\bibinfo  {journal} {Icarus}\ }\textbf
  {\bibinfo {volume} {282}},\ \bibinfo {pages} {183} (\bibinfo {year}
  {2017})},\ \Eprint {http://arxiv.org/abs/1602.07350} {arXiv:1602.07350
  [astro-ph.EP]} \BibitemShut {NoStop}%
\bibitem [{\citenamefont {Blanchet}\ and\ \citenamefont
  {Damour}(1986)}]{BlDa.86}%
  \BibitemOpen
  \bibfield  {author} {\bibinfo {author} {\bibfnamefont {L.}~\bibnamefont
  {Blanchet}}\ and\ \bibinfo {author} {\bibfnamefont {T.}~\bibnamefont
  {Damour}},\ }\href@noop {} {\bibfield  {journal} {\bibinfo  {journal} {Phil.
  Trans. Roy. Soc. Lond. A}\ }\textbf {\bibinfo {volume} {320}},\ \bibinfo
  {pages} {379} (\bibinfo {year} {1986})}\BibitemShut {NoStop}%
\bibitem [{\citenamefont {Gralla}(2018)}]{Gr.18}%
  \BibitemOpen
  \bibfield  {author} {\bibinfo {author} {\bibfnamefont {S.~E.}\ \bibnamefont
  {Gralla}},\ }\href@noop {} {\bibfield  {journal} {\bibinfo  {journal} {Class.
  Quant. Grav.}\ }\textbf {\bibinfo {volume} {35}},\ \bibinfo {pages} {085002}
  (\bibinfo {year} {2018})},\ \Eprint {http://arxiv.org/abs/1710.11096}
  {arXiv:1710.11096 [gr-qc]} \BibitemShut {NoStop}%
\bibitem [{\citenamefont {Fang}\ and\ \citenamefont
  {Lovelace}(2005)}]{FaLo.05}%
  \BibitemOpen
  \bibfield  {author} {\bibinfo {author} {\bibfnamefont {H.}~\bibnamefont
  {Fang}}\ and\ \bibinfo {author} {\bibfnamefont {G.}~\bibnamefont
  {Lovelace}},\ }\href@noop {} {\bibfield  {journal} {\bibinfo  {journal}
  {Phys. Rev. D}\ }\textbf {\bibinfo {volume} {72}},\ \bibinfo {pages} {124016}
  (\bibinfo {year} {2005})},\ \Eprint
  {http://arxiv.org/abs/arXiv:gr-qc/0505156} {arXiv:gr-qc/0505156} \BibitemShut
  {NoStop}%
\bibitem [{\citenamefont {Page}(1976)}]{Pa.76}%
  \BibitemOpen
  \bibfield  {author} {\bibinfo {author} {\bibfnamefont {D.~N.}\ \bibnamefont
  {Page}},\ }\href@noop {} {\bibfield  {journal} {\bibinfo  {journal} {Phys.
  Rev. D}\ }\textbf {\bibinfo {volume} {13}},\ \bibinfo {pages} {198} (\bibinfo
  {year} {1976})}\BibitemShut {NoStop}%
\bibitem [{\citenamefont {Thorne}(1980)}]{Th.80}%
  \BibitemOpen
  \bibfield  {author} {\bibinfo {author} {\bibfnamefont {K.~S.}\ \bibnamefont
  {Thorne}},\ }\href@noop {} {\bibfield  {journal} {\bibinfo  {journal} {Rev.
  Mod. Phys.}\ }\textbf {\bibinfo {volume} {52}},\ \bibinfo {pages} {299}
  (\bibinfo {year} {1980})}\BibitemShut {NoStop}%
\bibitem [{\citenamefont {Simon}\ and\ \citenamefont {Beig}(1983)}]{SiBe.83}%
  \BibitemOpen
  \bibfield  {author} {\bibinfo {author} {\bibfnamefont {W.}~\bibnamefont
  {Simon}}\ and\ \bibinfo {author} {\bibfnamefont {R.}~\bibnamefont {Beig}},\
  }\href@noop {} {\bibfield  {journal} {\bibinfo  {journal} {J. Math. Phys.}\
  }\textbf {\bibinfo {volume} {24}},\ \bibinfo {pages} {1163} (\bibinfo {year}
  {1983})}\BibitemShut {NoStop}%
\bibitem [{\citenamefont {Poisson}(2004)}]{Po2.04}%
  \BibitemOpen
  \bibfield  {author} {\bibinfo {author} {\bibfnamefont {E.}~\bibnamefont
  {Poisson}},\ }\href@noop {} {\bibfield  {journal} {\bibinfo  {journal} {Phys.
  Rev. D}\ }\textbf {\bibinfo {volume} {70}},\ \bibinfo {pages} {084044}
  (\bibinfo {year} {2004})},\ \Eprint
  {http://arxiv.org/abs/arXiv:gr-qc/0407050} {arXiv:gr-qc/0407050} \BibitemShut
  {NoStop}%
\bibitem [{\citenamefont {Poisson}\ and\ \citenamefont
  {Vlasov}(2010)}]{PoVl.10}%
  \BibitemOpen
  \bibfield  {author} {\bibinfo {author} {\bibfnamefont {E.}~\bibnamefont
  {Poisson}}\ and\ \bibinfo {author} {\bibfnamefont {I.}~\bibnamefont
  {Vlasov}},\ }\href@noop {} {\bibfield  {journal} {\bibinfo  {journal} {Phys.
  Rev. D}\ }\textbf {\bibinfo {volume} {81}},\ \bibinfo {pages} {024029}
  (\bibinfo {year} {2010})},\ \Eprint {http://arxiv.org/abs/0910.4311}
  {arXiv:0910.4311 [gr-qc]} \BibitemShut {NoStop}%
\bibitem [{\citenamefont {Chatziioannou}\ \emph {et~al.}(2013)\citenamefont
  {Chatziioannou}, \citenamefont {Poisson},\ and\ \citenamefont
  {Yunes}}]{Ch.al.13}%
  \BibitemOpen
  \bibfield  {author} {\bibinfo {author} {\bibfnamefont {K.}~\bibnamefont
  {Chatziioannou}}, \bibinfo {author} {\bibfnamefont {E.}~\bibnamefont
  {Poisson}}, \ and\ \bibinfo {author} {\bibfnamefont {N.}~\bibnamefont
  {Yunes}},\ }\href@noop {} {\bibfield  {journal} {\bibinfo  {journal} {Phys.
  Rev. D}\ }\textbf {\bibinfo {volume} {87}},\ \bibinfo {pages} {044022}
  (\bibinfo {year} {2013})},\ \Eprint {http://arxiv.org/abs/1211.1686}
  {arXiv:1211.1686 [gr-qc]} \BibitemShut {NoStop}%
\bibitem [{\citenamefont {Poisson}(2015)}]{Po4.15}%
  \BibitemOpen
  \bibfield  {author} {\bibinfo {author} {\bibfnamefont {E.}~\bibnamefont
  {Poisson}},\ }\href@noop {} {\bibfield  {journal} {\bibinfo  {journal} {Phys.
  Rev. D}\ }\textbf {\bibinfo {volume} {91}},\ \bibinfo {pages} {044004}
  (\bibinfo {year} {2015})},\ \Eprint {http://arxiv.org/abs/1411.4711}
  {arXiv:1411.4711 [gr-qc]} \BibitemShut {NoStop}%
\bibitem [{\citenamefont {Poisson}\ \emph {et~al.}(2011)\citenamefont
  {Poisson}, \citenamefont {Pound},\ and\ \citenamefont {Vega}}]{Po.al.11}%
  \BibitemOpen
  \bibfield  {author} {\bibinfo {author} {\bibfnamefont {E.}~\bibnamefont
  {Poisson}}, \bibinfo {author} {\bibfnamefont {A.}~\bibnamefont {Pound}}, \
  and\ \bibinfo {author} {\bibfnamefont {I.}~\bibnamefont {Vega}},\ }\href@noop
  {} {\bibfield  {journal} {\bibinfo  {journal} {Living Rev. Relativity}\
  }\textbf {\bibinfo {volume} {14}},\ \bibinfo {pages} {7} (\bibinfo {year}
  {2011})},\ \Eprint {http://arxiv.org/abs/1102.0529} {arXiv:1102.0529 [gr-qc]}
  \BibitemShut {NoStop}%
\bibitem [{\citenamefont {Zhang}(1986)}]{Zh.86}%
  \BibitemOpen
  \bibfield  {author} {\bibinfo {author} {\bibfnamefont {X.-H.}\ \bibnamefont
  {Zhang}},\ }\href@noop {} {\bibfield  {journal} {\bibinfo  {journal} {Phys.
  Rev. D}\ }\textbf {\bibinfo {volume} {34}},\ \bibinfo {pages} {991} (\bibinfo
  {year} {1986})}\BibitemShut {NoStop}%
\bibitem [{\citenamefont {Pani}(2013)}]{Pa.13}%
  \BibitemOpen
  \bibfield  {author} {\bibinfo {author} {\bibfnamefont {P.}~\bibnamefont
  {Pani}},\ }\href@noop {} {\bibfield  {journal} {\bibinfo  {journal} {Int. J.
  Mod. Phys. A}\ }\textbf {\bibinfo {volume} {28}},\ \bibinfo {pages} {1340018}
  (\bibinfo {year} {2013})},\ \Eprint {http://arxiv.org/abs/1305.6758}
  {arXiv:1305.6758 [gr-qc]} \BibitemShut {NoStop}%
\bibitem [{\citenamefont {Teukolsky}(1974)}]{Te.74}%
  \BibitemOpen
  \bibfield  {author} {\bibinfo {author} {\bibfnamefont {S.~A.}\ \bibnamefont
  {Teukolsky}},\ }\emph {\bibinfo {title} {Perturbations of a rotating black
  hole}},\ \href@noop {} {Ph.D. thesis},\ \bibinfo  {school} {California
  Institute of Technology} (\bibinfo {year} {1974})\BibitemShut {NoStop}%
\bibitem [{\citenamefont {Wald}(1973)}]{Wa.73}%
  \BibitemOpen
  \bibfield  {author} {\bibinfo {author} {\bibfnamefont {R.}~\bibnamefont
  {Wald}},\ }\href@noop {} {\bibfield  {journal} {\bibinfo  {journal} {J. Math.
  Phys.}\ }\textbf {\bibinfo {volume} {14}},\ \bibinfo {pages} {1453} (\bibinfo
  {year} {1973})}\BibitemShut {NoStop}%
\bibitem [{\citenamefont {Merlin}\ \emph {et~al.}(2016)\citenamefont {Merlin},
  \citenamefont {Ori}, \citenamefont {Barack}, \citenamefont {Pound},\ and\
  \citenamefont {{van de Meent}}}]{Me.al.16}%
  \BibitemOpen
  \bibfield  {author} {\bibinfo {author} {\bibfnamefont {C.}~\bibnamefont
  {Merlin}}, \bibinfo {author} {\bibfnamefont {A.}~\bibnamefont {Ori}},
  \bibinfo {author} {\bibfnamefont {L.}~\bibnamefont {Barack}}, \bibinfo
  {author} {\bibfnamefont {A.}~\bibnamefont {Pound}}, \ and\ \bibinfo {author}
  {\bibfnamefont {M.}~\bibnamefont {{van de Meent}}},\ }\href@noop {}
  {\bibfield  {journal} {\bibinfo  {journal} {Phys. Rev. D}\ }\textbf {\bibinfo
  {volume} {94}},\ \bibinfo {pages} {104066} (\bibinfo {year} {2016})},\
  \Eprint {http://arxiv.org/abs/1609.01227} {arXiv:1609.01227 [gr-qc]}
  \BibitemShut {NoStop}%
\bibitem [{\citenamefont {{van de Meent}}(2017)}]{vdM2.17}%
  \BibitemOpen
  \bibfield  {author} {\bibinfo {author} {\bibfnamefont {M.}~\bibnamefont {{van
  de Meent}}},\ }\href@noop {} {\bibfield  {journal} {\bibinfo  {journal}
  {Class. Quant. Grav.}\ }\textbf {\bibinfo {volume} {34}},\ \bibinfo {pages}
  {124003} (\bibinfo {year} {2017})},\ \Eprint
  {http://arxiv.org/abs/1702.00969} {arXiv:1702.00969 [gr-qc]} \BibitemShut
  {NoStop}%
\bibitem [{\citenamefont {Chandrasekhar}(1983)}]{Cha}%
  \BibitemOpen
  \bibfield  {author} {\bibinfo {author} {\bibfnamefont {S.}~\bibnamefont
  {Chandrasekhar}},\ }\href@noop {} {\emph {\bibinfo {title} {The mathematical
  theory of black holes}}}\ (\bibinfo  {publisher} {Oxford University Press},\
  \bibinfo {address} {Oxford},\ \bibinfo {year} {1983})\BibitemShut {NoStop}%
\bibitem [{\citenamefont {Berti}\ \emph {et~al.}(2006)\citenamefont {Berti},
  \citenamefont {Cardoso},\ and\ \citenamefont {Casals}}]{Be.al2.06}%
  \BibitemOpen
  \bibfield  {author} {\bibinfo {author} {\bibfnamefont {E.}~\bibnamefont
  {Berti}}, \bibinfo {author} {\bibfnamefont {V.}~\bibnamefont {Cardoso}}, \
  and\ \bibinfo {author} {\bibfnamefont {M.}~\bibnamefont {Casals}},\
  }\href@noop {} {\bibfield  {journal} {\bibinfo  {journal} {Phys. Rev. D}\
  }\textbf {\bibinfo {volume} {73}},\ \bibinfo {pages} {024013} (\bibinfo
  {year} {2006})},\ \bibinfo {note} {\textit{{E}rratum:} Phys. Rev. D
  \textbf{73}, 109902 (2006)},\ \Eprint
  {http://arxiv.org/abs/arXiv:gr-qc/0511111} {arXiv:gr-qc/0511111} \BibitemShut
  {NoStop}%
\bibitem [{\citenamefont {Yunes}\ and\ \citenamefont
  {Gonz{\'a}lez}(2006)}]{YuGo.06}%
  \BibitemOpen
  \bibfield  {author} {\bibinfo {author} {\bibfnamefont {N.}~\bibnamefont
  {Yunes}}\ and\ \bibinfo {author} {\bibfnamefont {J.~A.}\ \bibnamefont
  {Gonz{\'a}lez}},\ }\href@noop {} {\bibfield  {journal} {\bibinfo  {journal}
  {Phys. Rev. D}\ }\textbf {\bibinfo {volume} {73}},\ \bibinfo {pages} {024010}
  (\bibinfo {year} {2006})},\ \bibinfo {note} {\textit{{E}rratum:} Phys. Rev. D
  \textbf{89}, 089902 (2014)},\ \Eprint
  {http://arxiv.org/abs/arXiv:gr-qc/0510076} {arXiv:gr-qc/0510076} \BibitemShut
  {NoStop}%
\bibitem [{\citenamefont {Hawking}\ and\ \citenamefont
  {Hartle}(1972)}]{HaHa.72}%
  \BibitemOpen
  \bibfield  {author} {\bibinfo {author} {\bibfnamefont {S.~W.}\ \bibnamefont
  {Hawking}}\ and\ \bibinfo {author} {\bibfnamefont {J.~B.}\ \bibnamefont
  {Hartle}},\ }\href@noop {} {\bibfield  {journal} {\bibinfo  {journal}
  {Commun. Math. Phys.}\ }\textbf {\bibinfo {volume} {27}},\ \bibinfo {pages}
  {283} (\bibinfo {year} {1972})}\BibitemShut {NoStop}%
\bibitem [{{\relax DLMF}()}]{NIST:DLMF}%
  \BibitemOpen
  {\relax DLMF},\ \href {http://dlmf.nist.gov/} {\enquote {\bibinfo {title}
  {{\it NIST Digital Library of Mathematical Functions}},}\ }\bibinfo
  {howpublished} {http://dlmf.nist.gov/, Release 1.1.1 of 2021-03-15},\
  \bibinfo {note} {{F.~W.~J.} Olver, A.~B. {Olde Daalhuis}, D.~W. Lozier, B.~I.
  Schneider, R.~F. Boisvert, C.~W. Clark, B.~R. Miller, B.~V. Saunders, H.~S.
  Cohl, and M.~A. McClain, eds.}\BibitemShut {Stop}%
\bibitem [{\citenamefont {Chandrasekhar}(1975)}]{Ch2.75}%
  \BibitemOpen
  \bibfield  {author} {\bibinfo {author} {\bibfnamefont {S.}~\bibnamefont
  {Chandrasekhar}},\ }\href@noop {} {\bibfield  {journal} {\bibinfo  {journal}
  {Proc. R. Soc. Lond. A}\ }\textbf {\bibinfo {volume} {343}},\ \bibinfo
  {pages} {289} (\bibinfo {year} {1975})}\BibitemShut {NoStop}%
\bibitem [{\citenamefont {Regge}\ and\ \citenamefont
  {Wheeler}(1957)}]{ReWh.57}%
  \BibitemOpen
  \bibfield  {author} {\bibinfo {author} {\bibfnamefont {T.}~\bibnamefont
  {Regge}}\ and\ \bibinfo {author} {\bibfnamefont {J.~A.}\ \bibnamefont
  {Wheeler}},\ }\href@noop {} {\bibfield  {journal} {\bibinfo  {journal} {Phys.
  Rev.}\ }\textbf {\bibinfo {volume} {108}},\ \bibinfo {pages} {1063} (\bibinfo
  {year} {1957})}\BibitemShut {NoStop}%
\bibitem [{\citenamefont {Zerilli}(1970)}]{Ze3.70}%
  \BibitemOpen
  \bibfield  {author} {\bibinfo {author} {\bibfnamefont {F.~J.}\ \bibnamefont
  {Zerilli}},\ }\href@noop {} {\bibfield  {journal} {\bibinfo  {journal} {Phys.
  Rev. Lett.}\ }\textbf {\bibinfo {volume} {24}},\ \bibinfo {pages} {737}
  (\bibinfo {year} {1970})}\BibitemShut {NoStop}%
\bibitem [{\citenamefont {Cunningham}\ \emph {et~al.}(1978)\citenamefont
  {Cunningham}, \citenamefont {Price},\ and\ \citenamefont
  {Moncrief}}]{Cu.al.78}%
  \BibitemOpen
  \bibfield  {author} {\bibinfo {author} {\bibfnamefont {C.~T.}\ \bibnamefont
  {Cunningham}}, \bibinfo {author} {\bibfnamefont {R.~H.}\ \bibnamefont
  {Price}}, \ and\ \bibinfo {author} {\bibfnamefont {V.}~\bibnamefont
  {Moncrief}},\ }\href@noop {} {\bibfield  {journal} {\bibinfo  {journal}
  {Astrophys. J.}\ }\textbf {\bibinfo {volume} {224}},\ \bibinfo {pages} {643}
  (\bibinfo {year} {1978})}\BibitemShut {NoStop}%
\bibitem [{\citenamefont {Stewart}(1979)}]{St.79}%
  \BibitemOpen
  \bibfield  {author} {\bibinfo {author} {\bibfnamefont {J.~M.}\ \bibnamefont
  {Stewart}},\ }\href@noop {} {\bibfield  {journal} {\bibinfo  {journal} {Proc.
  R. Soc. Lond. A}\ }\textbf {\bibinfo {volume} {367}},\ \bibinfo {pages} {527}
  (\bibinfo {year} {1979})}\BibitemShut {NoStop}%
\bibitem [{\citenamefont {Wald}(1978)}]{Wa.78}%
  \BibitemOpen
  \bibfield  {author} {\bibinfo {author} {\bibfnamefont {R.~M.}\ \bibnamefont
  {Wald}},\ }\href@noop {} {\bibfield  {journal} {\bibinfo  {journal} {Phys.
  Rev. Lett.}\ }\textbf {\bibinfo {volume} {41}},\ \bibinfo {pages} {203}
  (\bibinfo {year} {1978})}\BibitemShut {NoStop}%
\bibitem [{\citenamefont {Keidl}\ \emph {et~al.}(2010)\citenamefont {Keidl},
  \citenamefont {Shah}, \citenamefont {Friedman}, \citenamefont {Kim},\ and\
  \citenamefont {Price}}]{Ke.al2.10}%
  \BibitemOpen
  \bibfield  {author} {\bibinfo {author} {\bibfnamefont {T.~S.}\ \bibnamefont
  {Keidl}}, \bibinfo {author} {\bibfnamefont {A.~G.}\ \bibnamefont {Shah}},
  \bibinfo {author} {\bibfnamefont {J.~L.}\ \bibnamefont {Friedman}}, \bibinfo
  {author} {\bibfnamefont {D.-H.}\ \bibnamefont {Kim}}, \ and\ \bibinfo
  {author} {\bibfnamefont {L.~R.}\ \bibnamefont {Price}},\ }\href@noop {}
  {\bibfield  {journal} {\bibinfo  {journal} {Phys. Rev. D}\ }\textbf {\bibinfo
  {volume} {82}},\ \bibinfo {pages} {124012} (\bibinfo {year} {2010})},\
  \Eprint {http://arxiv.org/abs/1004.2276} {arXiv:1004.2276 [gr-qc]}
  \BibitemShut {NoStop}%
\bibitem [{\citenamefont {{van de Meent}}\ and\ \citenamefont
  {Shah}(2015)}]{vdMSh.15}%
  \BibitemOpen
  \bibfield  {author} {\bibinfo {author} {\bibfnamefont {M.}~\bibnamefont {{van
  de Meent}}}\ and\ \bibinfo {author} {\bibfnamefont {A.~G.}\ \bibnamefont
  {Shah}},\ }\href@noop {} {\bibfield  {journal} {\bibinfo  {journal} {Phys.
  Rev. D}\ }\textbf {\bibinfo {volume} {92}},\ \bibinfo {pages} {064025}
  (\bibinfo {year} {2015})},\ \Eprint {http://arxiv.org/abs/1506.04755}
  {arXiv:1506.04755 [gr-qc]} \BibitemShut {NoStop}%
\bibitem [{\citenamefont {Lousto}\ and\ \citenamefont
  {Whiting}(2002)}]{LoWh.02}%
  \BibitemOpen
  \bibfield  {author} {\bibinfo {author} {\bibfnamefont {C.}~\bibnamefont
  {Lousto}}\ and\ \bibinfo {author} {\bibfnamefont {B.}~\bibnamefont
  {Whiting}},\ }\href@noop {} {\bibfield  {journal} {\bibinfo  {journal} {Phys.
  Rev. D}\ }\textbf {\bibinfo {volume} {66}},\ \bibinfo {pages} {024026}
  (\bibinfo {year} {2002})},\ \Eprint
  {http://arxiv.org/abs/arXiv:gr-qc/0203061} {arXiv:gr-qc/0203061} \BibitemShut
  {NoStop}%
\bibitem [{\citenamefont {Ori}(2003)}]{Or.03}%
  \BibitemOpen
  \bibfield  {author} {\bibinfo {author} {\bibfnamefont {A.}~\bibnamefont
  {Ori}},\ }\href@noop {} {\bibfield  {journal} {\bibinfo  {journal} {Phys.
  Rev. D}\ }\textbf {\bibinfo {volume} {67}},\ \bibinfo {pages} {124010}
  (\bibinfo {year} {2003})},\ \Eprint
  {http://arxiv.org/abs/arXiv:gr-qc/0207045} {arXiv:gr-qc/0207045} \BibitemShut
  {NoStop}%
\bibitem [{\citenamefont {Price}\ \emph {et~al.}(2007)\citenamefont {Price},
  \citenamefont {Shankar},\ and\ \citenamefont {Whiting}}]{Pr.al.07}%
  \BibitemOpen
  \bibfield  {author} {\bibinfo {author} {\bibfnamefont {L.~R.}\ \bibnamefont
  {Price}}, \bibinfo {author} {\bibfnamefont {K.}~\bibnamefont {Shankar}}, \
  and\ \bibinfo {author} {\bibfnamefont {B.~F.}\ \bibnamefont {Whiting}},\
  }\href@noop {} {\bibfield  {journal} {\bibinfo  {journal} {Class. Quant.
  Grav.}\ }\textbf {\bibinfo {volume} {24}},\ \bibinfo {pages} {2367} (\bibinfo
  {year} {2007})},\ \Eprint {http://arxiv.org/abs/arXiv:gr-qc/0611070}
  {arXiv:gr-qc/0611070} \BibitemShut {NoStop}%
\bibitem [{\citenamefont {Boyer}\ and\ \citenamefont
  {Lindquist}(1967)}]{BoLi.67}%
  \BibitemOpen
  \bibfield  {author} {\bibinfo {author} {\bibfnamefont {R.~H.}\ \bibnamefont
  {Boyer}}\ and\ \bibinfo {author} {\bibfnamefont {R.~W.}\ \bibnamefont
  {Lindquist}},\ }\href@noop {} {\bibfield  {journal} {\bibinfo  {journal} {J.
  Math. Phys.}\ }\textbf {\bibinfo {volume} {8}},\ \bibinfo {pages} {265}
  (\bibinfo {year} {1967})}\BibitemShut {NoStop}%
\bibitem [{\citenamefont {Olver}(1974)}]{Olv}%
  \BibitemOpen
  \bibfield  {author} {\bibinfo {author} {\bibfnamefont {F.~W.~J.}\
  \bibnamefont {Olver}},\ }\href@noop {} {\emph {\bibinfo {title} {Asymptotics
  and special functions}}}\ (\bibinfo  {publisher} {Academic Press},\ \bibinfo
  {address} {New York},\ \bibinfo {year} {1974})\BibitemShut {NoStop}%
\bibitem [{\citenamefont {Ernst}(1968)}]{Er.68}%
  \BibitemOpen
  \bibfield  {author} {\bibinfo {author} {\bibfnamefont {F.~J.}\ \bibnamefont
  {Ernst}},\ }\href@noop {} {\bibfield  {journal} {\bibinfo  {journal} {Phys.
  Rev.}\ }\textbf {\bibinfo {volume} {167}},\ \bibinfo {pages} {1175} (\bibinfo
  {year} {1968})}\BibitemShut {NoStop}%
\bibitem [{\citenamefont {Fodor}\ \emph {et~al.}(1989)\citenamefont {Fodor},
  \citenamefont {Hoenselaers},\ and\ \citenamefont {Perj{\'e}s}}]{Fo.al.89}%
  \BibitemOpen
  \bibfield  {author} {\bibinfo {author} {\bibfnamefont {G.}~\bibnamefont
  {Fodor}}, \bibinfo {author} {\bibfnamefont {C.}~\bibnamefont {Hoenselaers}},
  \ and\ \bibinfo {author} {\bibfnamefont {Z.}~\bibnamefont {Perj{\'e}s}},\
  }\href@noop {} {\bibfield  {journal} {\bibinfo  {journal} {J. Math. Phys.}\
  }\textbf {\bibinfo {volume} {30}},\ \bibinfo {pages} {2252} (\bibinfo {year}
  {1989})}\BibitemShut {NoStop}%
\bibitem [{\citenamefont {Thorne}\ and\ \citenamefont
  {Hartle}(1985)}]{ThHa.85}%
  \BibitemOpen
  \bibfield  {author} {\bibinfo {author} {\bibfnamefont {K.~S.}\ \bibnamefont
  {Thorne}}\ and\ \bibinfo {author} {\bibfnamefont {J.~B.}\ \bibnamefont
  {Hartle}},\ }\href@noop {} {\bibfield  {journal} {\bibinfo  {journal} {Phys.
  Rev. D}\ }\textbf {\bibinfo {volume} {31}},\ \bibinfo {pages} {1815}
  (\bibinfo {year} {1985})}\BibitemShut {NoStop}%
\bibitem [{\citenamefont {G{\"u}rsel}(1983)}]{Gu.83}%
  \BibitemOpen
  \bibfield  {author} {\bibinfo {author} {\bibfnamefont {Y.}~\bibnamefont
  {G{\"u}rsel}},\ }\href@noop {} {\bibfield  {journal} {\bibinfo  {journal}
  {Gen. Rel. Grav.}\ }\textbf {\bibinfo {volume} {15}},\ \bibinfo {pages} {737}
  (\bibinfo {year} {1983})}\BibitemShut {NoStop}%
\bibitem [{\citenamefont {Buonanno}\ and\ \citenamefont
  {Sathyaprakash}(2015)}]{BuSa.15}%
  \BibitemOpen
  \bibfield  {author} {\bibinfo {author} {\bibfnamefont {A.}~\bibnamefont
  {Buonanno}}\ and\ \bibinfo {author} {\bibfnamefont {B.~S.}\ \bibnamefont
  {Sathyaprakash}},\ }in\ \href@noop {} {\emph {\bibinfo {booktitle} {General
  relativity and gravitation: A centennial perspective}}},\ \bibinfo {editor}
  {edited by\ \bibinfo {editor} {\bibfnamefont {A.}~\bibnamefont {Ashtekar}},
  \bibinfo {editor} {\bibfnamefont {B.~K.}\ \bibnamefont {Berger}}, \bibinfo
  {editor} {\bibfnamefont {J.}~\bibnamefont {Isenberg}}, \ and\ \bibinfo
  {editor} {\bibfnamefont {M.}~\bibnamefont {{MacCallum}}}}\ (\bibinfo
  {publisher} {Cambridge University Press},\ \bibinfo {address} {Cambridge},\
  \bibinfo {year} {2015})\ p.\ \bibinfo {pages} {287},\ \Eprint
  {http://arxiv.org/abs/1410.7832} {arXiv:1410.7832 [gr-qc]} \BibitemShut
  {NoStop}%
\bibitem [{\citenamefont {Blanchet}(2014)}]{Bl.14}%
  \BibitemOpen
  \bibfield  {author} {\bibinfo {author} {\bibfnamefont {L.}~\bibnamefont
  {Blanchet}},\ }\href@noop {} {\bibfield  {journal} {\bibinfo  {journal}
  {Living Rev. Relativity}\ }\textbf {\bibinfo {volume} {17}},\ \bibinfo
  {pages} {2} (\bibinfo {year} {2014})},\ \Eprint
  {http://arxiv.org/abs/1310.1528} {arXiv:1310.1528 [gr-qc]} \BibitemShut
  {NoStop}%
\bibitem [{\citenamefont {Damour}\ and\ \citenamefont {Nagar}(2016)}]{DaNa}%
  \BibitemOpen
  \bibfield  {author} {\bibinfo {author} {\bibfnamefont {T.}~\bibnamefont
  {Damour}}\ and\ \bibinfo {author} {\bibfnamefont {A.}~\bibnamefont {Nagar}},\
  }\href@noop {} {\emph {\bibinfo {title} {The effective-one-body approach to
  the general relativistic two body problem}}},\ \bibinfo {series} {Lecture
  Notes in Physics}, Vol.\ \bibinfo {volume} {905}\ (\bibinfo  {publisher}
  {Springer},\ \bibinfo {address} {New York},\ \bibinfo {year}
  {2016})\BibitemShut {NoStop}%
\bibitem [{\citenamefont {Steinhoff}\ and\ \citenamefont
  {Puetzfeld}(2010)}]{StPu.10}%
  \BibitemOpen
  \bibfield  {author} {\bibinfo {author} {\bibfnamefont {J.}~\bibnamefont
  {Steinhoff}}\ and\ \bibinfo {author} {\bibfnamefont {D.}~\bibnamefont
  {Puetzfeld}},\ }\href@noop {} {\bibfield  {journal} {\bibinfo  {journal}
  {Phys. Rev. D}\ }\textbf {\bibinfo {volume} {81}},\ \bibinfo {pages} {044019}
  (\bibinfo {year} {2010})},\ \Eprint {http://arxiv.org/abs/0909.3756}
  {arXiv:0909.3756 [gr-qc]} \BibitemShut {NoStop}%
\bibitem [{\citenamefont {Ramond}\ and\ \citenamefont {{Le
  Tiec}}(2021)}]{RaLe.21}%
  \BibitemOpen
  \bibfield  {author} {\bibinfo {author} {\bibfnamefont {P.}~\bibnamefont
  {Ramond}}\ and\ \bibinfo {author} {\bibfnamefont {A.}~\bibnamefont {{Le
  Tiec}}},\ }\href@noop {} {\bibfield  {journal} {\bibinfo  {journal} {accepted
  in Class. Quant. Grav.}\ } (\bibinfo {year} {2021})},\ \Eprint
  {http://arxiv.org/abs/2005.00602} {arXiv:2005.00602 [gr-qc]} \BibitemShut
  {NoStop}%
\bibitem [{\citenamefont {Porto}(2016)}]{Po.16}%
  \BibitemOpen
  \bibfield  {author} {\bibinfo {author} {\bibfnamefont {R.}~\bibnamefont
  {Porto}},\ }\href@noop {} {\bibfield  {journal} {\bibinfo  {journal} {Phys.
  Rept.}\ }\textbf {\bibinfo {volume} {633}},\ \bibinfo {pages} {1} (\bibinfo
  {year} {2016})},\ \Eprint {http://arxiv.org/abs/1601.04914} {arXiv:1601.04914
  [gr-qc]} \BibitemShut {NoStop}%
\bibitem [{\citenamefont {Levi}(2020)}]{Le.20}%
  \BibitemOpen
  \bibfield  {author} {\bibinfo {author} {\bibfnamefont {M.}~\bibnamefont
  {Levi}},\ }\href@noop {} {\bibfield  {journal} {\bibinfo  {journal} {Rep.
  Prog. Phys.}\ }\textbf {\bibinfo {volume} {83}},\ \bibinfo {pages} {075901}
  (\bibinfo {year} {2020})},\ \Eprint {http://arxiv.org/abs/1807.01699}
  {arXiv:1807.01699 [hep-th]} \BibitemShut {NoStop}%
\bibitem [{\citenamefont {Alvi}(2001)}]{Al.01}%
  \BibitemOpen
  \bibfield  {author} {\bibinfo {author} {\bibfnamefont {K.}~\bibnamefont
  {Alvi}},\ }\href@noop {} {\bibfield  {journal} {\bibinfo  {journal} {Phys.
  Rev. D}\ }\textbf {\bibinfo {volume} {64}},\ \bibinfo {pages} {104020}
  (\bibinfo {year} {2001})},\ \Eprint
  {http://arxiv.org/abs/arXiv:gr-qc/0107080} {arXiv:gr-qc/0107080} \BibitemShut
  {NoStop}%
\bibitem [{\citenamefont {Porto}(2008)}]{Po.08}%
  \BibitemOpen
  \bibfield  {author} {\bibinfo {author} {\bibfnamefont {R.~A.}\ \bibnamefont
  {Porto}},\ }\href@noop {} {\bibfield  {journal} {\bibinfo  {journal} {Phys.
  Rev. D}\ }\textbf {\bibinfo {volume} {77}},\ \bibinfo {pages} {064026}
  (\bibinfo {year} {2008})},\ \Eprint {http://arxiv.org/abs/0710.5150}
  {arXiv:0710.5150 [gr-qc]} \BibitemShut {NoStop}%
\bibitem [{\citenamefont {Damour}\ and\ \citenamefont
  {Lecian}(2009)}]{DaLe.09}%
  \BibitemOpen
  \bibfield  {author} {\bibinfo {author} {\bibfnamefont {T.}~\bibnamefont
  {Damour}}\ and\ \bibinfo {author} {\bibfnamefont {O.~M.}\ \bibnamefont
  {Lecian}},\ }\href@noop {} {\bibfield  {journal} {\bibinfo  {journal} {Phys.
  Rev. D}\ }\textbf {\bibinfo {volume} {80}},\ \bibinfo {pages} {044017}
  (\bibinfo {year} {2009})},\ \Eprint {http://arxiv.org/abs/0906.3003}
  {arXiv:0906.3003 [gr-qc]} \BibitemShut {NoStop}%
\bibitem [{\citenamefont {Landry}\ and\ \citenamefont
  {Poisson}(2014)}]{LaPo.14}%
  \BibitemOpen
  \bibfield  {author} {\bibinfo {author} {\bibfnamefont {P.}~\bibnamefont
  {Landry}}\ and\ \bibinfo {author} {\bibfnamefont {E.}~\bibnamefont
  {Poisson}},\ }\href@noop {} {\bibfield  {journal} {\bibinfo  {journal} {Phys.
  Rev. D}\ }\textbf {\bibinfo {volume} {89}},\ \bibinfo {pages} {124011}
  (\bibinfo {year} {2014})},\ \Eprint {http://arxiv.org/abs/1404.6798}
  {arXiv:1404.6798 [gr-qc]} \BibitemShut {NoStop}%
\bibitem [{\citenamefont {Vega}\ \emph {et~al.}(2011)\citenamefont {Vega},
  \citenamefont {Poisson},\ and\ \citenamefont {Massey}}]{Ve.al.11}%
  \BibitemOpen
  \bibfield  {author} {\bibinfo {author} {\bibfnamefont {I.}~\bibnamefont
  {Vega}}, \bibinfo {author} {\bibfnamefont {E.}~\bibnamefont {Poisson}}, \
  and\ \bibinfo {author} {\bibfnamefont {R.}~\bibnamefont {Massey}},\
  }\href@noop {} {\bibfield  {journal} {\bibinfo  {journal} {Class. Quant.
  Grav.}\ }\textbf {\bibinfo {volume} {28}},\ \bibinfo {pages} {175006}
  (\bibinfo {year} {2011})},\ \Eprint {http://arxiv.org/abs/1106.0510}
  {arXiv:1106.0510 [gr-qc]} \BibitemShut {NoStop}%
\bibitem [{\citenamefont {O'Sullivan}\ and\ \citenamefont
  {Hughes}(2014)}]{OsHu.14}%
  \BibitemOpen
  \bibfield  {author} {\bibinfo {author} {\bibfnamefont {S.}~\bibnamefont
  {O'Sullivan}}\ and\ \bibinfo {author} {\bibfnamefont {S.~A.}\ \bibnamefont
  {Hughes}},\ }\href@noop {} {\bibfield  {journal} {\bibinfo  {journal} {Phys.
  Rev. D}\ }\textbf {\bibinfo {volume} {90}},\ \bibinfo {pages} {124039}
  (\bibinfo {year} {2014})},\ \bibinfo {note} {\textit{{E}rratum:} Phys. Rev. D
  \textbf{91}, 109901 (2015)},\ \Eprint {http://arxiv.org/abs/1407.6983}
  {arXiv:1407.6983 [gr-qc]} \BibitemShut {NoStop}%
\bibitem [{\citenamefont {O'Sullivan}\ and\ \citenamefont
  {Hughes}(2016)}]{OsHu.16}%
  \BibitemOpen
  \bibfield  {author} {\bibinfo {author} {\bibfnamefont {S.}~\bibnamefont
  {O'Sullivan}}\ and\ \bibinfo {author} {\bibfnamefont {S.~A.}\ \bibnamefont
  {Hughes}},\ }\href@noop {} {\bibfield  {journal} {\bibinfo  {journal} {Phys.
  Rev. D}\ }\textbf {\bibinfo {volume} {94}},\ \bibinfo {pages} {044057}
  (\bibinfo {year} {2016})},\ \Eprint {http://arxiv.org/abs/1505.03809}
  {arXiv:1505.03809 [gr-qc]} \BibitemShut {NoStop}%
\bibitem [{\citenamefont {Penna}\ \emph {et~al.}(2017)\citenamefont {Penna},
  \citenamefont {Hughes},\ and\ \citenamefont {O'Sullivan}}]{Pe.al.17}%
  \BibitemOpen
  \bibfield  {author} {\bibinfo {author} {\bibfnamefont {R.~F.}\ \bibnamefont
  {Penna}}, \bibinfo {author} {\bibfnamefont {S.~A.}\ \bibnamefont {Hughes}}, \
  and\ \bibinfo {author} {\bibfnamefont {S.}~\bibnamefont {O'Sullivan}},\
  }\href@noop {} {\bibfield  {journal} {\bibinfo  {journal} {Phys. Rev. D}\
  }\textbf {\bibinfo {volume} {96}},\ \bibinfo {pages} {064030} (\bibinfo
  {year} {2017})},\ \Eprint {http://arxiv.org/abs/1704.05471} {arXiv:1704.05471
  [gr-qc]} \BibitemShut {NoStop}%
\bibitem [{\citenamefont {Goldberger}\ \emph {et~al.}(2020)\citenamefont
  {Goldberger}, \citenamefont {Li},\ and\ \citenamefont
  {Rothstein}}]{Go.al.20}%
  \BibitemOpen
  \bibfield  {author} {\bibinfo {author} {\bibfnamefont {W.~D.}\ \bibnamefont
  {Goldberger}}, \bibinfo {author} {\bibfnamefont {J.}~\bibnamefont {Li}}, \
  and\ \bibinfo {author} {\bibfnamefont {I.~Z.}\ \bibnamefont {Rothstein}},\
  }\href@noop {} {\  (\bibinfo {year} {2020})},\ \Eprint
  {http://arxiv.org/abs/2012.14869} {arXiv:2012.14869 [gr-qc]} \BibitemShut
  {NoStop}%
\bibitem [{\citenamefont {Charalambous}\ \emph {et~al.}(2021)\citenamefont
  {Charalambous}, \citenamefont {Dubovsky},\ and\ \citenamefont
  {Ivanov}}]{Ch.al.21}%
  \BibitemOpen
  \bibfield  {author} {\bibinfo {author} {\bibfnamefont {P.}~\bibnamefont
  {Charalambous}}, \bibinfo {author} {\bibfnamefont {S.}~\bibnamefont
  {Dubovsky}}, \ and\ \bibinfo {author} {\bibfnamefont {M.~M.}\ \bibnamefont
  {Ivanov}},\ }\href@noop {} {\  (\bibinfo {year} {2021})},\ \Eprint
  {http://arxiv.org/abs/2102.08917} {arXiv:2102.08917 [gr-qc]} \BibitemShut
  {NoStop}%
\bibitem [{\citenamefont {Ogilvie}(2014)}]{Og.14}%
  \BibitemOpen
  \bibfield  {author} {\bibinfo {author} {\bibfnamefont {G.~I.}\ \bibnamefont
  {Ogilvie}},\ }\href@noop {} {\bibfield  {journal} {\bibinfo  {journal} {Annu.
  Rev. Astron. Astrophys.}\ }\textbf {\bibinfo {volume} {52}},\ \bibinfo
  {pages} {171} (\bibinfo {year} {2014})},\ \Eprint
  {http://arxiv.org/abs/1406.2207} {arXiv:1406.2207 [astro-ph.SR]} \BibitemShut
  {NoStop}%
\bibitem [{\citenamefont {Newman}\ and\ \citenamefont
  {Penrose}(1966)}]{NePe.66}%
  \BibitemOpen
  \bibfield  {author} {\bibinfo {author} {\bibfnamefont {E.~T.}\ \bibnamefont
  {Newman}}\ and\ \bibinfo {author} {\bibfnamefont {R.}~\bibnamefont
  {Penrose}},\ }\href@noop {} {\bibfield  {journal} {\bibinfo  {journal} {J.
  Math. Phys.}\ }\textbf {\bibinfo {volume} {7}},\ \bibinfo {pages} {863}
  (\bibinfo {year} {1966})}\BibitemShut {NoStop}%
\end{thebibliography}%

\end{document}